\documentclass[longauth]{aa}  

\usepackage{graphicx}
\usepackage{txfonts}
\usepackage{amsmath}
\usepackage{multirow}
\usepackage{xcolor}

\newcommand{\lya}{Ly$\alpha$}
\newcommand{\civ}{\ion{C}{IV}}
\newcommand{\nv}{\ion{N}{V}}
\newcommand{\siiv}{\ion{Si}{IV}}
\newcommand{\pv}{\ion{P}{V}}

\begin{document}

   \title{The WISSH quasars project}

\subtitle{XIII. A multi-epoch study of ultra-fast broad absorption line outflows at cosmic noon}

    \author{A. Deconto-Machado\inst{1,}\thanks{Corresponding author: alice.deconto@inaf.it}
          \and
          G. Vietri\inst{1}
          \and
          E. Piconcelli\inst{2}
          \and
          S. Bisogni\inst{1}
          \and
          T. Misawa\inst{3}
          \and
          A. Gargiulo\inst{1}
          \and
          G. Lanzuisi\inst{4}
          \and
          A. Travascio\inst{5,6}
          \and
          M. Bischetti\inst{6,7}
          \and
          L. Zappacosta\inst{2}
          \and
          E. Bertola\inst{8}
          \and
          G. Cresci\inst{8}
          \and
          F. la Franca\inst{9}
          \and
          M. Gaspari\inst{10}
          \and
          F. Salvestrini\inst{6,11}
          \and
          C. Vignali\inst{12,4}       
          \and
          A. Luminari\inst{13,2}
          \and
          E. Marini\inst{2}
          \and
          F. Tombesi\inst{14,2,15}
          \and
          A. Bongiorno\inst{2}
          \and
          S. Carniani\inst{16}
          \and
          C. Feruglio\inst{6,11}
          \and
          F. Fiore\inst{6,11}
          \and
          E. Glikman\inst{17}
          \and
          V. Testa\inst{2}         
                    }

   \institute{INAF - Istituto di Astrofisica Spaziale e Fisica Cosmica Milano, Via
A. Corti 12, 20133 Milan, Italy 
        \and
   INAF - Osservatorio Astronomico di Roma, Via Frascati 33, I-00040 Monte Porzio Catone, Italy
         \and
   Center for General Education, Shinshu University, 3-1-1 Asahi, Matsumoto, Nagano 390-8621, Japan
         \and
   INAF - Osservatorio di Astrofisica e Scienza dello Spazio di Bologna, Via Gobetti, 93/3, 40129 Bologna, Italy
         \and
   Dipartimento di Fisica ``G. Occhialini'', Università degli Studi di Milano-Bicocca, Piazza della Scienza 3, I-20126 Milano, Italy
        \and
   INAF - Osservatorio Astronomico di Trieste, Via G. B. Tiepolo 11, I-34131 Trieste, Italy
        \and 
   Dipartimento di Fisica "Enrico Fermi", Università di Pisa, Largo Bruno Pontecorvo 3, Pisa, I-56127, Italy 
        \and
   INAF - Osservatorio Astrofisico di Arcetri, largo E. Fermi 5, 50127, Firenze, Italy
        \and
   Dipartimento di Matematica e Fisica, Università degli Studi Roma Tre, via della Vasca Navale 84, I-00146 Roma, Italy
        \and
   Department of Physics, Informatics and Mathematics, University of Modena and Reggio Emilia, 41125 Modena, Italy
        \and
    IFPU, Institute for Fundamental Physics of the Universe, Via Beirut 2, 34014 Trieste, Italy
    \and 
  Dipartimento di Fisica e Astronomia ``Augusto Righi'', Università degli Studi di Bologna, Via P. Gobetti 93/2, 40129 Bologna, Italy
        \and
  INAF - Istituto di Astrofisica e Planetologia Spaziali, Via del Fosso del Caveliere 100, I-00133 Roma, Italy
        \and
  Physics Department, Tor Vergata University of Rome, Via della Ricerca Scientifica 1, 00133 Rome, Italy
        \and
  INFN - Rome Tor Vergata, Via della Ricerca Scientifica 1, 00133 Rome, Italy
        \and
  Scuola Normale Superiore, Piazza dei Cavalieri 7, I-56126 Pisa, Italy
        \and
  Department of Physics, Middlebury College, Middlebury, VT 05753, USA     
  }

   \date{September 11, 2026}

 
  \abstract
   {{Ultra-fast broad absorption line outflows (uBAL, $\sim 0.1-0.2c$) represent some of the most extreme manifestations of quasar-driven winds. With velocities reaching a significant fraction of the speed of light, they are prime candidates for efficient feedback. However, their physical structure and location remain poorly constrained, particularly at high luminosities and redshifts. In this context, variability studies offer a valuable avenue to probe the physical conditions and geometry of these extreme outflows.}}
   {We aim to investigate the variability of uBALs through their characterisation across different epochs and multiple ionic species. {The main goal is} to constrain their physical properties, locations, and kinetic powers to evaluate their potential impact on the respective quasar host galaxies. }
   {We performed a multi-epoch analysis of three hyper-luminous quasars from the WISSH sample, namely {WISSH53, WISSH56, and WISSH71}, at redshifts $z= 3.628$, 4.101, and 3.567, respectively. {New  and archival spectra}, spanning up to $\sim$ 23 years of monitoring in the observed frame, were used in the analysis. {We modeled absorption associated with the \civ\ transition} with Gaussian components, accounting for covering factor variations. A conservative analysis of multi-ion profiles (\pv, \lya, \nv, \siiv) was also performed, considering the same uBAL velocity as \civ. We derived ionic column density, as well as ionisation parameters, which allowed us to estimate lower and upper limits on distances and kinetic powers of the outflows.}
   {The \civ\ uBALs in the three sources exhibit velocities up to $\sim 0.2 c$  and show significant variability across epochs, likely driven by changes in ionisation state and/or transverse gas motion. The coordinated analysis of the multiple troughs highlights the dominant mechanisms causing variability and enables the derivation of robust lower and upper limits of the physical and energetic properties of the outflows. Outflow distances are constrained from the broad line region (BLR) radius ($\sim$ 1 pc) up to a few hundred parsecs, and kinetic powers of individual troughs reach up to $\sim$ 25\% of the quasar bolometric luminosity, in the most extreme cases. Our results demonstrate that extreme outflows such as the uBALs are able to inject sufficient kinetic energy to affect the host {medium}.}
   {}


   \maketitle
%
\section{Introduction}

\par Quasar-driven {winds} are expected to represent the most extreme outflows during the Cosmic Noon, around redshifts $z \sim $ 2-4, when supermassive black holes (SMBHs) were growing at {fast rates} and the accretion luminosity was the highest \citep{Madau_2014}. These active galactic nuclei (AGN) outflows, launched from the accretion disk, span multiple scales and are capable of injecting kinetic power and momentum into the host interstellar medium and suppressing star formation activity \citep[e.g.,][]{King_pounds_2015, Fiore_2017, Fluetsch_2019}. Observationally, these outflows can be detected in many different spectral ranges. Particularly, outflows originating from the SMBH immediate vicinity are typically detected as blueshifted absorption lines in the X-ray and {ultraviolet (UV)} spectra of quasars {\citep{Arav_1999, Crenshaw_2003,  Misawa_2007,Laha_2021, Yamada_2024}.}
\par {In the X-ray band, absorption features in the soft X-ray portion (< 2 keV) of the spectrum are typically associated with moderately ionised outflowing gas {\citep[i.e., the {\it warm absorbers}, e.g.][]{Blustin_2005, McKernan_2007, Laha_2021, Yamada_2024, Middei_2026}} with velocities in the range $10^{3}$-$10^{4}$ km s$^{-1}$. {In contrast, ultra-fast outflows (UFOs) are traced by highly-ionised iron absorption lines (e.g.,  \ion{Fe}{XXV}, \ion{Fe}{XXVI}) at $\sim$ 7-10 keV with blueshift velocities typically in the range 0.1-0.3$c$, reaching up to $\sim$ 0.6$c$ in the most extreme cases {\citep[see e.g.,][and references therein]{Tombesi_2010, Chartas_2021, Matzeu_2023,Yamada_2024, XRISM_2025}}.}} In the UV range, absorption lines {such as \civ{} and \siiv{}} are usually classified into three categories, depending on the width of their velocity range ($\Delta {\rm v}={\rm v}_{\rm max}- {\rm v}_{\rm min}$): broad absorption lines (BALs) with $\Delta {\rm v}$ $\gtrsim$ 2000 km s$^{-1}$, mini-BALs {(500 $\lesssim$ $\Delta {\rm v}$ $\lesssim$ 2000 km s$^{-1}$)}, and narrow absorption lines (NALs) with $\Delta {\rm v}$ $\lesssim$ 500 km s$^{-1}$. BALs {are traditionally identified within velocity ranges between 5000 and 25000 km s$^{-1}$, although these limits are motivated by the historical observational definition of BALs rather than by physical constraints on quasar outflows and were adopted to avoid contamination from associated absorbers at low velocities and from Si IV contamination of C IV BALs at high velocities \citep{Weymann_1991, Gibson_2009}. Their} main properties (i.e., velocity, width, and absorption strength) are known to vary across different observational epochs, with some features even undergoing complete disappearance/reappearance \citep[e.g.,][]{Gibson_2010,Capellupo_2012, FilizAk_2013, Trevese_2013, De_Cicco_2018, Green_2023}.

\par A particularly extreme class of BAL outflows are the ultra-fast BALs (uBALs), which are characterized by broad absorption features in the UV quasar spectra with blueshift velocities {typically ranging} $\sim 0.1$-$0.2c${ {\citep[][and \citealt{Bischetti_2022, Bischetti_2023} for uBALs at higher redshifts]{Rodriguez-Hidalgo_2011, Rogerson_2016,Bruni_2019,  Aromal_2021, Vietri_2022, Vietri_2025}}}. {Recently, \citet{Seaton_2026} discovered an UV outflow with $v \sim 0.3c$, suggesting that the velocity distribution of uBALs may extend to even higher values at its upper end.} 
UV uBALs, together with the X-ray UFOs, likely represent the most energetic phase of quasar outflows. {Despite their potential importance for AGN feedback, systematic, dedicated studies on uBALs remain very limited. Most existing studies focus on individual quasars \citep[and references therein]{Vietri_2025}, while only a limited number analysed {larger samples}. In particular,} \citet{Rodriguez-Hidalgo_2020} presented the first (and, to date, only) systematic analysis of uBALs by using high-quality {Sloan Digital Sky Survey (SDSS)} data, {referring to them as ``extremely high-velocity outflows'' (EHVOs)}. In a sample of $\sim 6700$ QSOs, they found 40 sources at $2 \lesssim z \lesssim 4.7$ showing \civ\ uBALs, which appeared to be more prevalent at higher luminosities. The authors also performed a search of \nv\ and \ion{O}{VI} uBALs, and found that 26 out of 40 quasars show confirmed or likely corresponding \nv\ outflows at velocities comparable to those of \civ, while 6 quasars display analogous \ion{O}{VI} features. {Notably, at $z \sim 6$, a luminosity-selected quasar sample showed a remarkably high fraction of extreme BALs, with 5 out of 14 sources displaying ${\rm v} > 30000$ km s$^{-1}$ and 10/14 exceeding ${\rm v} > 25000$ km s$^{-1}$ \citep{Bischetti_2022, Bischetti_2023}.}

\par Detailed multi-epoch studies of individual uBALs reveal strong variability on timescales ranging from months up to a few years in the quasar rest-frame. Using SDSS spectra, \citet{Bruni_2019} identified two UV uBALs in the Wide-Field Infrared Survey Explorer (WISE)/Sloan Digital Sky Survey (SDSS)-selected Hyper-luminous quasar (WISSH) quasars at $z \sim 3$–3.6, with velocities up to $\sim 0.16c$. \citet{Vietri_2022} performed a follow-up spectroscopic analysis of one of these targets spanning 17 yr in the observed frame and discovered a highly variable system of \civ\ troughs. Through the analysis of the variability, combined with the characterization of the \civ\ absorption and the search for corresponding outflows at similar velocities in other ionic species such as \lya, \nv, and \siiv, the authors were able to constrain the column density {$N_{\rm H}$}, ionisation parameter {$U$}, and radial distance {$r_{\rm uBAL}$} of the outflowing gas from the central SMBH.

\par {BAL variability can be produced by different physical mechanisms. The two primary cases are changes in the ionization state of the gas, driven by variations in the continuum \citep[e.g.,][]{Hamann_2011,Trevese_2013}, and the motion of absorbing material across the line of sight (gas transverse motion; e.g., \citealt{Misawa_2005}). In the first case, fluctuations in the ionizing flux modify the strength of the absorption without requiring bulk gas displacement, while in the second case, the covering factor changes as clumps move across the line of sight of the background source. Identifying the {cause of such variations} is important because the inferred distance of the absorber $R_{\rm uBAL}$ depends differently on the variability time scale: 
$R_{\rm uBAL} \propto \Delta t_{\rm rest}^{2}$ for gas motion and 
$R_{\rm uBAL} \propto \sqrt{\Delta t_{\rm rest}}$ for ionization changes, where $\Delta t_{\rm rest}$ is the time between epochs in the quasar rest frame {\citep[e.g.,][and references therein]{Capellupo_2013, Trevese_2013}}. As a result, short-term observations are key to tracing rapid variability and better constraining the physical properties and structure of uBALs.}

\par {In a broader physical context, rapid outflow variability is expected if SMBH fueling is intrinsically intermittent. In precipitation-driven scenarios such as Chaotic Cold Accretion (CCA), multiphase clouds and filaments condense out of a turbulently stirred hot medium and undergo recurrent interactions in
the inner region, driving stochastic accretion-rate fluctuations and hence bursty radiative and mechanical
output. This weather-like cycle provides a natural framework for both coordinated ionization-driven
changes in absorption strength and, in some cases, covering-factor variability as clumpy structures cross
the line of sight \citep[e.g.][]{Gaspari_2013, Gaspari_2017}.}


\par Building on the results of \citet{Vietri_2022}, and after visually inspecting all {the available epochs of the} WISSH SDSS spectra, {we identified {{$\sim 10$}} sources showing }uBALs. Several monitoring campaigns were initiated to follow up these uBALs using facilities such as {the Large Binocular Telescope (LBT) equipped with the Multi-Object Double Spectrograph (MODS) and the Very Large Telescope (VLT) equipped with the Ultraviolet and Visual Echelle Spectrograph (UVES)}. For the first time, we present a detailed {multi-epoch spectroscopic analysis of the first three WISSH quasars that present uBALs:}  {WISSH53, WISSH56, and WISSH71}. {{The complete fraction of uBALs within the WISSH sample will be presented by Vietri et al. (in prep.) once the ongoing observational campaigns are completed.}} {In order to explore the temporal and ion-dependent variability of uBALs, we combine new (VLT/UVES, LBT/MODS) and archival data from the SDSS, the Baryon Oscillation Spectroscopic Survey (BOSS), the VLT/XSHOOTER spectrograph, the VLT Multi Unit Spectroscopic Explorer (VLT/MUSE), the VLT/UVES, the LBT/MODS, and the Subaru Faint Object Camera and Spectrograph (Subaru/FOCAS). }


\par The paper is structured as follows: in Section \ref{sec:sample}, we describe the sample, the spectroscopic observations, and the data reduction. In Section \ref{sec:results}, we analyse the \civ\ troughs and discuss the uBAL systems of each source. Section \ref{sec:ionising} discusses the possible detection of uBAL in other ionic species, with the same velocities found for \civ\ and discusses derived estimates of the ionising structure (i.e., column density and ionisation). In Section \ref{sec:radius}, we present lower and upper limits of the outflow locations. Section \ref{sec:kin_power} contains the estimates of the uBAL energetics. The summary and conclusions of this work are presented in Section \ref{sec:summary}. Throughout this paper, velocities are given relative to the longer-wavelength component of the \civ{} doublet, assuming the relativistic Doppler correction. We adopt a flat $\Lambda$CDM cosmology with $H_0 = 70$ km s$^{-1}$ Mpc$^{-1}$ and $\Omega_\Lambda = 0.7$ {\citep{Komatsu_2011}}.

\begin{table}[]
    \centering
    \caption{Summary of the sample.}
    \resizebox{\linewidth}{!}{
    \begin{tabular}{lcccc}
    \hline
    \hline
    \noalign{\smallskip}
    WISSH ID & SDSS ID &  $z$ & $\log L_{\rm{bol}}$ & $\log M_{\rm{BH}}$\\
    & & & [erg s$^{-1}$] & [M$_{\odot}$]\\
    (1) & (2) & (3) & (4) & (5)\\
    \noalign{\smallskip}
    \hline
    \noalign{\smallskip}
    WISSH53 & J124957.24-015928.8     &   3.628 & 47.59 & 10.01\\
    WISSH56$^\star$     & J130502.28+052151.1  &   4.101 & 47.53 & - \\
    WISSH71 & J153830.55+085517.0  &  3.567 & 47.83 & 9.74\\
    \noalign{\smallskip}
    \hline
    \end{tabular}
    }
    \label{tab:sample}
    \tablefoot{(1) Identification of the source in the WISSH catalog. (2) Identification in the SDSS catalog. (3) {Redshift from \citet{Saccheo_2023}.} (4) Bolometric luminosity as determined by \citet{Saccheo_2023}. (5) H$\beta$-based black hole mass from \citet{Vietri_2018} and Vietri et al. (in prep.). $^\star$No $M_{\rm BH}$ estimates based on H$\beta$/\ion{Mg}{II} are available for this QSO.}
\end{table}

\section{Sample and data}
\label{sec:sample}
\par Our sample consists of three quasars from the WISSH survey that exhibit time-variable uBAL features in the multi-epoch spectroscopic observations, {namely  WISSH53, WISSH56, and WISSH71}. We present new spectroscopic observations of the three sources obtained with VLT/UVES \citep{Dekker_2000}, complemented by new LBT/MODS data for two targets and archival spectra from various instruments, covering a time span of $\sim 20$ ($\sim 4.5$–5) years in the observed (rest) frame. WISSH53 was observed at seven epochs, providing a {broad} temporal baseline. For WISSH56, four epochs are available, including new observations from VLT/UVES in 2023 and LBT/MODS in 2025. WISSH71, a well-studied hyper-luminous quasar with prominent outflows \citep[e.g.,][ and particularly \citealt{Vietri_2022}]{Vietri_2018, Bruni_2019, Travascio_2020, Deconto-Machado_2023}, was analysed over nine epochs, comprising one new observation from VLT/UVES in 2023 and five new observations from LBT/MODS in 2023 and 2024. Table \ref{tab:sample} lists the main quasars properties.

\subsection*{New observations and data reduction}
\begin{itemize}
    \item[-] \textit{VLT/UVES:} {observations} were carried out between October 2022 and March 2023 as part of the {European Southern Observatiory (ESO)} program ID 110.240J.001 (PI G. Vietri). For WISSH53 and WISSH71, observations were performed using dichroic \#1, providing spectral coverage of $\sim$3000–4000 \AA\ in the blue arm and $\sim$5000–8000 \AA\ in the red arms. For WISSH56, dichroic \#2 was used, covering $\sim$4000–5000 \AA\  and $\sim$6000–10000 \AA\ in the blue and red arms, respectively. For the three sources, a slit width of 1.0\arcsec\ (R $\sim$ 40000) was selected. {The data were reduced with the UVES ESO pipeline (version 6.4.6) using the \textsc{esorex} (version 3.13.7).} {For flux calibration, the standard stars Feige 67, CD-32 9927, and LTT6248 were observed and reduced using the same calibration data as the science frames.} For WISSH56, two VLT/UVES observations taken one month apart were made {to investigate short-term variability and therefore better constrain the spatial scales of the absorbing outflow (see Sect. \ref{sec:radius}).} Each of these observations was independently reduced following the same approach. As no significant differences were identified between them, the observations were co-added to improve the signal-to-noise ratio. 

\item[-] \textit{{LBT/MODS:}} For both WISSH56 and WISSH71, we obtained new observations with the MODS1 blue and red gratings, using a slit of 0.6\arcsec {(program IDs: IT-2022B-035 and IT-2023B-031, PI: G. Vietri)}. WISSH56 was observed on the nights of April 26 and April 29, 2025. Since no variation was found between the two spectra, they were combined into a single, co-added spectrum. WISSH71 was observed on six dates: May 28 and June 25, 2023, and February 14, February 19, April 10, and June 9, 2024. No significant differences were detected between the spectra obtained on February 14 and 19; therefore, these two epochs were combined, resulting in five independent LBT/MODS observations in total. In all cases, we performed the standard data reduction using the {Spectroscopic Interactive Pipeline and Graphical Interface (SIPGI)} pipeline \citep{Gargiulo_2022}, {designed for LBT data, including MODS.} The procedure includes bias subtraction, flat-field correction, wavelength calibration, sky subtraction, and 1D spectral extraction. The spectrophotometric standard star BD+33 2642 was used within SIPGI for the flux calibration, and we performed an additional absolute flux calibration using our own python routine to ensure consistency across epochs.
    
\end{itemize}

\begin{figure*}[ht!]
    \centering
    \includegraphics[width=\linewidth]{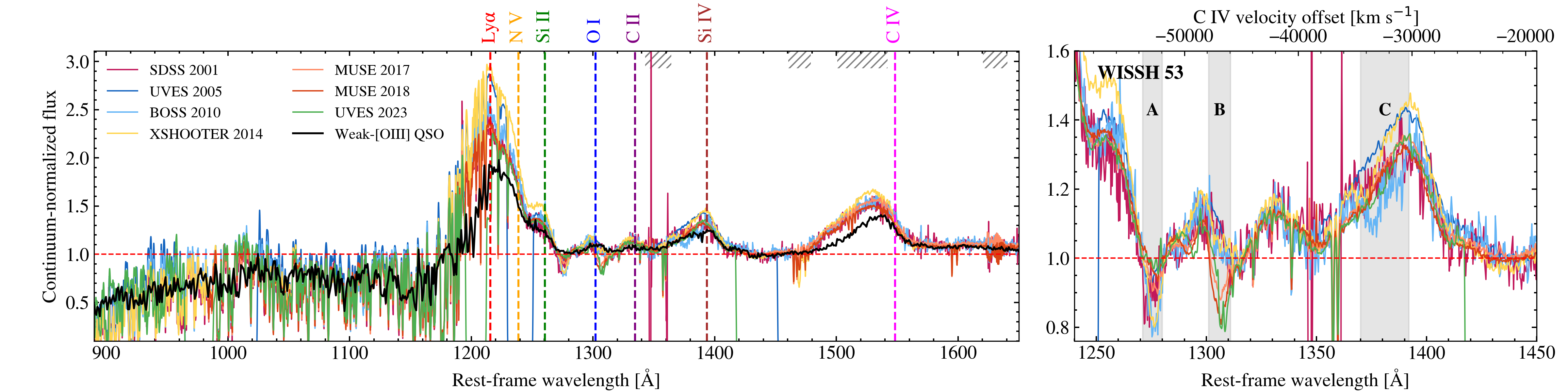}
    \includegraphics[width=\linewidth]{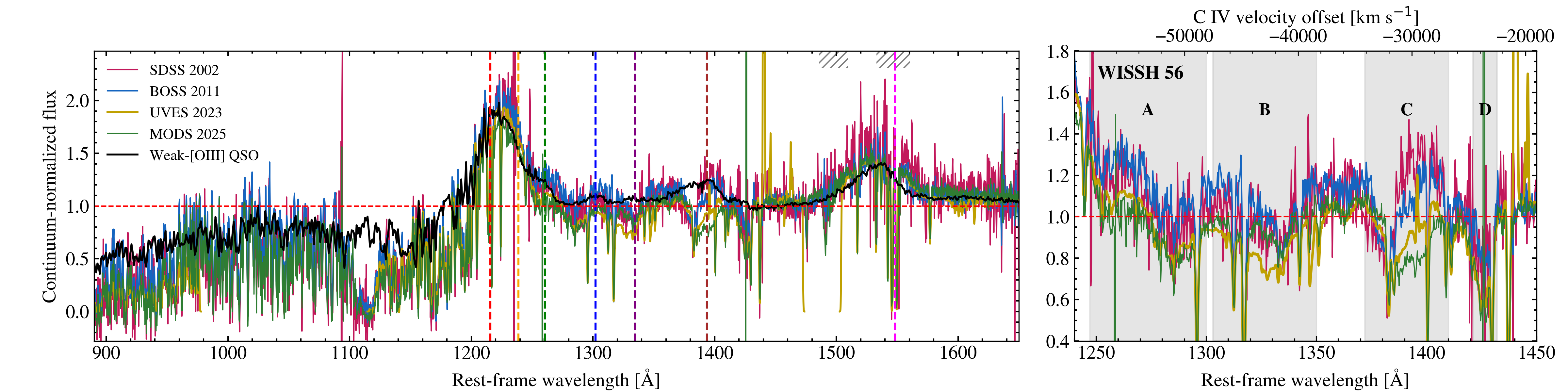}
    \includegraphics[width=\linewidth]{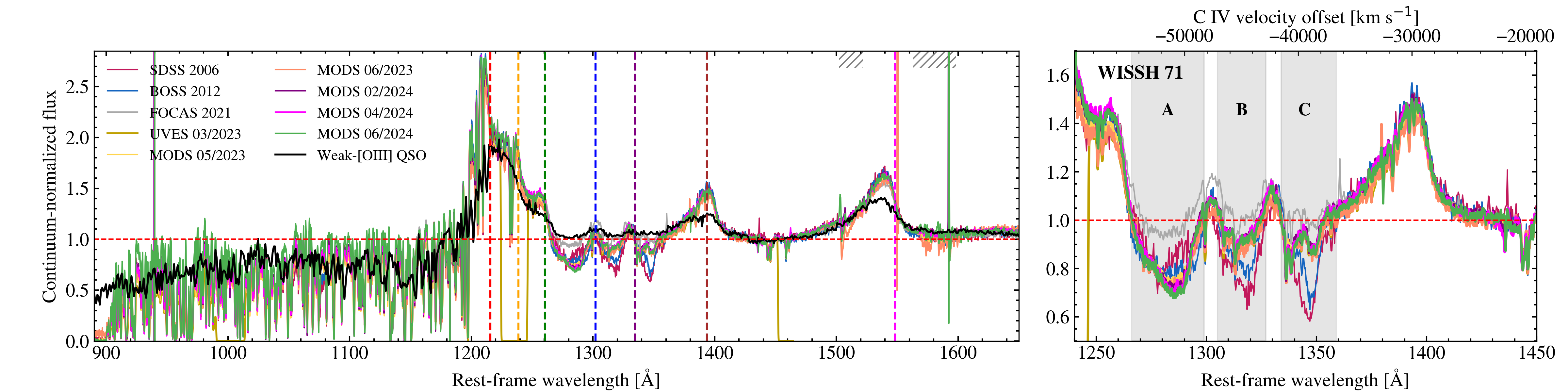}
    
    \caption{\textit{Left panels:} Overview of the available spectra of the three quasars analysed in this work. The weak-[O III] (EW([\ion{O}{III}]) $<5$\AA) quasar template from the WISSH sample (Vietri et al., in prep.) is shown in black for comparison. Vertical dashed lines indicate rest-frame wavelength of the main emission lines between \lya{} and \civ{}. {Hatched bands at the top of the panels indicate wavelength regions contaminated by telluric absorption.} \textit{Right panels:} Zoom on the \civ{} absorption regions, emphasizing the shape and variability of the troughs at different epochs. {The regions where the troughs are observed are indicated with gray-shaded areas.} For a better comparison, all spectra were resampled to the same resolution. Red horizontal lines indicate the continuum level. {Each spectrum was normalized by dividing it by the corresponding best-fit continuum model.}}
    \label{fig:multiepoch}
\end{figure*}

\par {Details of the archival data are provided in Appendix \ref{archival}, and a summary of all the observations used in this paper is listed in Table \ref{table:obs}.}

\section{Results}
\label{sec:results}
\subsection{Characterizing \ion{C}{IV} troughs}
\label{sec:civ_troughs}
\begin{figure}[t!]
    \centering
    \includegraphics[width=\linewidth]{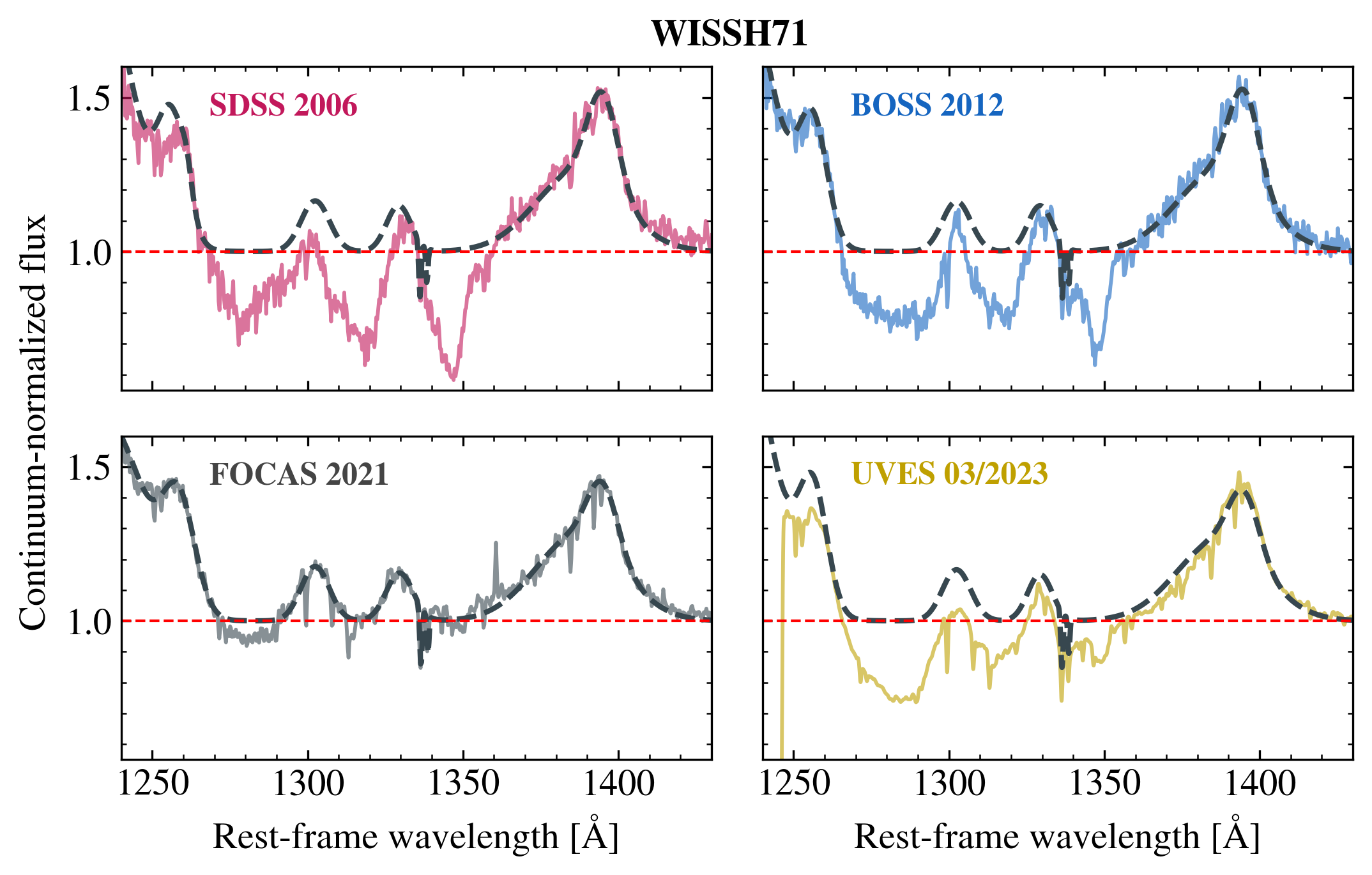}

    \caption{Continuum + emission line spectral fitting of the first four epochs of WISSH71. The FOCAS 2021 spectrum, which is the least affected by absorption in this source, was adopted as the best representation of the intrinsic (unabsorbed) emission and used as the reference model. This reference fit was then applied to the other epochs to reconstruct their underlying emission profiles. Black dashed line represents the best fit of FOCAS 2021. {An additional weak (i.e., with low equivalent width) Gaussian emission component was included when necessary to accurately fit the strong emission lines in spectral regions unaffected by absorption, ensuring a proper subtraction of the emission-line contribution (e.g., C IV and Ly$\alpha$).} Red horizontal line indicates the continuum level. }
    \label{fig:J1538_example_emission_fitting}

\end{figure}

\par Fig. \ref{fig:multiepoch} shows all the {spectra (obtained at different epochs)} considered in this paper for our three targets. In the three quasars, significant absorptions are found in the spectral region between \lya\ and \siiv. {These features initially suggest the presence of \siiv\ classical BAL. However, the absence of corresponding \civ\ absorption at similar velocity \citep[see e.g.,][and references therein]{Bruni_2019, Bischetti_2023} rules out this interpretation. The observed absorption features are therefore unambiguously identified as \civ\ uBALs \citep{FilizAk_2013, Rodriguez-Hidalgo_2020}. }

\par The \civ{} troughs with velocities around 0.1-0.2$c$ fall in the spectral region between \lya{} and \siiv{} emission lines. In order to isolate the \civ{} absorption features, we fitted the continuum and emission lines in this region. The continuum was modeled using a power-law function, while the most prominent emission lines, namely \lya, \nv, \ion{Si}{II}, \siiv, \ion{O}{IV}, \civ, \ion{Al}{II}, \ion{O}{III}], and those associated with the {\ion{C}{III}] complex}, {were fitted  using an empirical parametrization consisting of one or more Gaussian components, adding as many components as required to reproduce the observed profile.} For the SDSS, BOSS, and XSHOOTER spectra, {the fittings of both the emission lines and the continuum} were performed over the rest-frame wavelength interval 1210-1970 \AA\ to avoid {the absorption due to the} \lya\ forest region blueward of 1210 \AA\ and iron emission redward of 1970 \AA. In the case of the UVES and FOCAS spectra, the fittings were restricted to 1210-1415 \AA\ and 1210-1600 \AA, respectively, due to the smaller available wavelength range. Additionally, the interval 1570-1631 \AA, which contains unidentified emission features, and the range 1687-1833 \AA, which includes a complex blend of lines as well as several \ion{Fe}{II} multiples \citep{Nagao_2006}, were systematically {masked in} the emission line fitting procedure, when present. This was done in order to minimise uncertainties in both continuum placement and emission line modeling.

\par For each source, we visually inspected the spectra of all the available epochs and selected the spectrum least affected by absorption to serve as a reference for the continuum and emission-line fitting (see Fig. \ref{fig:J1538_example_emission_fitting}). The continuum slope and normalization were initially fixed to match those of the reference spectrum, although small adjustments (see below) in both parameters were allowed when necessary. The emission lines were always kept fixed in amplitude, centroid, and width. Specifically, for WISSH53 we adopted the UVES 2005 spectrum as the reference between \lya{} and \siiv{}, for WISSH56 the BOSS 2011 spectrum, and for WISSH71 the FOCAS 2021 spectrum. In cases where significant variations from the reference were observed (such as the \siiv{} profiles in WISSH53, as shown in Fig. \ref{fig:J1249_profiles_troughs}), the Gaussian amplitudes were allowed to vary or additional narrow Gaussians were added to account for the total emission flux. 
\par After modeling the continuum and the emission lines, {we divided the observed spectra by the resulting best-fit model} in order to isolate the absorption features. For all identified absorption troughs that were associated with \civ, we estimated minimum and maximum velocities, as well as the balnicity index (BI) at 90\% of the continuum level, defined as:
\begin{equation}
    \mathrm{BI} = -\int_{{60\,000}}^{{30\,000}} \left[1 - \frac{f({\rm v})}{0.9} \right] C \, \mathrm{d}{\rm v}
\end{equation}
where $f(\rm{v})$ is the continuum-normalized flux as a function of velocity, and $C$ is 1 if the value inside the brackets remains continuously positive over a chosen velocity interval, and 0 otherwise. The division by 0.9 is applied to exclude shallow absorptions, considering only those deeper than 10\% below the normalized continuum. In our analysis, we define \civ\ uBAL troughs as absorption features that reach below 90\% of the continuum level and extend over a minimum width of 1000 km s$^{-1}$.

\par The \civ\ absorption troughs were fitted using the continuum as the reference level, i.e., the absorption was measured relative to the continuum flux. Both components of the 1548 \AA\ and 1550 \AA\ doublet were {fitted under a partial coverage formalism}\footnote{{A  2:1 optical depth ratio was adopted between the two components of the \civ\ doublet at 1548 \AA\ and 1550 \AA.}},  using the following expression for the continuum-normalized flux \citep{Arav_2005}:

\begin{equation}
f(v) = (1 - C_{\rm f}) + C_{\rm f}  e^{-\tau({\rm v})},
\end{equation}
where $\tau(v)$ denotes the optical depth, assumed to follow a Gaussian profile characterized by the central optical depth, centroid velocity, and Doppler parameter. The term $C_{\rm f}$ represents the covering factor, i.e., the fraction of the background emission source that is obscured by the absorbing gas \citep[$0 < C_{\rm f} < 1$;][]{Hamann_1999}. A value of $C_{\rm f} = 1$ corresponds to {the absorber fully covering the continuum source}, while $C_{\rm f} < 1$ indicates partial coverage. In our fits, the covering factor was varied from $C_{\rm f} = 0.1$ to $C_{\rm f} = 1.0$ in steps of 0.1, and the corresponding $\chi^2$ values were evaluated. In all cases, the best-fit solution occurred at $C_{\rm f} = 1.0$ (hereafter, $C_{\rm f,max}$), which minimized $\chi^2$ and was therefore adopted as our fiducial model. We also evaluated the Bayesian Information Criterion (BIC) for the different covering factors. While $C_{\rm f} = 1.0$ minimizes the BIC, models with $\Delta \rm BIC < 10$  are statistically indistinguishable and cannot be formally excluded. We define $C_{\rm f,min}$ as the lowest covering factor that satisfies this criterion, typically in the range 0.3–0.4. We estimated the column density $N_{\rm H}$ for both $C_{\rm f,max}$ and $C_{\rm f,min}$. {We note that this approach assumes the \civ\ doublet is close to optically thin. In the majority of the troughs, the doublet ratios appear consistent with this assumption. Additionally, we stress that the covering factor is treated as constant within each trough; while some BAL systems can show velocity-dependent covering, a single $C_{\rm f}$ per component provides a robust first-order characterization given the blending and the S/N of the multi-ion region.} {The absorption fits obtained assuming $C_{\rm f}=C_{\rm f,max}$ for all epochs are shown in the Appendices} (Figs. \ref{fig:WISSH53_absorptions}, \ref{fig:WISSH56_absorptions}, and \ref{fig:WISSH71_absorptions}, for WISSH53, WISSH56, and WISSH71, respectively).

\begin{table}[t!]
    \centering
    \caption{Properties of the \civ{} troughs of WISSH53 for the different epochs.}
    \resizebox{0.9\linewidth}{!}{
    \begin{tabular}{lccccc}
    \hline
    \hline
    \noalign{\smallskip}
    Comp. & Epoch & BI & ${\rm v}_\textrm{min}$ & ${\rm v}_\textrm{max}$ & $\Delta t$ \\
    & & [km s$^{-1}$] & [km s$^{-1}$] & [km s$^{-1}$] & [yr]\\
    (1) & (2) & (3) & (4) & (5) & (6) \\
    \noalign{\smallskip}
    \hline
    \noalign{\smallskip}
    \multicolumn{6}{c}{WISSH53}\\
    \noalign{\smallskip}
    \hline
    \noalign{\smallskip}
    A & 2001 & $50_{+30}^{-10}$ & 52460$_{+30}^{-90}$ & 53710$_{+140}^{-80}$ & - \\
    &  2005 & - & - & - & 0.88 \\
    &  2010 & $110_{+40}^{-70}$ & 52060$_{+90}^{-40}$ & 53540$_{+40}^{-90}$ & 1.00 \\
    & 2014 & $100^{-10}_{+10}$&  52020$^{-10}_{+10}$ & 53370$^{-10}_{+20}$ &  0.86 \\
    & {2017} & - & - & - & 0.72\\
    & {2018} & - & - & - & 0.13\\
    &  2023 & - & - & - & 1.10 \\
    \noalign{\smallskip}
    B & 2001 & - & - & - & -\\
    &  2005 & - & - & - & 0.88 \\
    &  2010  & - & - & - & 1.00\\
    & 2014 & - & - & - & 0.86 \\
    & {2017} & $100^{-20}_{+30}$ & 46410$_{+30}^{-140}$ & 47910$_ {+180}^{-50}$ & 0.72\\
    & {2018} & $190^{-20}_{+30}$ & 46280$_{+40}^{-460}$ & 48090$_{+80}^{-30}$ & 0.13\\
    &  2023 & $180^{-10}_{+10}$ & 45910$^{-10}_{+10}$ & 47800$^{-10}_{+20}$ & 1.10\\
    \noalign{\smallskip}
    C & 2001 & - & - & - & - \\
    &  2005 & - & - & - & 0.88 \\
    &  2010 & $150^{-40}_{+90}$ & 30710$_{+40}^{-110}$ & 34400$^{-40}_{+100}$ & 1.00 \\
    & 2014 & - & - & - & 0.86 \\
    & {2017} & - & - & - & 0.72\\
    & {2018} & - & - & - & 0.13\\
    &  2023 &  - & - & - & 1.10 \\
    \noalign{\smallskip}
    \hline
    \noalign{\smallskip}
    \end{tabular}
    }
    \label{tab:tab_J1249}
    \tablefoot{(1) Absorption component. (2) Epoch of observation. (3) Balnicity index. (4) Minimum velocity. (5) Maximum velocity. (6) Rest-frame time difference respect to the previous observation. The reported uncertainties represent the 16th–84th percentile range of the estimated parameters.}
\end{table}

\subsection{Analysis of the individual sources}
\subsubsection{WISSH53}

\par We identified three absorption components in WISSH53, labeled A, B, C, which span the wavelength range  $\sim$ 1270-1390 \AA, corresponding to outflow velocities within the range $\sim$ 31000-54000 km s$^{-1}$ (see Fig. \ref{fig:multiepoch}). Trough A exhibited depths exceeding 90\% of the continuum level during most of the earliest epochs and showed pronounced variability: it weakened from 2001 to 2005, then deepened again in 2010 and 2014, reaching flux levels below 90\% of the continuum. Across the three epochs in which it is clearly detected, trough A maintained a roughly constant velocity range, varying between $\sim$ 52000 and 53700 km s$^{-1}$, and presenting a maximum BI value of 110 km s$^{-1}$ in the BOSS 2010 epoch (see Table \ref{tab:tab_J1249}). In the later epochs (2017–2023), trough A gradually faded and disappeared. 
In contrast, trough B was detected only in the  most recent epochs (2017-2023), where it became progressively deeper with time, showing a width range of $\sim 46000$ and $48000$ km s$^{-1}$. Trough C, which falls at the top of the \siiv{} emission line, was detected to be below 90\% of the continuum flux only in the 2010 spectrum, {when this feature showed its broadest velocity} ($\sim 30700$-34400 km s$^{-1}$), then became weaker in 2014 and completely disappeared by 2023. It should be noted, as this feature lies on the blue side of the emission line \siiv{}, isolating the real contribution of the absorption features is difficult, therefore, any interpretation should be considered with care.

\par
\begin{figure}[t!]
    \centering
    \includegraphics[width=\linewidth]{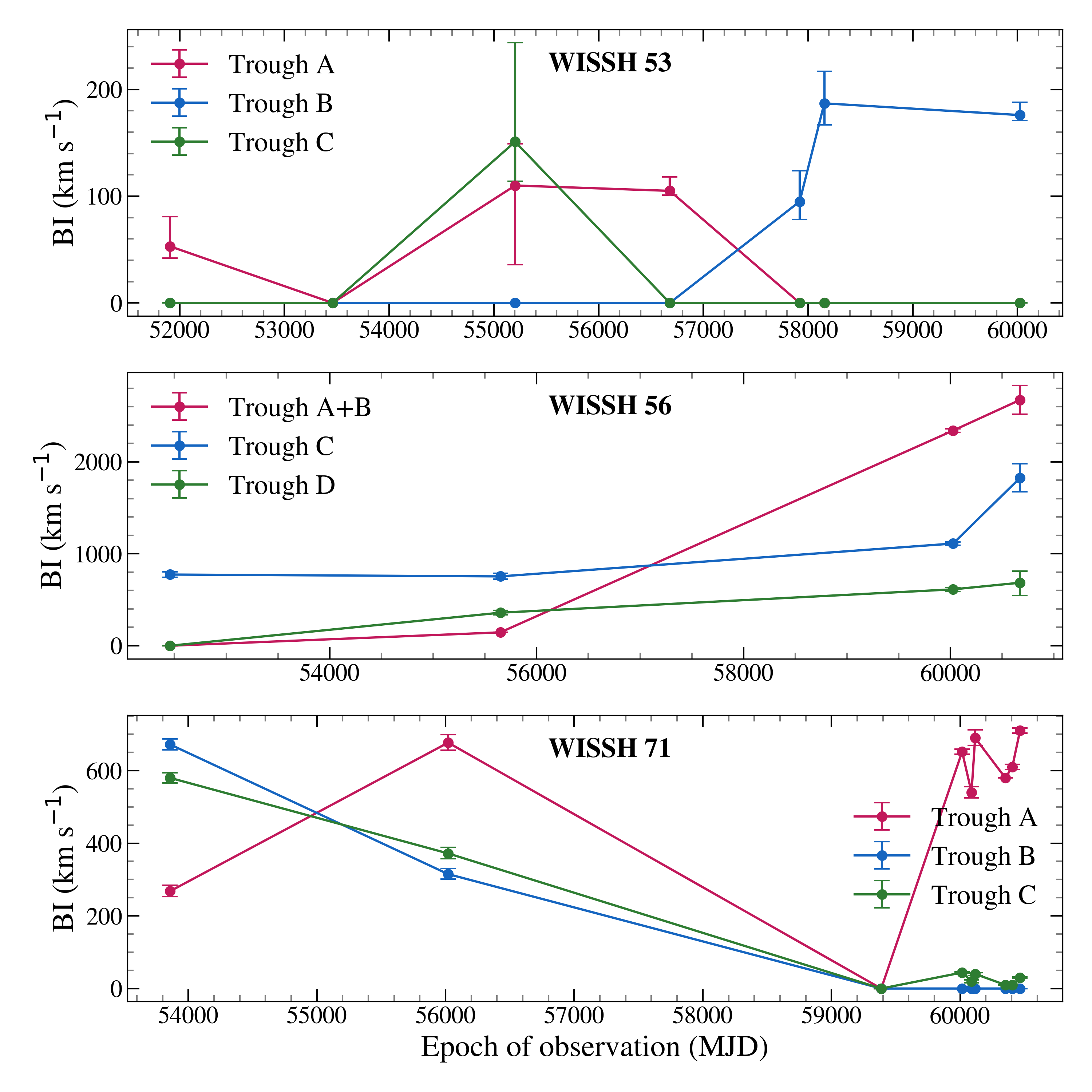}
    \caption{{Balnicity index variation throughout the different epochs for the three sources. \textit{From top to bottom:} WISSH53, WISSH56, and WISSH71. Error bars indicate the 95\% percentile-based confidence interval.}}
    \label{fig:balnicity}
\end{figure}

 Figure \ref{fig:balnicity} shows the evolution of the BI parameter across different epochs. In the case of WISSH53, troughs A and C exhibited a coherent variability pattern, both strengthened and weakened in phase (see also Fig. \ref{fig:J1249_profiles_troughs}). {The coordinated behavior of absorption components at different velocities is more naturally explained by changes in the ionization state of the gas driven by continuum variability, rather than by a gas-motion scenario.}  In contrast, trough B showed a distinct temporal evolution, being detected only when the other troughs were not observed. Such behavior is more naturally explained by the gas-motion scenario, where the appearance/disappearance of the absorption is likely due to a cloud crossing the line of sight to the continuum-emitting region, rather than to variations in ionisation.

\begin{table}[t!]
    \centering
    \caption{Properties of the \civ{} troughs of WISSH56 for the different epochs.}
    \resizebox{0.9\linewidth}{!}{
    \begin{tabular}{lccccc}
    \hline
    \hline
    \noalign{\smallskip}
    Comp. & Epoch & BI & ${\rm v}_\textrm{min}$ & ${\rm v}_\textrm{max}$ & $\Delta t$ \\
    & & [km s$^{-1}$] & [km s$^{-1}$] & [km s$^{-1}$] & [yr]\\
    (1) & (2) & (3) & (4) & (5) & (6) \\
    \noalign{\smallskip}
    \hline
    \noalign{\smallskip}
    \multicolumn{6}{c}{WISSH56}\\
    \noalign{\smallskip}
    \hline
    \noalign{\smallskip}
    A+B & 2002 & - & - & - & - \\
    & 2011 & $150^{-70}_{+80}$ & 40840$_{+60}^{-160}$ & 41910$^{-50}_{+130}$ & 1.75 \\
    & 2023 & $2340^{-30}_{+30}$ & 40940$_{+20}^{-20}$ & 52960$^{-20}_{+20}$ & 2.32\\
    & 2025 & $2670_{+270}^{-300}$& 41060$^{-90}_{+30}$ & 58190$_{+80}^{-40}$ & 0.42\\
    \noalign{\smallskip}
    C & 2002 & $770^{-30}_{+20}$ & 31560$^{-280}_{+60}$ & 34630$^{-90}_{+190}$ &  -\\
    & 2011 & $750^{-150}_{+150}$ & 29820$^{-260}_{+100}$ & 33590$_{+200}^{-40}$ & 1.75\\
    & 2023 & $1110^{-30}_{+30}$ & 29730$_{+10}^{-20}$ & 34130$^{-10}_{+20}$ & 2.32\\
    & 2025 & $1830_{+230}^{-220}$ & 26250$_{+30}^{-90}$ & 33830$_{+90}^{-40}$ & 0.42\\
    \noalign{\smallskip}
    D & 2002 & - & - & - & - \\
    & 2011 & $340^{-20}_{+30}$ & 22900$^{-180}_{+170}$ & 24360$^{-80}_{+130}$ & 1.75\\
    & 2023 & $610^{-20}_{+20}$ & 22900$_{+10}^{-10}$ & 24900$_{+10}^{-10}$ & 2.32\\
    & 2025 & $690_{+130}^{-140}$ & 22890$^{-100}_{+60}$ & 25320$^{-40}_{+90}$ & 0.42\\
    \noalign{\smallskip}
    \hline
    \noalign{\smallskip}
    \end{tabular}
    }
    \tablefoot{Same description as Table \ref{tab:tab_J1249}.}
    
    \label{tab:tab_J1305}
\end{table}

\subsubsection{WISSH56}
\par For this source, a system of four \civ{} absorption troughs was identified, with very significant variations in all the observed epochs. They were located within 1270-1430 \AA\ in the rest-frame wavelength and the associated outflow velocities ranged from $\sim$ 23000 up to 58000 km s$^{-1}$ (corresponding to $\sim$ 0.07 up to 0.19$c$). Fig. \ref{fig:J1305_profile_troughs} shows the variation of the region between \lya\ and \siiv\ for WISSH56. Trough A appeared to be absent in the earliest available spectra (2002 and 2011), and showed a progressive increase in depth in the more recent observations from 2023 and 2025. However, we note that this absorption feature is likely not solely attributable to \civ{}, but rather represents a blend of \civ{} and \siiv{} contributions, as the wavelength region where \siiv\ absorption is expected to fall within the same rest-frame interval as the observed \civ\ troughs. The trough B showed a gradual deepening over time from 2002 to 2023. In 2002, it barely reached the 0.9 continuum level, although this may be partially attributed to the low S/N ratio of the SDSS spectrum. In 2011, the feature was more clearly detected with a width $\lesssim$ 1000 km s$^{-1}$. The feature became more pronounced in the 2023 spectrum, indicating an increase in absorption strength. However, in the most recent epoch (2025), the trough appeared to weaken again, suggesting a possible fading of the outflow signature. During 2023 and 2024 epochs, troughs A and B appeared to span similar velocity ranges, and are therefore treated as a single component in our analysis. Trough C showed weaker absorption in 2011 and 2023 compared to 2002, but in the 2025 spectrum it appeared to broaden while its depth increases again. Trough D was clearly detected from 2011 to 2025, deepened between 2011 and 2023, and then remained roughly constant from 2023 to 2025.

\par Table \ref{tab:tab_J1305} lists the variation of the BI and the minimum and maximum velocities. Both troughs A+B and C were extremely broad, with BI $=2670$ and $1820$ km s$^{-1}$ in the last epoch (MODS 2025). The exceptionally high BI value of the A+B trough, in particular, places it among the largest reported so far for uBALs with velocities exceeding 35000 km s$^{-1}$ \citep{Rodriguez-Hidalgo_2020}. In the MODS 2025 spectrum, this absorption feature showed a minimum velocity of $\sim$41000 km s$^{-1}$ and a maximum of 58200 km s$^{-1}$, spanning a total velocity range of nearly 20000 km s$^{-1}$. As shown in Fig. \ref{fig:balnicity} (middle panel), the BI of the three \civ{} troughs A+B, C, and D exhibited a progressive increase from 2002 through 2025, suggesting an overall strengthening of these absorption features over time. The fact that all three troughs appeared to vary in a similar way suggests that a common mechanism could be the prime responsible for the observed variability. Such a coordinated behavior among multiple absorption components suggests that the observed changes are primarily driven by variations in the ionising continuum, consistent with photoionisation-driven scenarios \citep[e.g.,][and references therein]{FilizAk_2013}.

\subsubsection{WISSH71}
\label{sec:WISSH71}
\par Three \civ\ absorption troughs were detected in WISSH71, spanning the wavelength range of $\sim$ 1270-1360 \AA\ (in velocity, $\sim 38000-55000$ km s$^{-1}$). As previously reported by \citet{Vietri_2022}, trough A showed a clear broadening trend between 2006 and 2012, which may reflect changes in the velocity structure or covering fraction of the absorbing material. Additionally, troughs B and C varied in a correlated manner despite exhibiting different outflow velocities, which suggests that they could originate from absorbers with comparable physical conditions, such as similar ionisation parameters or densities. All the three features disappeared in the 2021 observation with Subaru/FOCAS (see Fig. \ref{fig:J1538_profiles_troughs}), indicating a possible temporary change in the outflow properties. However, the most recent UVES and MODS spectra obtained in 2023 and 2024 revealed the reappearance of trough A, now deeper than in previous epochs, while troughs B and C remained significantly weaker, reaching only about 90\% of the continuum flux. The observations between 2023 and 2024 LBT/MODS did not show strong variations, with troughs A, B, and C exhibiting a behaviour largely consistent with the UVES 03/2023 epoch. Nevertheless, a change in the trough A was measured between the May and June 2023 MODS spectra, corresponding to only $\sim 0.02$ yr ($\sim$ 7 days) in the rest frame. {Such an extremely short-timescale variation was already observed in other cases} \citep[e.g.,][]{Capellupo_2013} and provides a direct evidence for rapid transverse motion in the absorbing gas. {We also note that trough C is composed of multiple sub-components, including two relatively narrow features at higher velocities and a broader, lower-velocity component. These structures did not appear to show significant changes observed in their profiles between the 03/2023 and 06/2024 epochs.}

\begin{table}[t!]
    \centering
    \caption{Properties of the \civ{} troughs of WISSH71. }
    \resizebox{0.9\linewidth}{!}{
    \begin{tabular}{lccccc}
    \hline
    \hline
    \noalign{\smallskip}
    Comp. & Epoch & BI & ${\rm v}_\textrm{min}$ & ${\rm v}_\textrm{max}$ & $\Delta t$ \\
    & & [km s$^{-1}$] & [km s$^{-1}$] & [km s$^{-1}$] & [yr]\\
    (1) & (2) & (3) & (4) & (5) & (6) \\
    \noalign{\smallskip}
    \hline
    \noalign{\smallskip}
    \multicolumn{6}{c}{WISSH71}\\
    \noalign{\smallskip}
    \hline
    \noalign{\smallskip}
     A & 2006 & $270^{-10}_{+10}$ & 50480$_{+50}^{-130}$ & 53390$_{+100}^{-70}$ & - \\
    & 2012 & $680^{-20}_{+20}$ & 48120$_{+33}^{-240}$ & 54800$_{+180}^{-60}$ & 1.32\\
    & 2021 & - & - & - & 2.00\\
    & 03/2023 & $650^{-10}_{+10}$ & 48700$_{+10}^{-20}$ & 54480$_{+30}^{-20}$ & 0.60\\
    & 05/2023 & $540^{-20}_{+40}$ & 49200$^{-40}_{+20}$ & 53900$^{-20}_{+40}$ & 0.04\\
    & 06/2023 & 690$^{-20}_{+40}$ & 48590$^{-40}_{+20}$ & 54140$^{-20}_{+50}$ & 0.02\\
    & 02/2024 & 580$^{-20}_{+30}$ & 49290$^{-40}_{+20}$ & 53850$^{-20}_{+50}$ & 0.14\\
    & 04/2024 & 610$^{-20}_{+40}$ & 49140$^{-50}_{+20}$ & 53850$^{-20}_{+50}$ & 0.03\\
    & 06/2024 & 710$^{-20}_{+40}$ & 49170$^{-50}_{+10}$ & 54250$_{+50}^{-20}$ & 0.04\\
    \noalign{\smallskip}
    B & 2006 & $670^{-20}_{+20}$ & 42910$_{+60}^{-110}$ & 48350$_{+150}^{-60}$ & - \\
    & 2012 & $320^{-20}_{+20}$ & 43200$_{+50}^{-110}$ & 47190$_{+150}^{-40}$ & 1.32\\
    & 2021 & - & - & - & 2.00\\
    & 03/2023 & - & - & - & 0.60\\
    & 05/2023 & - & - & - & 0.04 \\
    & 06/2023 & - & - & - & 0.02\\
    & 02/2024 & - & - & - & 0.14\\
    & 04/2024 & - & - & - & 0.03\\
    & 06/2024 & - & - & - & 0.04\\
    \noalign{\smallskip}
    C & 2006 & $580^{-10}_{+10}$ & 37950$_{+60}^{-230}$ & 40830$^{-230}_{+220}$ & - \\
    & 2012 & $370^{-20}_{+20}$ & 38130$_{+50}^{-220}$ & 41190$_{+170}^{-40}$ & 1.32\\
    & 2021 & - & - & - & 2.00\\
    & 03/2023 & $40^{-30}_{+30}$ & 38360$_{+30}^{-20}$ & 39330$_{+60}^{-30}$ & 0.60\\
    & 05/2023 & 20$^{-10}_{+20}$ & 38400$^{-50}_{+20}$ & 39270$^{-20}_{+40}$ & 0.04\\
    & 06/2023 & 40$^{-10}_{+20}$ & 38260$^{-50}_{+20}$ & 39210$^{-20}_{+40}$ & 0.02\\
    & 02/2024 & $10^{-10}_{+20}$ & 38460$^{-50}_{+20}$ & 39040$^{-10}_{+50}$ & 0.14\\
    & 04/2024 & $10^{-10}_{+20}$ & 38430$^{-40}_{+20}$ & 39240$^{-20}_{+50}$ & 0.03\\
    & 06/2024 & $30^{-10}_{+20}$ & 38370$^{-50}_{+20}$ & 39210$^{-30}_{+40}$ & 0.04\\
    \noalign{\smallskip}
    \hline
    \noalign{\smallskip}
    \end{tabular}}
    \label{tab:tab_J1538}
    \tablefoot{Same description as Table \ref{tab:tab_J1249}.}

\end{table}

\par The variation of the BI parameter through the different epochs is shown in Fig. \ref{fig:balnicity}. Troughs B and C exhibited similar trends in both BI and velocity, gradually weakened over time. In contrast, trough A showed a distinct pattern, with its BI varying strongly (from 270 to 680 km s$^{-1}$) while the velocity range remained broad and nearly constant between $\sim$48000 and 54800 km s$^{-1}$ (see Table \ref{tab:tab_J1538}). In agreement with \citet{Vietri_2022}, {the coordinated variability of troughs B and C, again observed in the new epochs (UVES and MODS, observed in 2023 and 2024)} is very likely driven mainly by changes in the ionising continuum. In contrast, the behavior of trough A of WISSH71 favors the scenario in which gas motion is the dominant driver of the variability.

\begin{figure*}[t!]
    \centering
    \includegraphics[width=\linewidth]{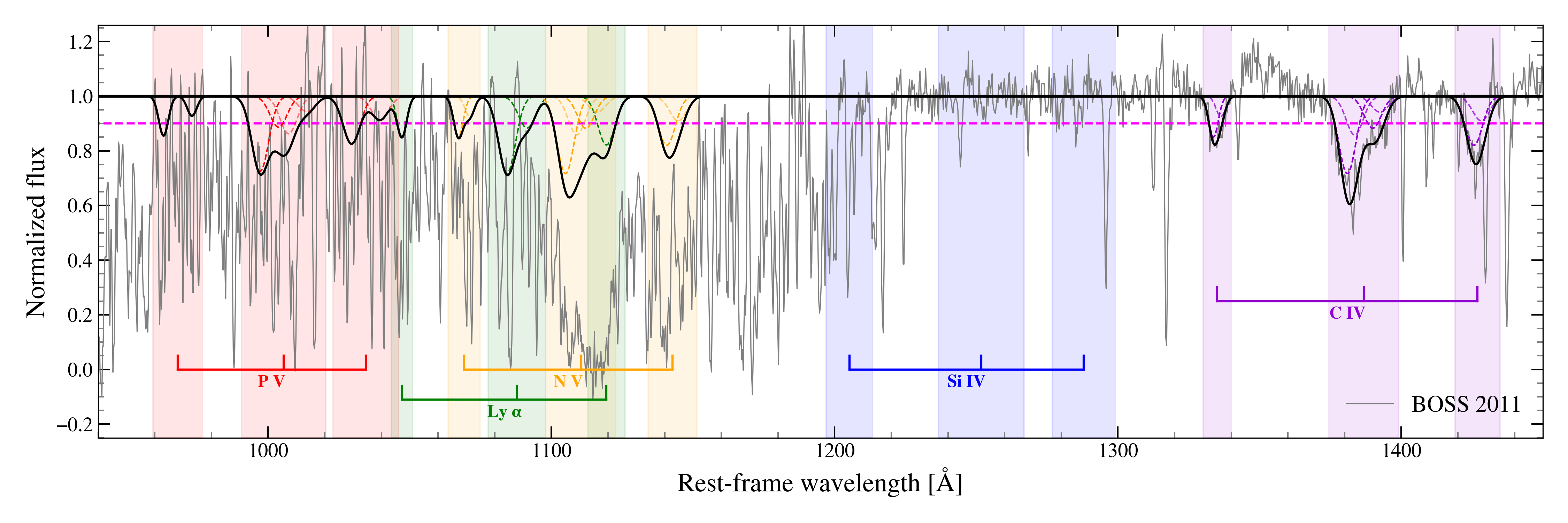}
    \caption{2011 BOSS spectrum of WISSH56, normalized by the continuum and emission lines, showing the fitted absorption troughs for each ion {assuming $C_{\rm f}=C_{\rm f,max}$}: \pv{} (red), \nv{} (orange), \lya{} (green), \siiv{} (blue), and \civ{} (purple). The best-fit model is shown in black, with the magenta dashed line indicating the 0.9 continuum level.}
    \label{fig:J1305_example_other_abs_fitting}
\end{figure*}

\begin{figure}[t!]
    \centering
    \includegraphics[width=\linewidth]{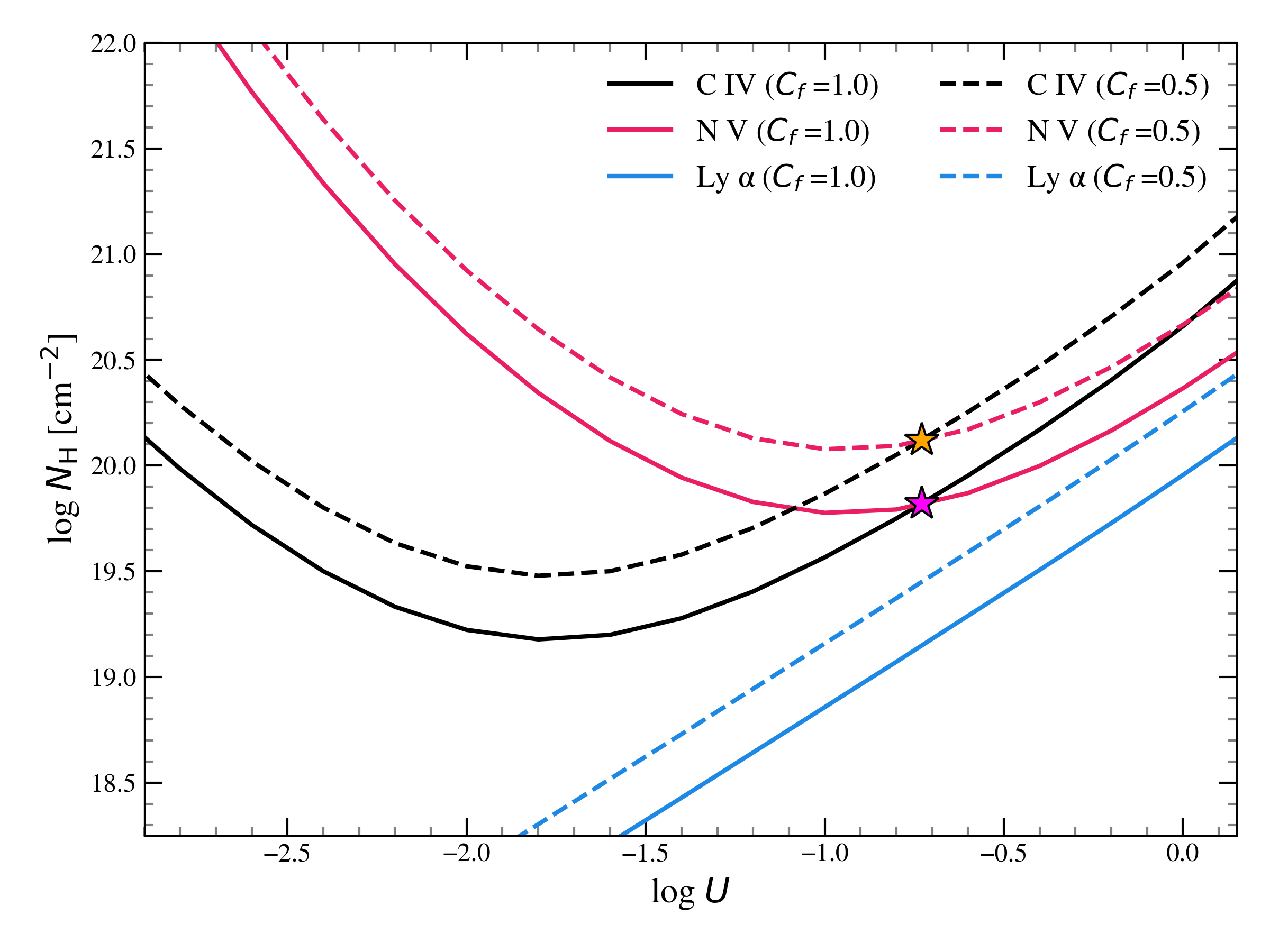}
    \caption{Relation between the theoretical values of the ionisation parameter $\log U$ and the computed column density $\log N_{\rm H}$ (see Sect. \ref{sec:ionising}), for trough D in the MODS 2025 spectrum of WISSH56. Dashed and solid lines indicate the $N_{\rm H}$ curves derived using $C_{\rm f,min}$ (in this case, 0.5) and $C_{\rm f,max}$, respectively. Black lines represent $N_{\rm H}$ constraint from \civ{}, while pink and blue lines show the upper limits from \nv{} and \lya{}, respectively. Orange and magenta stars indicate the most likely solution for $N_{\rm H}$ and $U$ for $C_{\rm f,min}$ and $C_{\rm f,max}$, respectively.}
    \label{fig:cloudy}
\end{figure}

\section{Ionising structure}
\label{sec:ionising}

\par Following the characterization of the \civ\ absorption troughs, we searched for corresponding absorption features in \pv, \lya, \nv, and \siiv\ at velocity offsets consistent with those derived for the \civ\ features. The absorption troughs of the different ions were fitted using the same velocity shift and width obtained from the \civ\ profile. The amplitudes of the Gaussian components were rescaled to preserve their relative proportions, allowing for variations up to the amplitude measured for \civ, which gave conservatives limits of the respective absorption depths, ensuring that the measured values represent the maximum plausible strength without overinterpreting the data due to noise or blending. {For doublet lines (\pv{}, \nv{}, and \siiv{}), a 2:1 optical depth ratio between the two components was assumed, and their wavelength separation was fixed according to the respective values for each line.} Additionally, the narrow absorption lines from the \lya\ forest were masked in the fitting procedure to avoid biasing the profile modeling. The multi-ion analysis was performed for the two limiting cases of the covering factor, $C_{\mathrm{f,min}}$ and $C_{\rm f,max}$, as determined in Sect.\ \ref{sec:civ_troughs}. An example of these fits done assuming $C_{\rm f,max}$ is shown in Fig.\ \ref{fig:J1305_example_other_abs_fitting}. 

\par  From the multi-ion fit, we derived the optical depths $\tau (v)$ for each ion. These values were used to calculate the ionic column density $N_{\rm ion}$, for each of these ions, following \citet{arav_2001}:

\begin{equation}
    N_{\rm ion}=\frac{3.7679 \times 10^{14}\ \textrm{cm}^{-2}}{\lambda f}\int\tau (v)\textrm{d}v,
\end{equation}
in which $\lambda$ is the laboratory wavelength and $f$ is the oscillator strength associated with the specific ionic transition.  
These measured $N_{\rm ion}$ values represent the observable constraints on the outflow's ionization state.

\par To translate $N_{\rm ion}$ into total hydrogen column densities $N_{\rm H}$ and ionization parameter $U$, we ran a grid of photoionisation models using the version C23 of \textsc{cloudy} \citep{Ferland_1998}. The models span a wide range of ionisation parameters ($-4 \leq \log U \leq 1$), assuming solar chemical abundances \citep{Lodders_2003}. The spectral energy distributions of each source used in the models were the ones derived by \citet{Saccheo_2023}. From these models, we extracted the predicted ionic fractions of each ion as a function of $\log U$, allowing us to compute the corresponding $\log N_{\rm H}$ as a function of $\log U$ for both minimum and maximum $C_{\rm f}$, taking into consideration the relativistic effects as reported in \citet{Luminari_2024}. The minimum covering factor, $C_{\mathrm{f,min}}$, provides upper limits on both $U$ and $N_{\rm H}$, since a smaller covering fraction requires higher ionic columns to reproduce the same absorption depth. Conversely, the full-covering case ($C_{\rm f} = 1$) yields lower limits on these quantities.

\par An example of the derived $\log N_{\rm H}$-$\log U$ curves is shown in Fig. \ref{fig:cloudy}. We considered as the most likely solution the values of $\log N_{\rm H}$ and $\log U$ where the curves of \civ{} and \nv{} intersected. The \civ{} curve provides a consistent lower limit on $\log N_{\rm H}$, since it is the strongest and most reliably measured line. Lower $N_{\rm H}$ values, such as those inferred from \lya{}, are therefore less likely to provide reliable $N_{\rm H}$ estimates.
Additionally, the curves of \pv{} and \siiv{} are not plotted in Fig. \ref{fig:cloudy} because they are highly uncertain, since the \siiv{} uBALs fall exactly in the \lya{} emission region and \pv{} is very difficult to detect.

\begin{table}[t!]
    \centering
    \caption{{Kinetic} properties of the \civ{} uBAL troughs.}
    \resizebox{\linewidth}{!}{
    \begin{tabular}{lcccccccc}
    \hline
    \hline
    \noalign{\smallskip}
    Source & Trough  & Scenario & $R_{\rm min}$/$R_{\rm max}$  & $\dot{E}_{\rm K,min}$/$\dot{E}_{\rm K,max}$ & $\dot{M}_{\rm min}$/$\dot{M}_{\rm max}$\\
    & & & [pc] & [$10^{44}$ erg s$^{-1}$] & [M$_{\odot}$ yr$^{-1}$]\\
    (1) & (2) & (3) & (4) & (5) & (6)\\
    \noalign{\smallskip}
    \hline
    \noalign{\smallskip}
    \multirow{3}{*}{WISSH53}& A &  ionisation &  -/420 & -/540 & -/60\\
    & B  & gas motion & 0.6/2 & 0.01/4 & 0.02/0.33\\
    & C & ionisation & 0.6/400 & 0.08/270 & 0.02/70\\
    \noalign{\smallskip}
    \hline
    \noalign{\smallskip}
    \multirow{3}{*}{WISSH56}& A+B & ionisation & 1.0/493 & 0.88/910 & 0.10/110\\
    & C & ionisation & 1.0/499 & 0.33/400 & 0.09/110\\
    & D & ionisation & 1.0/496 & 0.06/70 & 0.03/40\\
    \noalign{\smallskip}
    \hline
    \noalign{\smallskip}
    \multirow{3}{*}{WISSH71}& A & gas motion & 0.03$^\star$ & $7.12\times 10^{-5}$$^\star$ &  $7.6 \times 10^{-3}$$^\star$\\
    & B & ionisation & 1.0/252 & 0.32/230 & 0.04/30\\
    & C & ionisation & -/124 & -/100 & -/20\\
    \noalign{\smallskip}
    \hline
    \noalign{\smallskip}
    
    \end{tabular}
    }
   \tablefoot{(1) Source name. (2) uBAL trough. (3) Most favored scenario driving the observed variability in the troughs{, from which the maximum outflow radius $R_{\rm max}$ is derived (see Sect. \ref{sec:radius}).} (4) Minimum and maximum radii, where $R_{\rm min}$ corresponds to $R_{\rm BLR}$ estimated from the \citet{Lira_2018} \siiv{} radius-luminosity relation, and the maximum radius is derived from the adopted variability scenario (see Sect. \ref{sec:radius}). (5) Mean lower and upper limits of $\dot{E}_{\rm K}$, computed assuming $C_{\rm f,max}$ and $R_{\rm uBAL} = R_{\rm BLR}$ (lower limit) and $C_{\mathrm{f,min}}$ and $R_{\rm uBAL} = R_{\rm max}$ (upper limit). (6) Mean lower and upper limits of $\dot{M}$.  {{$^\star$Although this upper limit appears smaller than the BLR radius inferred from the \citet{Lira_2018} relation (adopted here as a lower limit), it is consistent with estimates of the inner BLR radius \citep[see e.g.,][]{gravity_2024}. Therefore, we can consider this value as a measurement.}}}
    \label{tab:kin_power}
\end{table}

\section{Determination of the outflow location}
\label{sec:radius}

\par As discussed in Section \ref{sec:civ_troughs}, the variability observed in the majority of the absorption troughs is primarily consistent with changes in the ionisation state of the outflowing gas. Only for the cases of trough B of WISSH53 and trough A of WISSH71, transverse gas motion seems to be the dominant factor. In the following, we describe the estimation of the distances of the absorbers under both scenarios, adopting the most plausible scenario for each case.

\subsection{Upper limits due to change in the ionisation scenario} 
\par
\begin{figure}
    \centering
    \includegraphics[width=\linewidth]{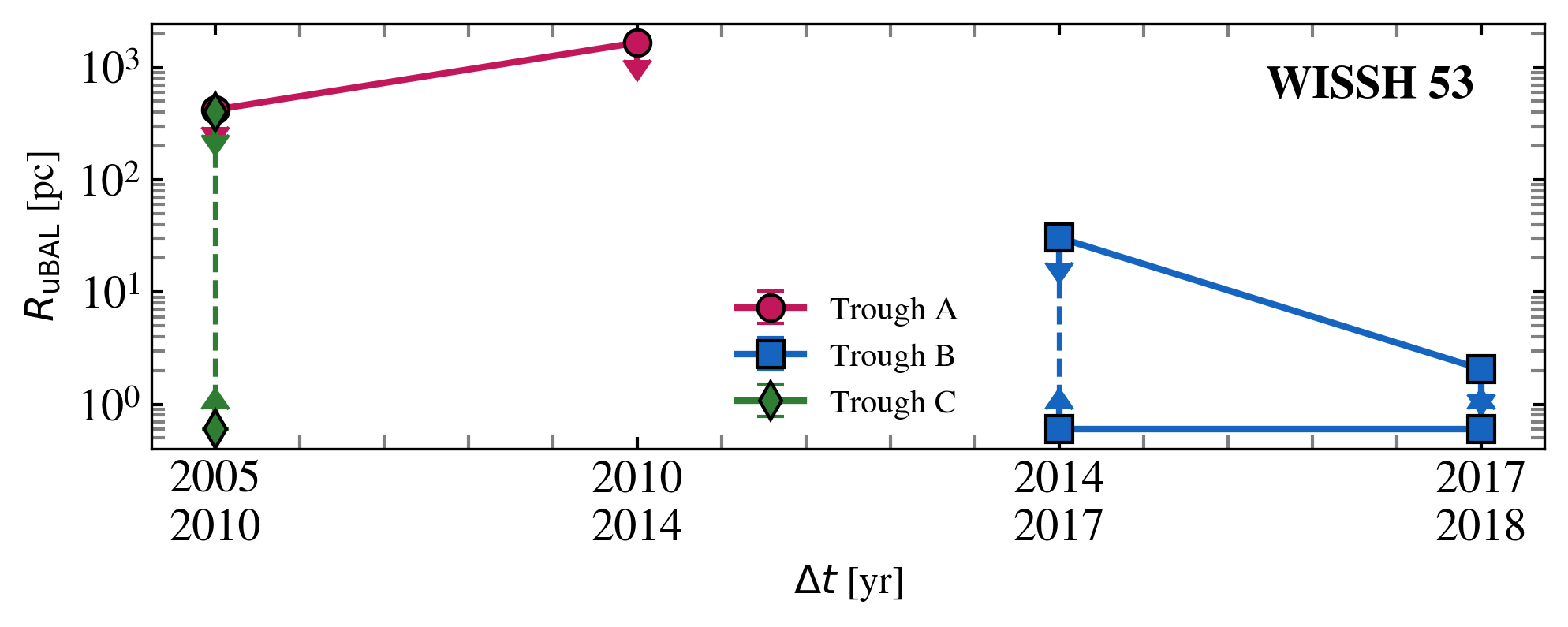}
    \includegraphics[width=\linewidth]{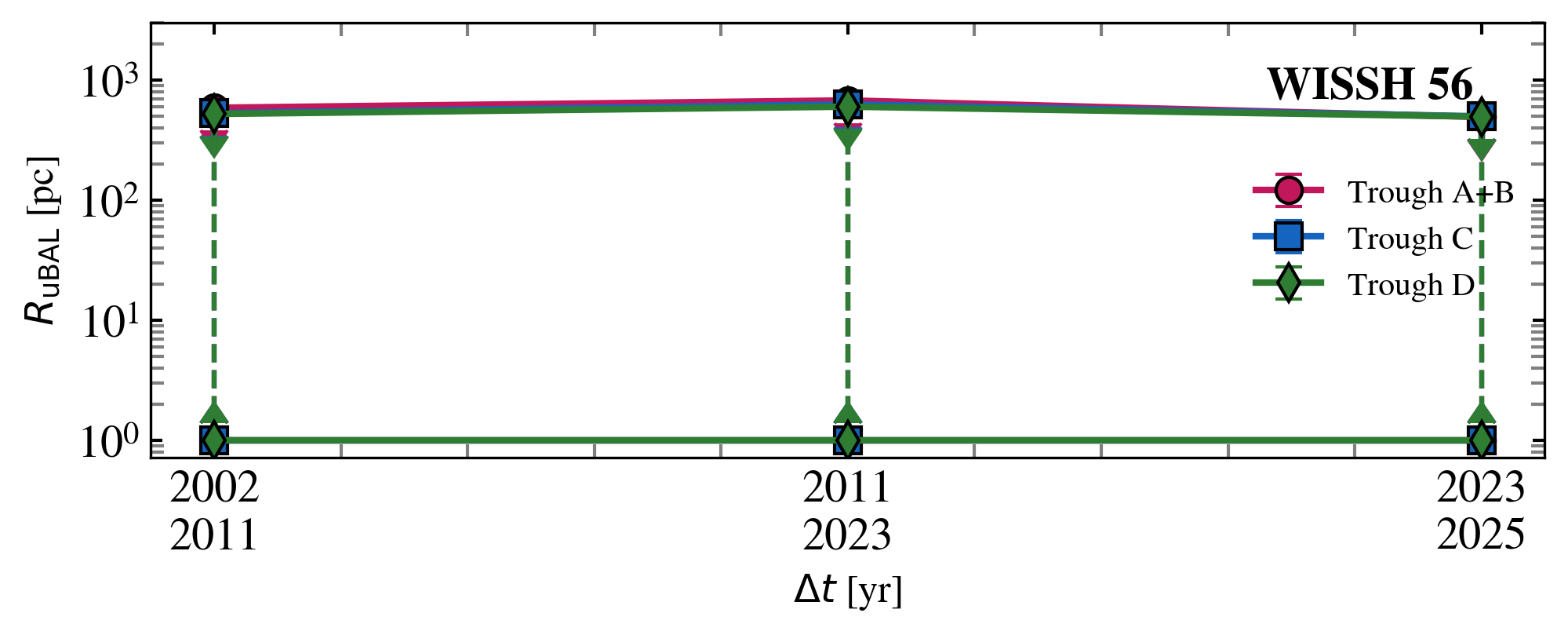}
    \includegraphics[width=\linewidth]{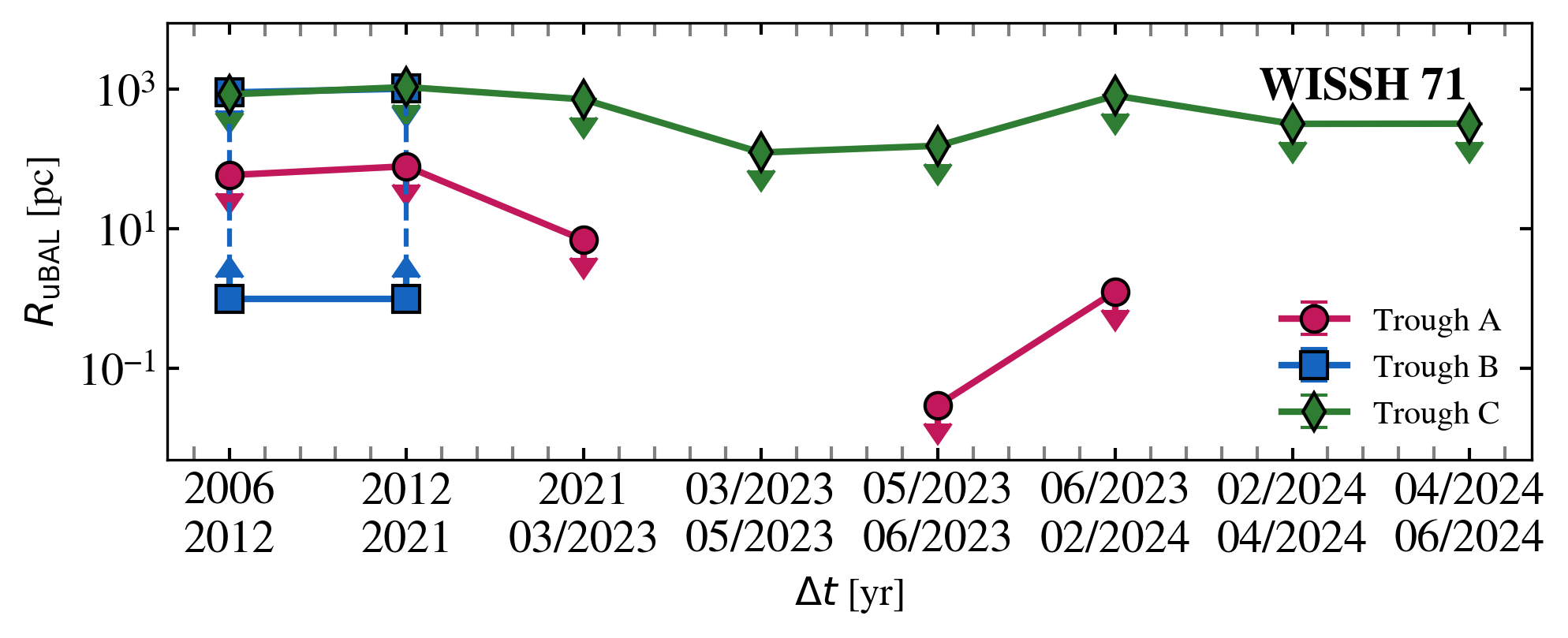}
    \caption{{Outflow radii as derived for each temporal baseline ($\Delta t$). Upper limits are derived from the variability analysis (see Sect.~\ref{sec:radius}), while lower limits are given by the BLR radius, estimated from the \citealt{Lira_2018} radius--luminosity relation (see Sect.~\ref{sec:radius_blr}). For each trough, the best-constrained upper limit corresponds to the smallest derived value of $R_{\rm uBAL}$, i.e. the one associated with the shortest temporal baseline ($\Delta t$). The different symbols indicate the different troughs. Vertical dashed lines connect upper and lower limits.}}
    \label{fig:radius_epochs}
\end{figure}

 According to \citet{Narayanan_2004}, the ionisation parameter $U$ is defined as

\begin{equation}
U = \frac{1}{4 \pi R_{\rm uBAL}^2 n_{\rm e} c} \int_0^{912  \text{ \AA}} \frac{\lambda L_\lambda}{hc}  d\lambda,
\end{equation}
where $R_{\rm uBAL}$ is the radial distance of the outflow, and $n_{\rm e}$ is the electron density that can be approximated as $n_{\rm e} \sim {1}/{\alpha_{\rm rec} t_{\rm rec}} \gtrsim {1}/{\alpha_{\rm rec}\Delta t}$,
in which $\alpha_{\rm rec}=2.8\times 10^{-12}\ \textrm{cm}^{-3}\ \textrm{s}^{-1}$ is the recombination coefficient \citep{Arnaud_1985} and $t_{\rm rec}$ is the recombination timescale. The largest possible value of $t_{\rm rec}$ is given by the time interval $\Delta t$ between two consecutive epochs. This approach allowed us to estimate upper limits on the distance $R_{\rm uBAL}$ of the absorbing gas responsible for the troughs, based on the theoretical values derived from the \textsc{cloudy} simulations (see Sect.~\ref{sec:ionising}). Fig. \ref{fig:radius_epochs} shows the $R_{\rm uBAL}$ variation depending on $\Delta t$. The preferred radius upper limits are those derived from the epochs separated by the shortest time interval, providing the most accurate estimate of the outflow distance.

\par Table \ref{tab:kin_power} reports the $R_{\rm uBAL}$ upper limits adopted for each trough. Under the assumption of ionisation-driven variability, the inferred upper limits on the absorber location are $R_{\rm uBAL}^{\rm upp}\lesssim$400 pc for WISSH53, $\lesssim$500 pc for WISSH56, and $\lesssim$ 250 pc for WISSH71. These upper limits lie within the radius distribution reported by \citet{He_2019}, who found typical outflow distances of tens to a few hundred parsecs in a large SDSS quasar sample (see their Fig. 2).

\subsection{Upper limits due to gas transverse motion scenario}
\label{sec:radius_gas_motion}
\par In this case,  the absorber distance was estimated assuming that the transverse speed of the absorbing cloud corresponds to the local Keplerian velocity around the SMBH. Following \citet{Capellupo_2013}, we derived the size of the continuum region at a given wavelength, $D_{\lambda}$, assuming a standard thin accretion disc emitting as a blackbody. Using Wien's law ($kT = hc/\lambda$) and following \citet{Morgan_2010}, the disc radius at the peak wavelength can be expressed as:  

\begin{equation}
R_{\lambda_\mathrm{max}} = 9.7 \times 10^{15} 
\left(\frac{\lambda_\mathrm{max}}{\mu\mathrm{m}}\right)^{4/3} 
\left(\frac{M_\mathrm{BH}}{10^9\,M_\odot}\right)^{2/3} 
\left(\frac{L_\mathrm{bol}}{\eta L_\mathrm{Edd}}\right)^{1/3} \mathrm{cm},
\end{equation}

where $M_\mathrm{BH}$ is the black hole mass, $L_\mathrm{bol}/L_\mathrm{Edd} = \lambda_\mathrm{Edd}$ the Eddington ratio, $\eta=0.10$ the radiative efficiency, and $\lambda_\mathrm{max}$ the peak emission wavelength. The size of the continuum is given by $D_{\lambda}=2 R_{\lambda_{\rm max}}$, where we assume $\lambda_{\rm max}=1300$ \AA\ for our sources, due to the location of the troughs.

\par To estimate the transverse motion of the absorption troughs, we measured the absorption strength $A_{\rm S}$, defined as the fraction of continuum-normalized flux removed by absorption ($0 \lesssim A_{\rm S} \lesssim 1 $) and its variation $\Delta A_{\rm S}$ between two consecutive observed epochs. $\Delta A_{\rm S}$ is computed as the difference in $A_{\rm S}$ measured within the velocity interval that satisfies two criteria: (i) a minimum width of 1200 km s$^{-1}$ and (ii) a flux variation of at least $4\sigma$ between the two epochs, where $\sigma$ is defined as in Eq. 1 of \citet{Capellupo_2012}. Figs. \ref{fig:J1249_delta_as} and \ref{fig:J1538_delta_as} show the regions considered in the $\Delta A_{\rm S}$ estimates for the trough B of WISSH53 and the trough A of WISSH71, respectively. Following the crossing-disc model of \citet{Capellupo_2013}, the linear distance traveled by the absorber can be expressed as $d = \sqrt{\Delta A_S}\, D_\lambda$, where $\Delta A_S$ is the change in absorption strength between two consecutive epochs and $D_\lambda$ is the size of the continuum source at the observed wavelength. The corresponding transverse velocity is ${v}_\mathrm{cross} = {d}/{\Delta t_\mathrm{rest}}$, with $\Delta t_\mathrm{rest}$ being the rest-frame time interval between the two epochs. Assuming that this transverse velocity corresponds to the Keplerian velocity around the central black hole, the location of the uBAL component is constrained by $R_\mathrm{uBAL} = {G M_\mathrm{BH}}/{{v}_\mathrm{cross}^2}$. 
\par Following this approach, the best-constrained upper limits on $R_{\rm uBAL}$ are $R_{\rm uBAL}^{\rm upp}\sim 2$ and $0.03$ pc for WISSH53 and WISSH71, respectively. In the latter case, the very short observed variability timescale ($\Delta t_{\rm rest} \simeq 0.02$ yr) between the MODS 05/2023 and MODS 06/2023 epochs leads to a radius smaller than the {estimated from \citet{Lira_2018}. However, such value is consistent with recent estimates of the BLR inner radius in similar targets. For instance, in the hyper-luminous quasar PDS 456, which has a luminosity comparable to our sources ($\log L_{\mathrm{bol}} \approx 47$ [erg s$^{-1}$]) at $z = 0.185$, the inner BLR radius has been estimated to be $\approx 0.01$ pc \citep{gravity_2024}. Furthermore, studies of quasars with high accretion rates \citep[e.g.][]{du_2019}, as the WISSH QSOs, have shown that their BLRs tend to be more compact than the standard radius–luminosity relationship would predict.}  
It was not possible to estimate such values for WISSH56, as we lack unbiased measurements of its black hole mass and Eddington ratio (see Table \ref{tab:sample}).
\par {Finally, we stress that the characteristic locations inferred from variability are proxies: cloud motions may deviate from purely circular Keplerian orbits due to turbulence and radial inflow/outflow components. If the effective transverse velocity differs from $\rm v_{K}$ by a factor $f$, the inferred characteristic radius rescales approximately as $f^2$, so our constraints are intended as physically motivated order-of-magnitude estimates.}

\subsection{Lower limits on the outflow distance}
\label{sec:radius_blr}
\par {{Since the majority of the \ion{C}{IV} absorption troughs lie in spectral regions dominated by BLR emission lines (e.g., Ly$\alpha$, \ion{N}{V}, \ion{O}{I}, \ion{C}{II}, and \ion{Si}{IV}), we assume that the absorber is located at radii comparable to or larger than the BLR. We therefore adopt the BLR radius as a lower limit to the outflow distance (see, however, \citealt{Turnshek_1988,Ganguly_2001} for an alternative interpretation in which smaller troughs do not fully cover the BLR, despite lying on top of the broad emission lines).}}

The BLR radius can be estimated by the empirical radius–luminosity relation of \citet{Lira_2018}, which relates the size of the \ion{Si}{IV}-emitting region to the ultraviolet continuum luminosity at 1350\AA. Here, $\lambda L_\lambda(1350\text{ \AA})$ was estimated from the fitted continuum in each epoch. This relation anchors the minimum distance of the absorber, assuming that the gas is located at or beyond the BLR. We adopted \ion{Si}{IV} instead of \ion{C}{IV} because the observed absorptions are located within the \ion{Si}{IV} emission line region, and in some cases they directly absorb parts of the \ion{Si}{IV} emission. In addition, due to the stratification of the BLR according to the ionisation potential of the lines, the \ion{Si}{IV} emission typically originates at slightly larger radii than \ion{C}{IV}, making it a conservative choice for estimating the minimum outflow distance. This approach results in $R_{\rm BLR} \simeq$ 0.6, 1.0, and 1.0 pc for WISSH53, WISSH56, and WISSH71, respectively.

\par In two cases, the trough A of WISSH53 and the trough C of WISSH71, the absorption features lie within regions dominated by continuum, where there are no broad emission lines. Consequently, the absorbing gas in these cases is unlikely to be associated with the line-emitting region and therefore $R_{\rm BLR}$ cannot be reliably used as a lower limit of the outflow distance in these troughs.

\section{Kinetic power estimates}
\label{sec:kin_power}
\par Following \citet{Hamann_2019}, we estimated the kinetic power associated with each uBAL trough using the expression:

\begin{equation}
\resizebox{0.95\columnwidth}{!}{$
\dot{E}_{\rm K,uBAL}=\frac{A}{t_{\rm flow}}\left(\frac{Q}{0.15}\right)\left(\frac{N_{\rm H}}{5 \times 10^{22} \,\textrm{cm}^{-2}}\right)\left(\frac{R_{\rm uBAL}}{1 \,\textrm{pc}}\right)^2\left(\frac{{\rm v}_{\rm uBAL}}{8000 \,\textrm{km s}^{-1}}\right)^2 \,\textrm{erg,}
$}
\end{equation}
where $A= 4.8\times 10^{53}$ and $t_{\rm flow}=R_{\rm uBAL}/{\rm v}_{\rm uBAL}$. $Q$ is the global covering factor \citep[$Q=0.15$, based on the incidence of \civ{} BALs in SDSS QSOs,][]{Gibson_2009,Hamann_2019}, $N_{\rm H}$ is the hydrogen column density, $R_{\rm uBAL}$ is the outflow radius, and ${\rm v}_{\rm uBAL}$ is the characteristic outflow velocity measured in the absorption troughs in the corresponding epochs. {Here, ${\rm v}_{\rm uBAL}$ is assumed to be the maximum velocity ${\rm v}_{\rm max}$ of each trough.} 

\begin{figure*}
    \centering
    \includegraphics[width=0.48\linewidth]{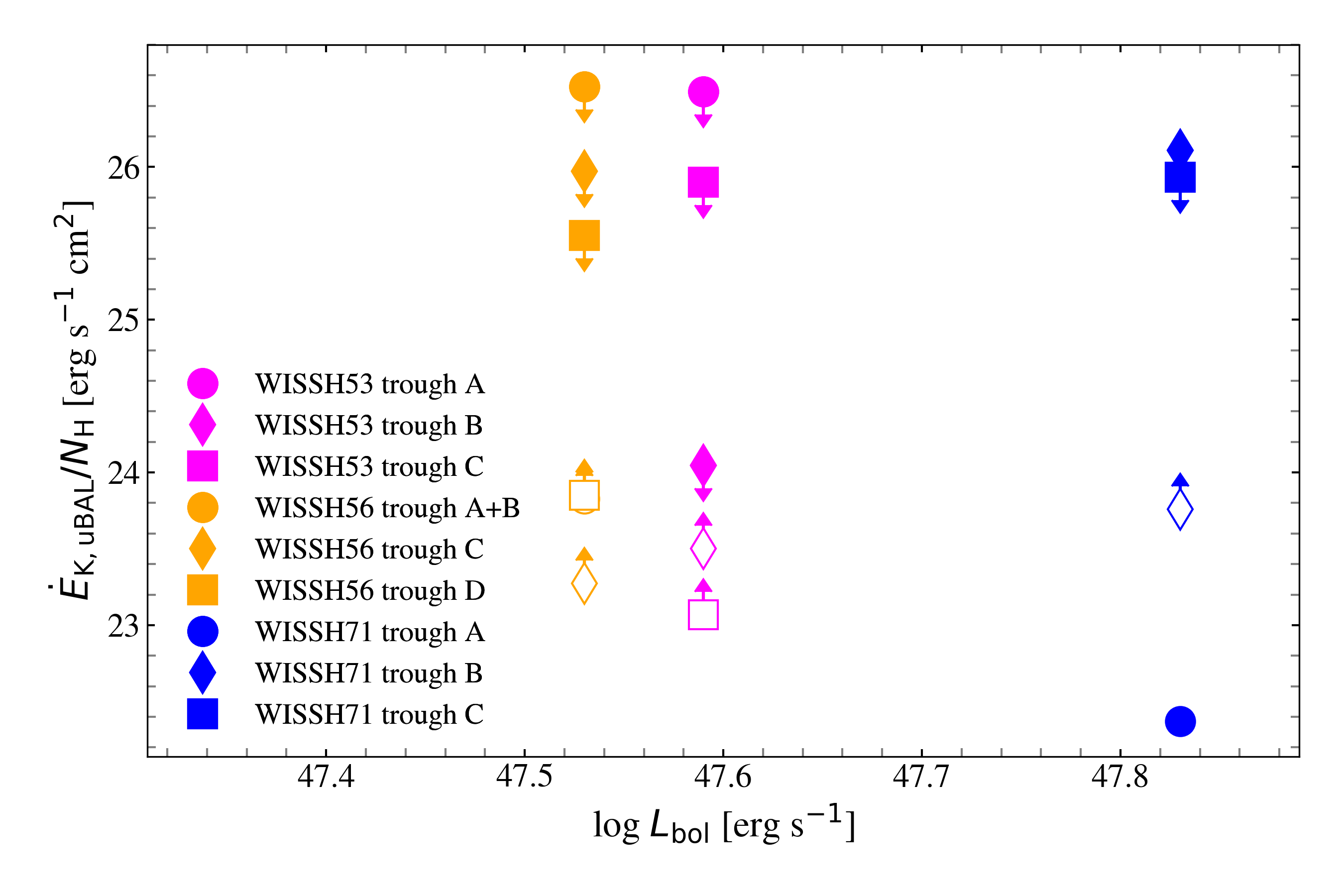}
    \includegraphics[width=0.48\linewidth]{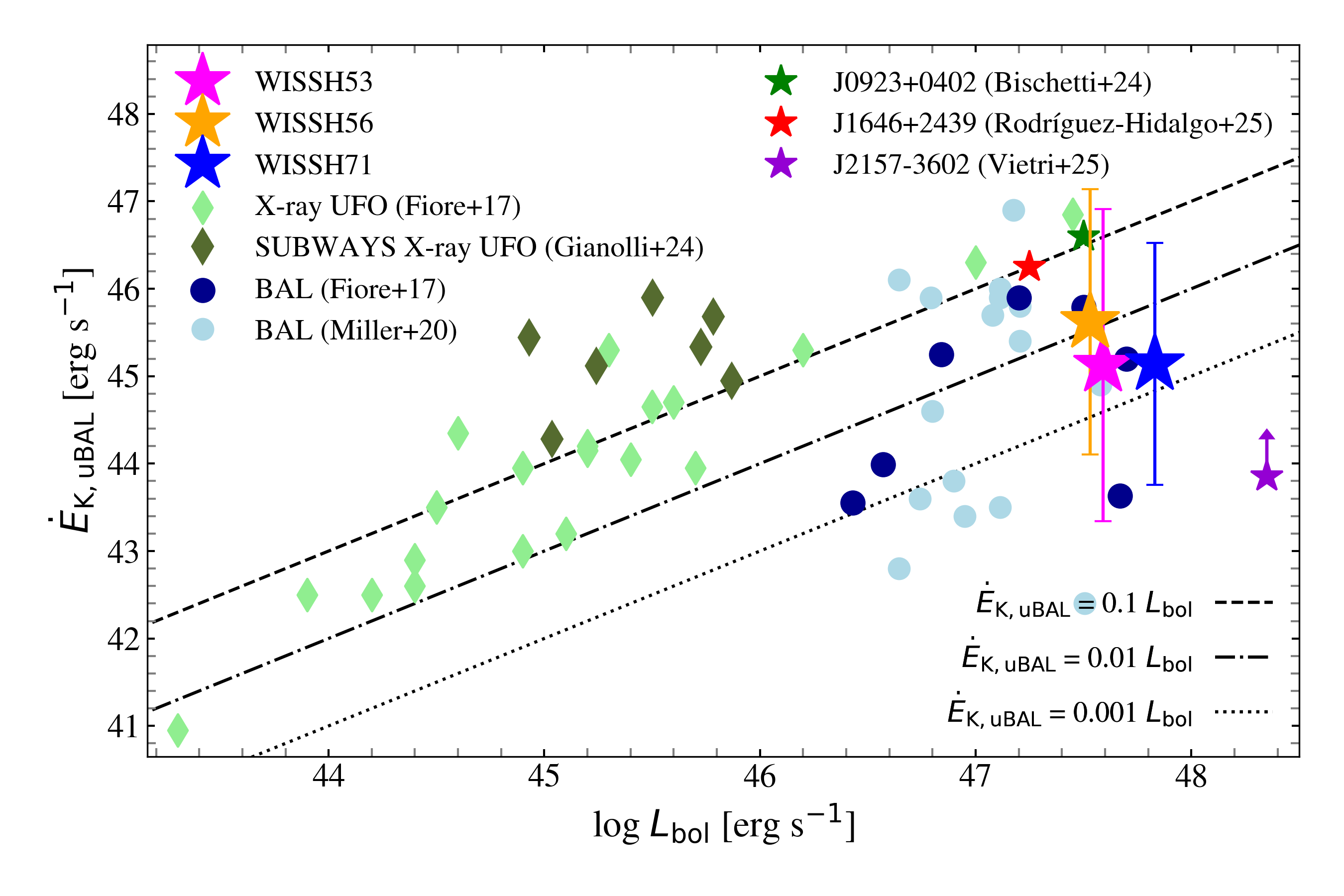}
    \caption{{Outflow energetics as a function of bolometric luminosity $L_{\rm bol}$.} \textit{Left panel:} {Zoom-in on the three WISSH sources, showing the $N_{\rm H}$-normalized lower and upper limits of $\dot{E}_{\rm K,uBAL}$ for each individual trough.} WISSH53, WISSH56, and WISSH71 troughs are represented by magenta, orange, and blue symbols, respectively, with distinct marker types for each trough. {Filled and open symbols indicate the upper and lower limits of $\dot{E}_{\rm K,uBAL}/N_{\rm H}$ (derived using $R_{\rm max}$ and $R_{\rm BLR}$), respectively.} {The error bars in the WISSH quasars indicate the range between the derived upper and lower limits.} \textit{Right panel:} Comparison of our measurements with other samples and individual sources from literature. {For each of our three sources, we plot the mean $\dot{E}_{\rm K,uBAL}$, obtained from the respective summed upper and lower limits.} Dark-blue circles and light-green diamonds indicate BAL and X-ray samples from \citet{Fiore_2017}, respectively. BAL quasars from \citet{Miller_2020} are shown as light-blue circles. {Geometric mean values of} {the Supermassive Black Hole Winds in X-Rays (SUBWAYS)} UFOs {\citep{Gianolli_2024}} are represented by dark-green diamonds. Dashed, dot-dashed, and dotted lines correspond to 0.1, 1, and 10\% of $L_{\rm bol}$, respectively.}
    \label{fig:ekin_range}
\end{figure*}

\par For each epoch, we estimated $\dot{E}_{\rm K,uBAL}$ taking into consideration the values of $N_{\rm H}$ and $R_{\rm uBAL}$ derived in the previous sections. 
{To illustrate how the time separation between two epochs affects  $\dot{E}_{\rm K,uBAL}$ (through its impact on the estimate of $R_{\rm uBAL}$, which is inferred from variability analysis), we estimated $\dot{E}_{\rm K,uBAL}/N_{\rm H}$, as shown in left plot of Fig. \ref{fig:ekin_range}. This normalization emphasizes how shorter and better-sampled monitoring can significantly improve constraints on the recombination timescales and, consequently, on the outflow radius. Indeed, the best constrained radii were found for trough A of WISSH71 and trough B of WISSH53, for which we derived $R_{\rm uBAL}^{\rm upp} \sim 0.03$ pc and $\sim 2$ pc, respectively. These tight limits are due to the short rest-frame time intervals ($\Delta t \sim 0.13$ and 0.02 yr). Instead, the largest radius limits were obtained for WISSH56, where the shortest $\Delta t$ is 0.42 yr (see Section \ref{sec:radius}). }

\par {Since $\dot{E}_{\rm K,uBAL}$ also depends on the column density, constraining $N_{\rm H}$ is equally important. \citet{Hamann_2019} detected \pv\ in every composite spectrum, including both mini-BALs and the strongest BALs. The detection of \pv\ implies $\log N_{\rm H} > 22.7$ cm$^{-2}$. Such high $N_{\rm H}$ would lead to significantly larger $R_{\rm uBAL}$ estimates compared to those inferred from other density tracers, and consequently to extremely high $\dot{E}_{\rm K,uBAL}$, typically up to two orders of magnitude larger than the values reported in Table \ref{tab:kin_power}. The detection of \pv\ in our targets is therefore expected in light of the results of \citet{Hamann_2019}. However, it falls within the Ly$\alpha$ region, making its identification uncertain.}

\par {Throughout our analysis, we use the $N_{\rm H}$ values derived following Section \ref{sec:ionising}. The lower limits were obtained assuming $C_{\rm f} = 1.0$ and $R_{\rm uBAL} = R_{\rm min}$, while the upper limits were derived using $C_{\rm f,min}$ and $R_{\rm uBAL} = R_{\rm max}$.} {Assuming that all the uBAL troughs originate from a single outflow episode, we summed the derived lower and upper limits of $\dot{E}_{\rm K,uBAL}$ across all troughs. The resulting range and mean for each target are shown in the right panel of Fig. \ref{fig:ekin_range}. The total $\dot{E}_{\rm K,uBAL}$ corresponds to  $ \sim 0.006$-$20\%$, $0.04$-$40\%$, and $0.008$-$5\%$ of $L_{\rm bol}$ for WISSH53, WISSH56, and WISSH71, respectively.} It is known that outflows with kinetic powers exceeding $\sim$ 0.5-5\% of $L_{\rm bol}$ are considered capable of significantly affecting their host galaxies by driving gas out of the nuclear regions, quenching star formation, or regulating the growth of the central SMBH \citep[e.g.,][]{DiMatteo_2005,Hopkins_2010}.
\par Our derived mean $\dot{E}_{\rm K,uBAL}$ values are slightly lower than what is found for the X-ray UFO reported in \citet{Fiore_2017} and the {mean} values estimated for the SUBWAYS survey \citep{Gianolli_2024}. Almost all of the X-ray UFOs shown in this figure, which present $ L_{\rm bol} \sim 10^{43}$ to $10^{46.5}$ erg s$^{-1}$, have kinetic efficiencies > 1\%, with roughly half of them reaching $\dot{E}_{\rm K,uBAL}/L_{\rm bol} > 10 \%$.  In contrast, the $\dot{E}_{\rm K,uBAL}$ values derived in the present work are consistent with the values measured for UV BAL outflows from \citet{Fiore_2017} and \citet{Miller_2020}, located at $L_{\rm bol} > 10^{46.5}$ erg s$^{-1}$ and with outflow velocities up to $\sim$ 20000 km s$^{-1}$. Our estimates also agree with the values found for individual sources showing BAL and uBAL  outflows with velocities $\gtrsim 30000$ km s$^{-1}$ \citep{Bischetti_2024,Rodriguez-Hidalgo_2025,Vietri_2025}.

\par {We also derived the mass outflow rates using \(\dot{M}_{\rm uBAL} = 2\dot{E}_{\rm K,uBAL}/{\rm v}_{\rm uBAL}^2\) (see Table~\ref{tab:kin_power}). To assess the relevance of these outflows in the context of black hole growth, we compared them to the accretion rate onto the SMBH, estimated as \(\dot{M}_{\rm acc} = L_{\rm bol} / \eta c^2\), assuming a typical radiative efficiency \(\eta = 0.1\) \citep[e.g.,][]{Davis_2011}. 
We find \(\dot{M}_{\rm uBAL}/\dot{M}_{\rm acc}\) ratios ranging from \(\sim 0.006\) to \(1.93\) for WISSH53, \(\sim 0.004\) to \(4.34\) for WISSH56, and \(\sim 0.0004\) to \(0.45\) for WISSH71, suggesting that the uBAL outflows can carry mass at rates comparable to $\dot{M}_{\rm acc}$. {Our values are generally smaller than those reported by \citet{Miller_2020}, who found $\dot{M}_{\rm out}$ from tens to several thousand $M_{\odot}$ yr$^{-1}$ and $\dot{M}_{\rm out}/\dot{M}_{\rm acc} \sim 0.04$ and $> 10^2$. This difference is mainly caused by the large radii derived for some of their sources. Our results are instead more consistent
with those reported by \citet{Fiore_2017}, who found $\dot{M}_{\rm out}$ of a few $M_{\odot}$ yr$^{-1}$ and $\dot{M}_{\rm out}/\dot{M}_{\rm acc} \sim $ 0.02-0.4. }}

\par While the exact $\dot{E}_{\rm K,uBAL}$  {and $\dot{M}_{\rm uBAL}$} values are sensitive to many assumptions about the physical conditions such as the radii and column densities, the derived ranges shown in Table \ref{tab:kin_power} suggest that, even under conservative assumptions, the uBALs in our sample are able to carry a significant fraction of $L_{\rm bol}$, with kinetic efficiencies approaching or, in some cases, exceeding the thresholds thought to drive the AGN feedback. This finding therefore highlights uBALs as prime candidates for powering large-scales feedback in luminous quasars. It is crucial to monitor statistical samples of uBALs and to investigate the connection between their kinetic output and galaxy-wide outflows, and to further link the nuclear uBALs with the properties of the gas on CGM scales, as both cosmological hydrodynamic simulations and observational studies suggest that small-scale outflows and large-scale nebulae may, in fact, be physically connected {\citep[e.g.][]{Gaspari_2020,Travascio_2020, Costa_2022}}.

\section{Summary and conclusions}
\label{sec:summary}
\par We presented a detailed multi-epoch spectroscopic study of three hyper-luminous quasars from the WISSH sample (WISSH53, WISSH56, and WISSH71), focusing on the variability of their uBALs in the rest-frame UV. We modeled the \civ{} absorption troughs and detected multiple components reaching velocities up to  0.2$c$, revealing complex variability patterns. The analysis of such variations allowed the estimate of the uBAL energetics. The main findings can be summarized as follows:
\begin{itemize}
    \item[-] WISSH53 exhibits three distinct \civ{} absorption troughs, spanning rest-frame velocities from $\sim 31000$ to $54000$ km s$^{-1}$. {The corresponding BI values are relatively small for all components, as reported in Table \ref{tab:tab_J1249}.} The variability analysis indicates that troughs A and C are likely driven by changes in the ionising continuum, while trough B is best explained by transverse motion of absorbing gas crossing the line of sight. {The inferred outflow distances differ significantly between these scenarios (see Table \ref{tab:kin_power}).} {The mean $\dot{E}_{\rm K,uBAL}$ of WISSH53 shown in Fig. \ref{fig:ekin_range} corresponds to $\sim$ 0.3\% of $L_{\rm bol}$, with trough A dominating the contribution.}

    \item[-] In WISSH56, four \civ{} absorption troughs are identified in a broad range of velocities{, reaching up to $58000$ km s$^{-1}$ (see Table \ref{tab:tab_J1305}).} The BI values increase considerably over the observed epochs, with the combined A+B trough reaching $\sim$ 2700 km s$^{-1}$ in 2025{, one of the largest BI reported so far for uBALs.} The troughs exhibit significant variability particularly over the epochs 2023 and 2025, {making this source an excellent candidate to further investigate the uBALs properties through intensive monitoring observations in the future years}. The coordinated variability of all troughs suggests that changes in the ionising continuum dominate the observed behavior of the features, leading to $r_{\rm out}\lesssim$ 500 pc. {The inferred $\dot{E}_{\rm K,uBAL}$ values are significant, with the mean across all troughs reaching $\sim$1.2\% of $L_{\rm bol}$.}

    \item[-] For WISSH71, three \civ{} troughs are detected, with velocities up to $55000$ km s$^{-1}$. {All three features temporarily disappear in 2021, but they reappeared with the new observations from VLT/UVES and LBT/MODS. Fig. \ref{fig:balnicity} shows that troughs B and C present correlated BI trends, gradually weakening over time.} An extremely short-timescale variation is observed between the May and June 2023 MODS spectra, corresponding to only $\sim 0.02$ yr ($\sim$ 7 days) in the rest frame{, allowing for a better constrained $R_{\rm uBAL}$ estimate}.  Under the gas-motion scenario, trough A is located at $\sim$ 0.03 pc, whereas the troughs B and C, whose variability are likely ionisation-driven{, are at larger distances (Table \ref{tab:kin_power})}. {The mean value of $\dot{E}_{\rm K,uBAL}$ of WISSH71 correspond to $\sim$ 0.2\% of $L_{\rm bol}$.} 
\end{itemize}

\par 
{As shown in Fig. \ref{fig:ekin_range}, the $\dot{E}_{\rm K,uBAL}$ values are comparable with the 0.5-5\% $L_{\rm bol}$ threshold commonly invoked for effective AGN feedback. The large uncertainties on $\dot{E}_{\rm K,uBAL}$ are primarily driven by the limited constraints currently available on $N_{\rm H}$ and $R_{\rm uBAL}$ of the outflowing gas. However, it is worth noting that in the case of trough B of WISSH53 we were able to constrain a $R_{\rm max}/R_{\rm min} \sim 3$. This highlights the} importance of multi-epoch, multi-ion analyses to accurately constrain their energetics. Future multi-epoch observations across different ionic species are essential to fully characterize the energetics and feedback impact of these extreme quasar outflows. {This study, limited to three sources, can be regarded as a pathfinder aimed at exploring the parameter space of the most extreme BALs through multi-epoch spectroscopy. Furthermore, a complete census of uBALs in the WISSH quasars (Vietri, in prep.) will enable this investigation to be extended both qualitatively and quantitatively through dedicated observational campaigns designed to probe the origins of variability in these outflows, to tightly constrain their physical properties, and to assess their potential role as an efficient feedback mechanism in luminous quasars.}

\begin{acknowledgements}
A.D.M., G.V., E.P., and A.G. acknowledge financial support from the Bando Ricerca Fondamentale  INAF 2023 Guest Observer Grant ``Assessing the role of ultra-fast outflows in hyper-luminous quasars at Cosmic Noon''. G.V. acknowledges financial support from the Bando Ricerca Fondamentale INAF 2022 Mini-grant ``Searching for UV ultra-fast outflow in AGN by exploiting widearea public spectroscopic surveys''. T.M. acknowledges financial support from the JSPS KAKENHI grant No. 25K01038 ``Comprehensive study of the feedback efficiency of AGN outflows to host galaxies''. E.B. acknowledges the support of the INAF GO grant ``A JWST/MIRI MIRACLE: Mid-IR Activity of Circumnuclear Line Emission'' and of the ``Ricerca Fondamentale 2024'' INAF program (mini-grant 1.05.24.07.01). F.S., L.Z., and M.B. acknowledge financial support from European Union - Next Generation EU, PRIN/MUR 2022 2022TKPB2P – BIG-z. F.S. and M.B. also acknowledge financial support from Ricerca Fondamentale INAF 2023 Data Analysis grant 1.05.23.03.04 ``ARCHIE ARchive Cosmic HI \& ISM  Evolution''. F.S. also acknowledges financial support from Ricerca Fondamentale INAF 2024 under project 1.05.24.07.01 MINI-GRANTS RSN1 ``ECHOS'' and Bando Finanziamento ASI CI-UCO-DSR-2022-43 CUP:C93C25004260005 project ``IBISCO: feedback and obscuration in local AGN''. L.Z. also acknowledges financial support from the Bando Ricerca Fondamentale INAF 2022 Large Grant ``Toward an holistic view of the Titans: multi-band observations of z > 6 QSOs powered by greedy supermassive black holes'' and Bando Ricerca Fondamentale INAF 2024 Large Grant ``The DEepest study of LUminous QSOs in X-ray at z=2-7 (DELUX)''. M.G. acknowledges support from the ERC Consolidator Grant \textit{BlackHoleWeather} (101086804). Based on observations collected at the European Organisation for Astronomical Research in the Southern Hemisphere under ESO programmes  110.240J.001, 075.A-0464(A), 189.A-0424(A), and 197.A-0384(A). This work is also based on observations collected by LBT under the programmes  IT-2022B-035 and IT-2023B-031.  The LBT is an international collaboration among institutions
in the United States, Italy and Germany. LBT Corporation partners are: Istituto
Nazionale di Astrofisica, Italy; The University of Arizona on behalf of the Arizona Board of Regents; LBT Beteiligungsgesellschaft, Germany, representing
the Max-Planck Society, The Leibniz Institute for Astrophysics Potsdam, and
Heidelberg University; The Ohio State University, representing OSU, University
of Notre Dame, University of Minnesota and University of Virginia. This paper
used data obtained with the MODS spectrographs built with funding from NSF
grant AST-9987045 and the NSF Telescope System Instrumentation Program
(TSIP), with additional funds from the Ohio Board of Regents and the Ohio
State University Office of Research. This research is based in part on data collected at the Subaru Telescope, which is operated by the National Astronomical Observatory of Japan. We are honored and grateful for the opportunity of observing the Universe from Maunakea, which has the cultural, historical, and natural significance in Hawaii.   
\end{acknowledgements}

\bibliographystyle{aa}
\bibliography{bib}

\begin{appendix}

\section{Archival spectroscopic data and details of the rest-frame UV observations}
\label{archival}
\par The archival spectroscopic data used in this paper are the following:
\begin{itemize}
    \item[-] \textit{SDSS/BOSS:} {For the three sources, we collected spectra from the archive \citep{Gunn_2006,Smee_2013}.} The data for WISSH53 were released in DR1 and DR9, for WISSH56 in DR2 and DR9, and for WISSH71 in DR6 and DR10.
    \item[-] \textit{VLT/UVES:} For WISSH53, we retrieved an additional spectrum from the ESO archive (programme ID 075.A-0464(A), PI: T.-S. Kim), observed in 2005. The observations were carried out with a 1.0\arcsec\ slit with exposures during April, May, and June 2005. The data reduction was also performed using \textsc{esorex}. As no significant differences were found among these three epochs, the data were combined into a single spectrum.
    \item[-] \textit{VLT/XSHOOTER:} For WISSH53, we also included the XQ-100 spectrum available in the ESO archive (programme ID 189.A-0424(A), PI: S. López), observed and reduced by \citet{Lopez_2016}. A slit width of 0.9\arcsec\ was used for the UVB arm, while a width of 1.0\arcsec\ was adopted for both the VIS and NIR arms, resulting in a spectral resolution ranging from 5100 to 8800.
    \item[-] \textit{{VLT/MUSE}}: For WISSH53, we also included the fully reduced data available in the ESO archive (programme 197.A-0384(A), PI: M. Fumagalli). The target was observed twice in 2017 and twice in 2018. No significant differences were found between observations taken within the same year, therefore the exposures were combined, resulting in a single spectrum for 2017 and another for 2018. 
    
    \item[-] \textit{Subaru/FOCAS}: For WISSH71, we also included the Subaru/FOCAS spectrum (PI: T. Misawa) presented in \citet{Vietri_2022}. 
\end{itemize}

\par A summary of the new and archival observations is listed in Table \ref{table:obs}.

\begin{table}[h!]
\caption{Details of the rest-frame UV observations.}            
\label{table:obs}      
\centering                      
\resizebox{\linewidth}{!}{
\begin{tabular}{lccccc}        
\hline\hline  
\noalign{\smallskip}
Obs. ID & Instrument & R & $\lambda$ & Obs. date & Seeing \\ 
& & & [\AA] & & ['']\\
(1) & (2) & (3) & (4) & (5) & (6)\\
\noalign{\smallskip}
\hline
\noalign{\smallskip}
\multicolumn{6}{c}{WISSH53}\\
\noalign{\smallskip}
\hline              
\noalign{\smallskip}
SDSS 2001 & SDSS & 2000 & 3800-9200 & 2001-03-31 & 2.5\\
UVES 2005 & VLT/UVES & 40000 & 3000-8000 & 2005-04-07 & 0.8\\
 &  & 40000 & 3000-8000 & 2005-04-09 & 1.0\\
 &  & 40000 & 3000-8000 & 2005-05-12 & 0.4\\
 &  & 40000 & 3000-8000 & 2005-05-14 & 0.8\\
 &  & 40000 & 3000-8000 & 2005-05-15 & 0.9\\
 &  & 40000 & 3000-8000 & 2005-06-03 & 0.5\\
BOSS 2010 & BOSS & 2000 & 3600-10300& 2010-01-26 & 2.5\\
XSHOOTER 2014 & VLT/XSHOOTER & 8800 & 3100-24800 & 2014-01-28 & 1.1\\
{MUSE 2017} & VLT/MUSE & 3000 & 4650-9300 & 2017-04-29 & 0.9\\
 &  & 3000 & 4650-9300 & 2017-06-20 & 0.9\\
{MUSE 2018} & VLT/MUSE & 3000 & 4650-9300 & 2018-01-24 & 0.7\\
 &  & 3000 & 4650-9300 & 2018-02-11 & 0.7\\

UVES 2023 & VLT/UVES & 40000 & 3000-8000 & 2023-03-15 & 0.7\\
\noalign{\smallskip}
\hline
\noalign{\smallskip}
\multicolumn{6}{c}{WISSH56}\\
\noalign{\smallskip}
\hline
\noalign{\smallskip}
SDSS 2002 & SDSS & 2000 & 3800-9200& 2002-06-14 & 2.9\\
BOSS 2011 & BOSS & 2000 & 3600-10300& 2011-05-05 & 1.3\\
UVES 2023 & VLT/UVES & 40000 & 4000-10000 & 2023-02-28 & 0.5\\
 & & 40000 & 4000-10000 & 2023-03-28 & 0.7\\
{MODS 2025} & LBT/MODS & 2000 & 3200-10000 & 2025-04-26 & 0.9\\
& & 2000 & 3200-10000 & 2025-04-29& 1.2\\
\noalign{\smallskip}
\hline
\noalign{\smallskip}
\multicolumn{6}{c}{WISSH71}\\
\noalign{\smallskip}
\hline
\noalign{\smallskip}
SDSS 2006 & SDSS & 2000 & 3800-9200 & 2006-04-05 & 1.4\\
BOSS 2012 & BOSS & 2000 & 3600-10300 & 2012-04-16& 1.4\\
FOCAS 2021 & Subaru/FOCAS & 2500 & 3700-6000 & 2021-06-19 & 0.7\\
UVES 03/2023 & VLT/UVES & 40000 & 3000-8000 & 2023-03-16 & 0.5\\
MODS 05/2023 & LBT/MODS & 2000 & 3200-10000 & 2023-05-28 & 1.1\\
MODS 06/2023 & LBT/MODS & 2000 & 3200-10000 & 2023-06-25 & 1.0\\
MODS 02/2024 & LBT/MODS & 2000 & 3200-10000 & 2024-02-14 & 1.3\\
& & 2000 & 3200-10000 & 2024-02-19 & 1.0\\
MODS 04/2024 & LBT/MODS & 2000 & 3200-10000 & 2024-04-10 & 1.0\\
MODS 06/2024 & LBT/MODS & 2000 & 3200-10000 & 2024-06-09 & 1.0\\
\hline
\end{tabular}
}
\tablefoot{(1) Identification of the observation. (2) Spectrograph. (3) {Spectral resolution.} (4) Observed wavelength range in \AA. (5) Date of observation. (6) Averaged seeing in arcsecs.}
\end{table}

\onecolumn
\section{WISSH53}
\label{J1249}

\begin{figure*}[h!]
    \centering
    \includegraphics[width=\linewidth]{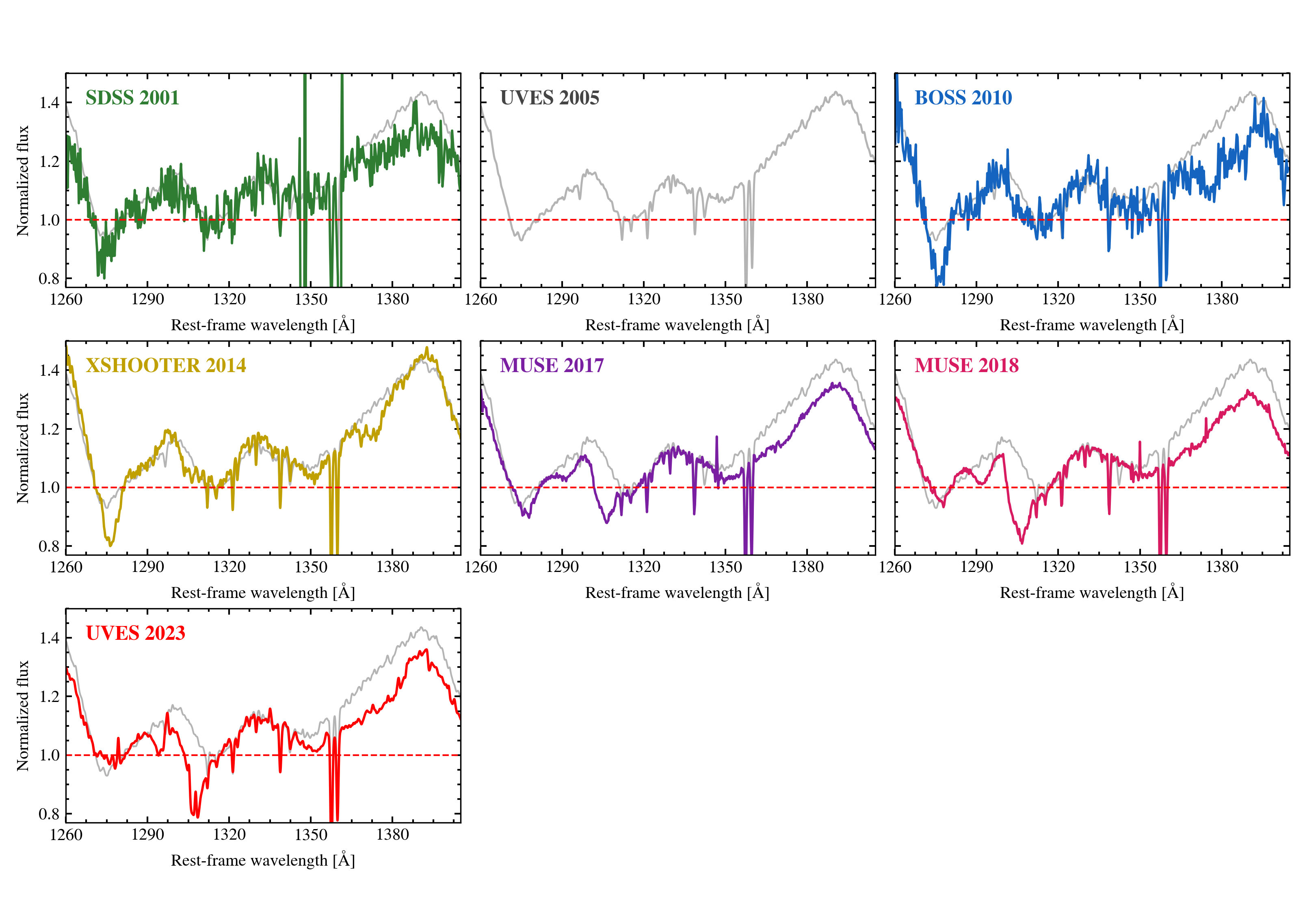}
    \caption{Variation of WISSH53 spectra within the region where the \civ{} uBALs are detected. The gray curve in all panels represents the UVES 2005 spectrum, adopted as the reference for the continuum and emission-line fitting up to \siiv{} (Sect. \ref{sec:civ_troughs}). For a better comparison, all spectra resampled to the same resolution. The red horizontal line indicates the continuum level.}
    \label{fig:J1249_profiles_troughs}
\end{figure*}

\begin{figure*}[h!]
   \includegraphics[width=0.33\linewidth]{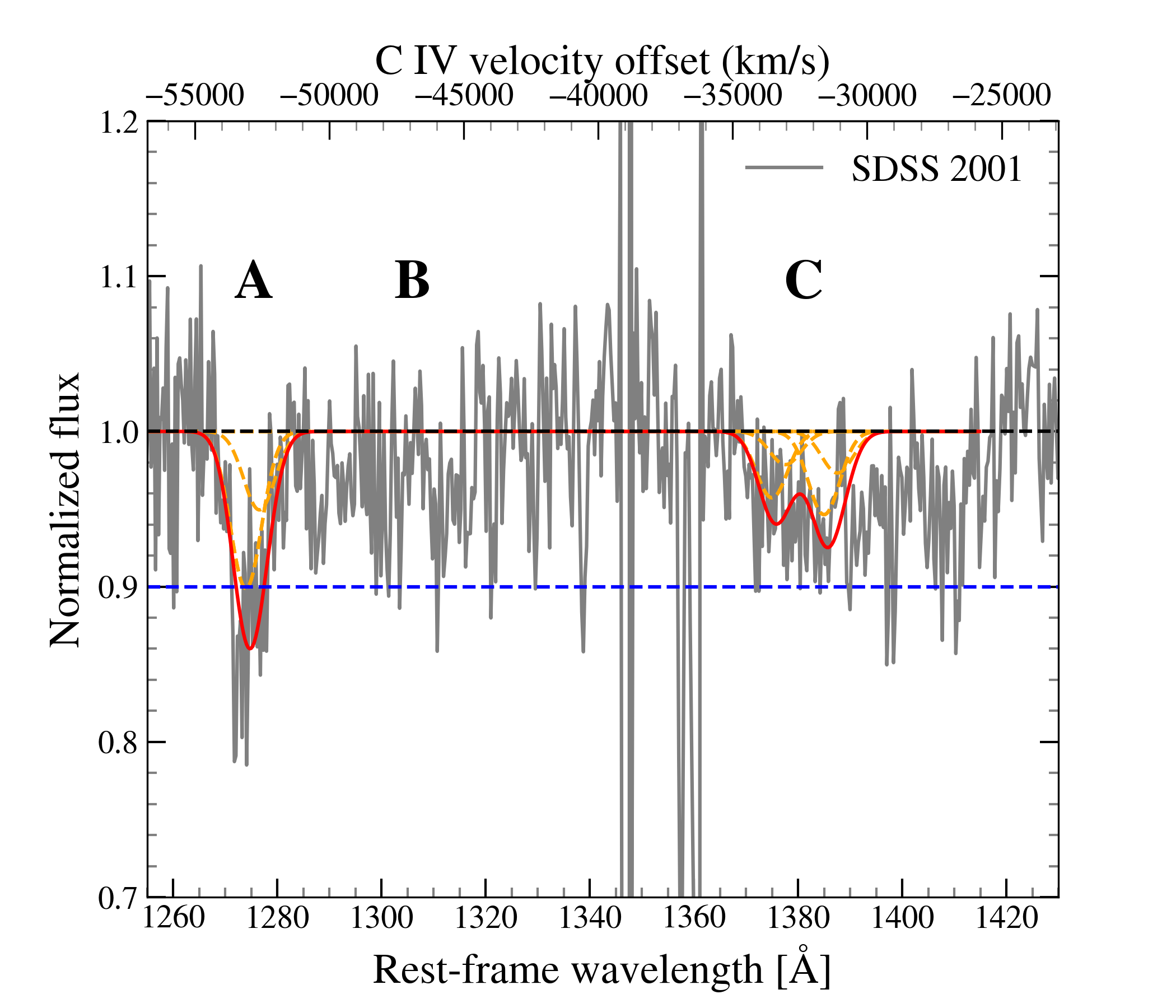}
   \includegraphics[width=0.33\linewidth]{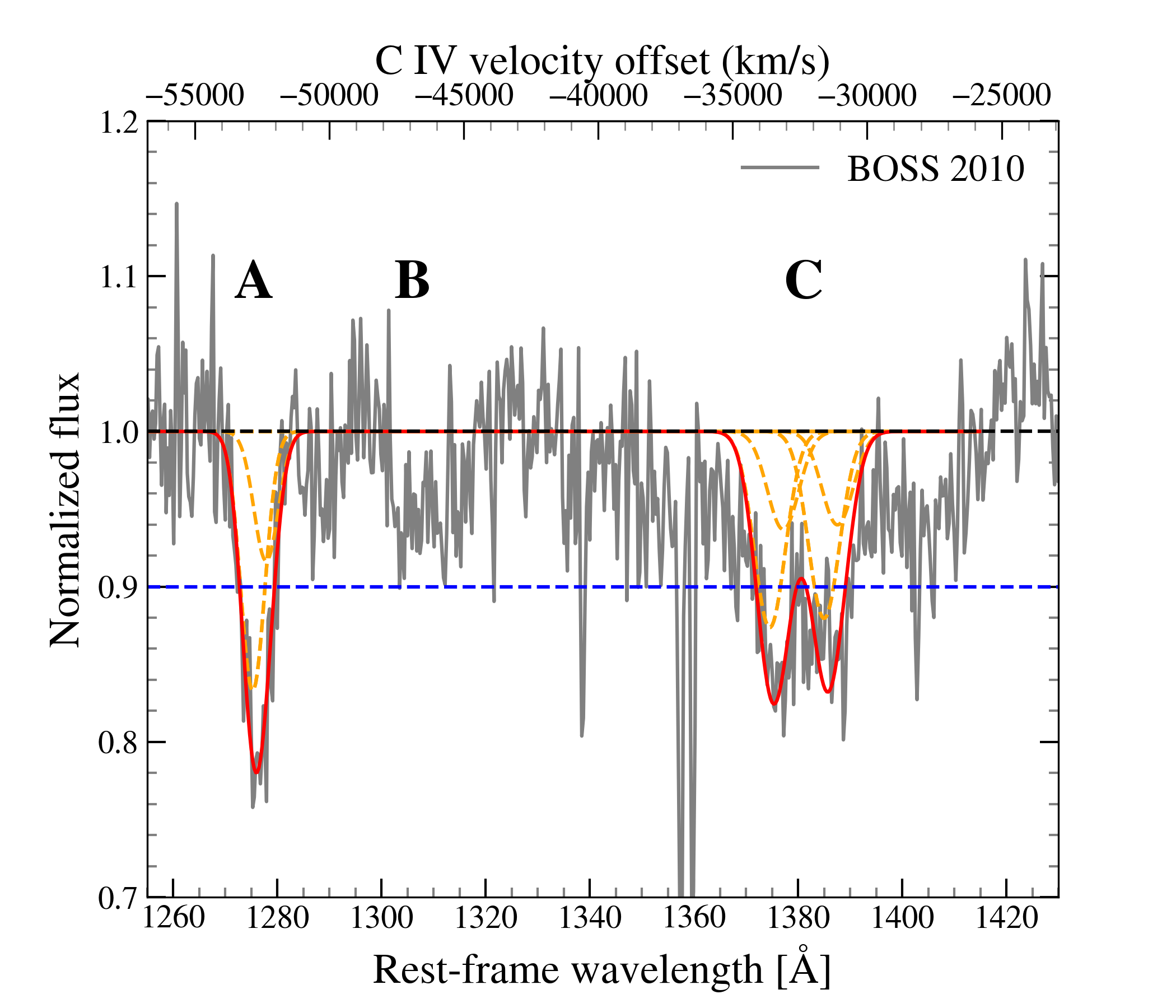}
   \includegraphics[width=0.33\linewidth]{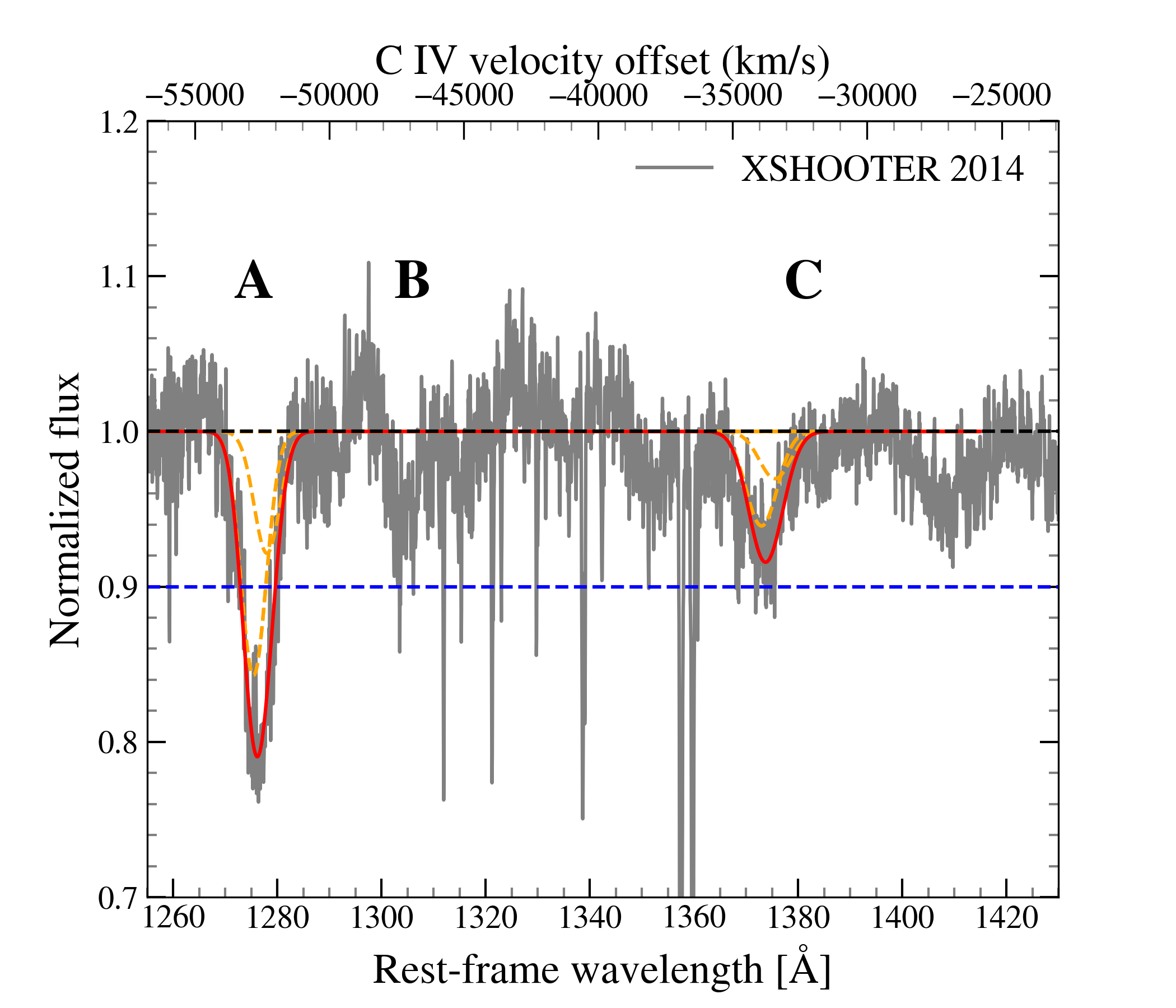}
   \includegraphics[width=0.33\linewidth]{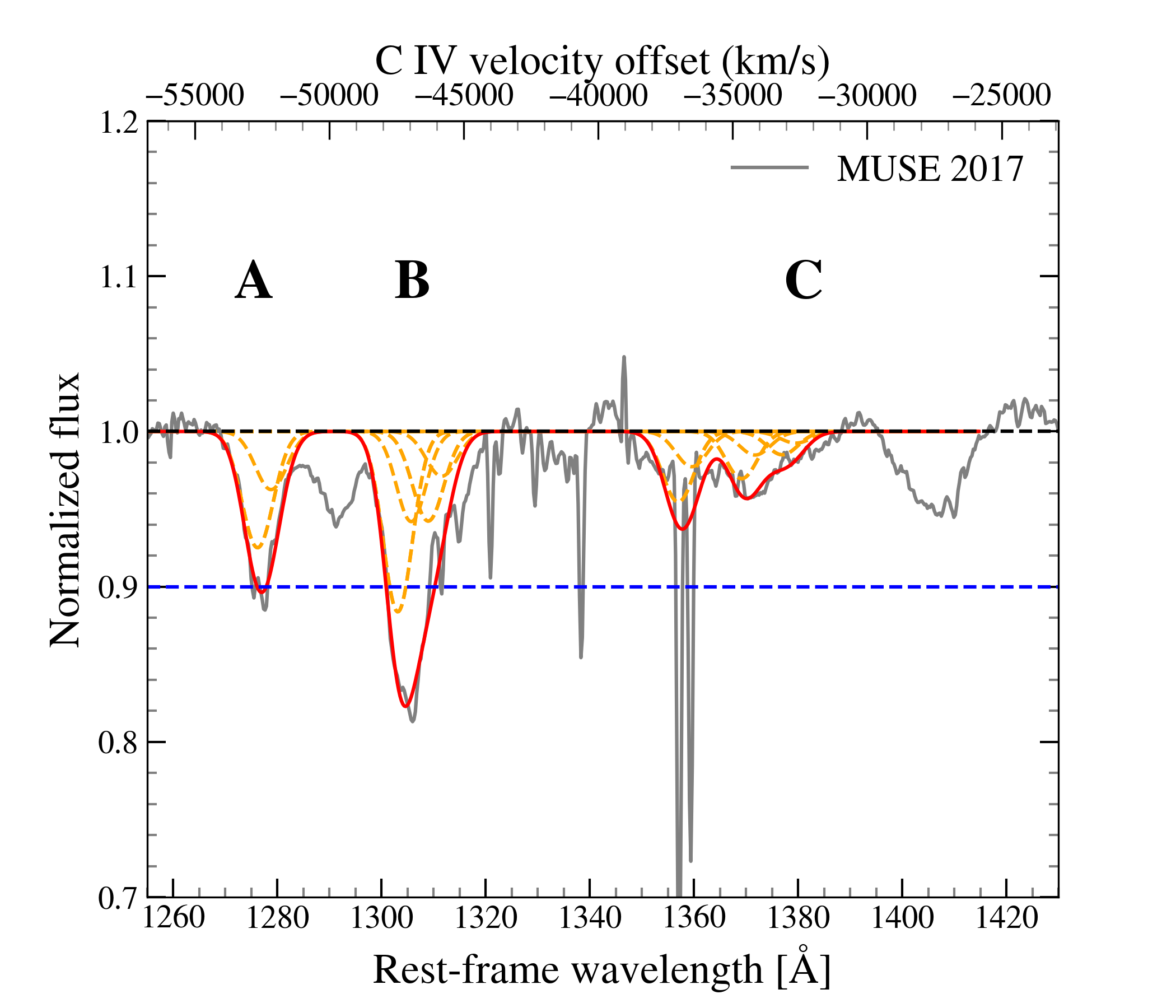}
   \includegraphics[width=0.33\linewidth]{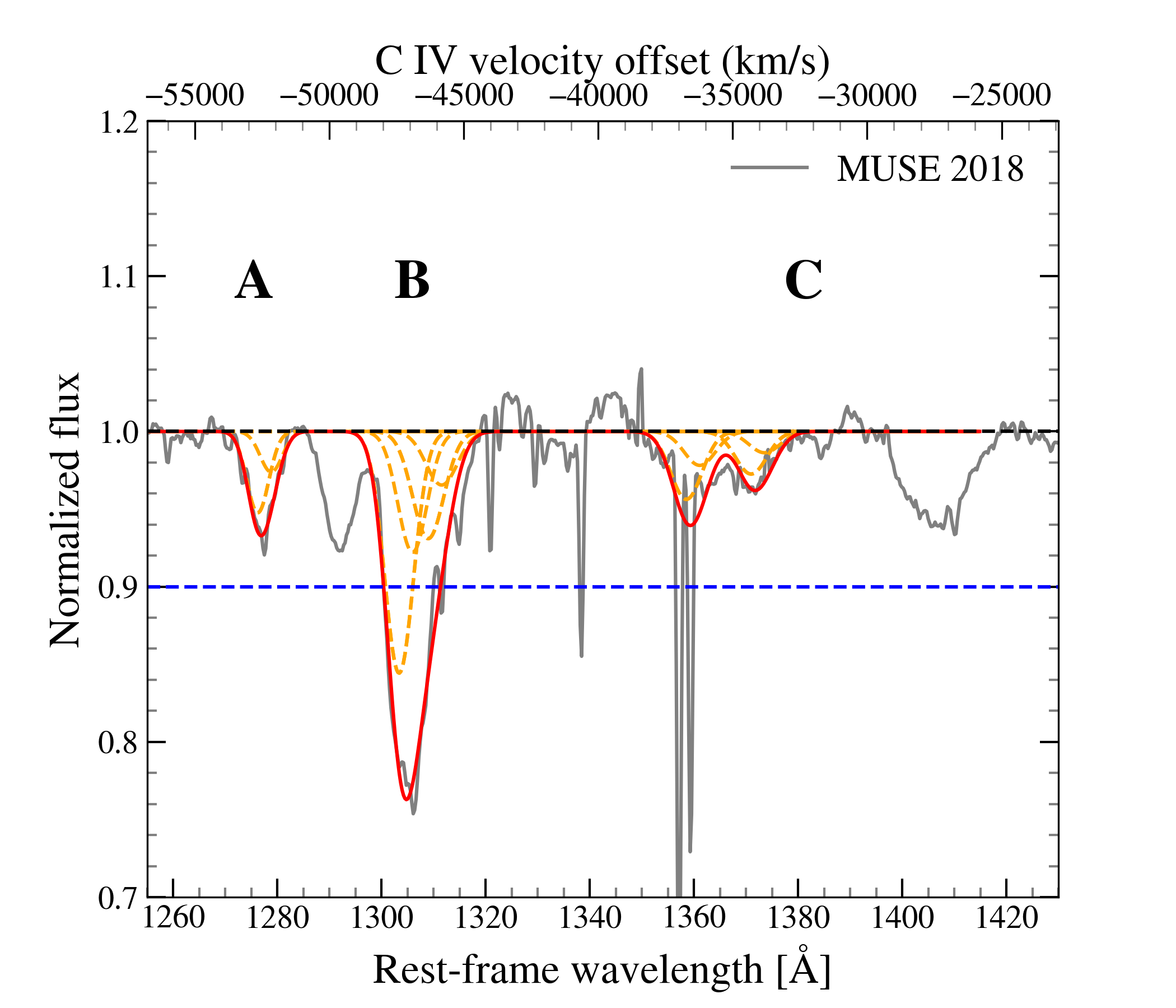}
   \includegraphics[width=0.33\linewidth]{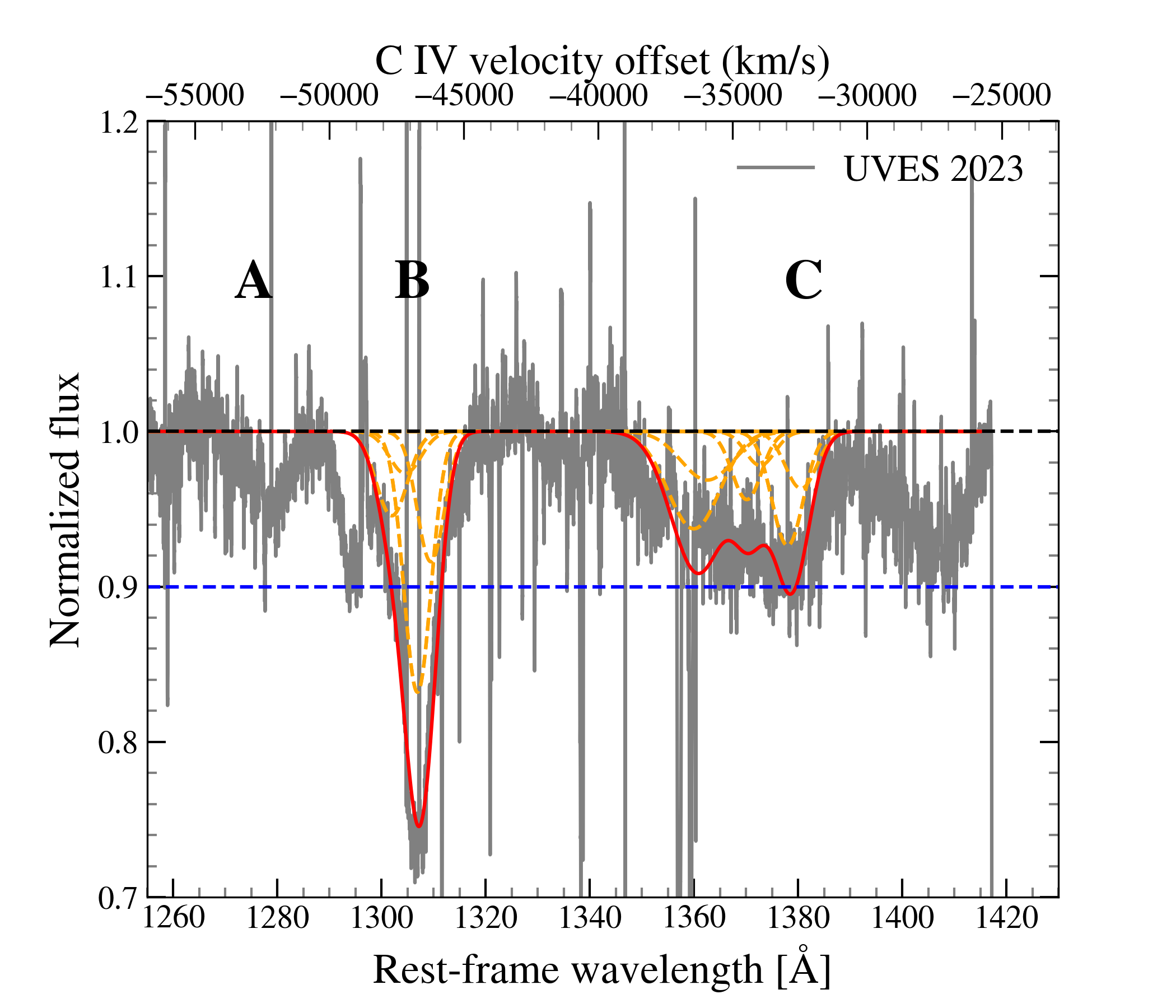}

    \caption{Fittings of the WISSH53 absorption features, done at the continuum level. Individual Gaussians are represented by dashed orange lines and the final fit is shown in red. Horizontal blue line indicates the 90\% of the continuum level. The fittings are done in all troughs that are deeper than 90\% of the continuum level at least in one epoch.  }
    \label{fig:WISSH53_absorptions}
\end{figure*}

\begin{figure}[th!]
    \centering
    \includegraphics[width=0.33\linewidth]{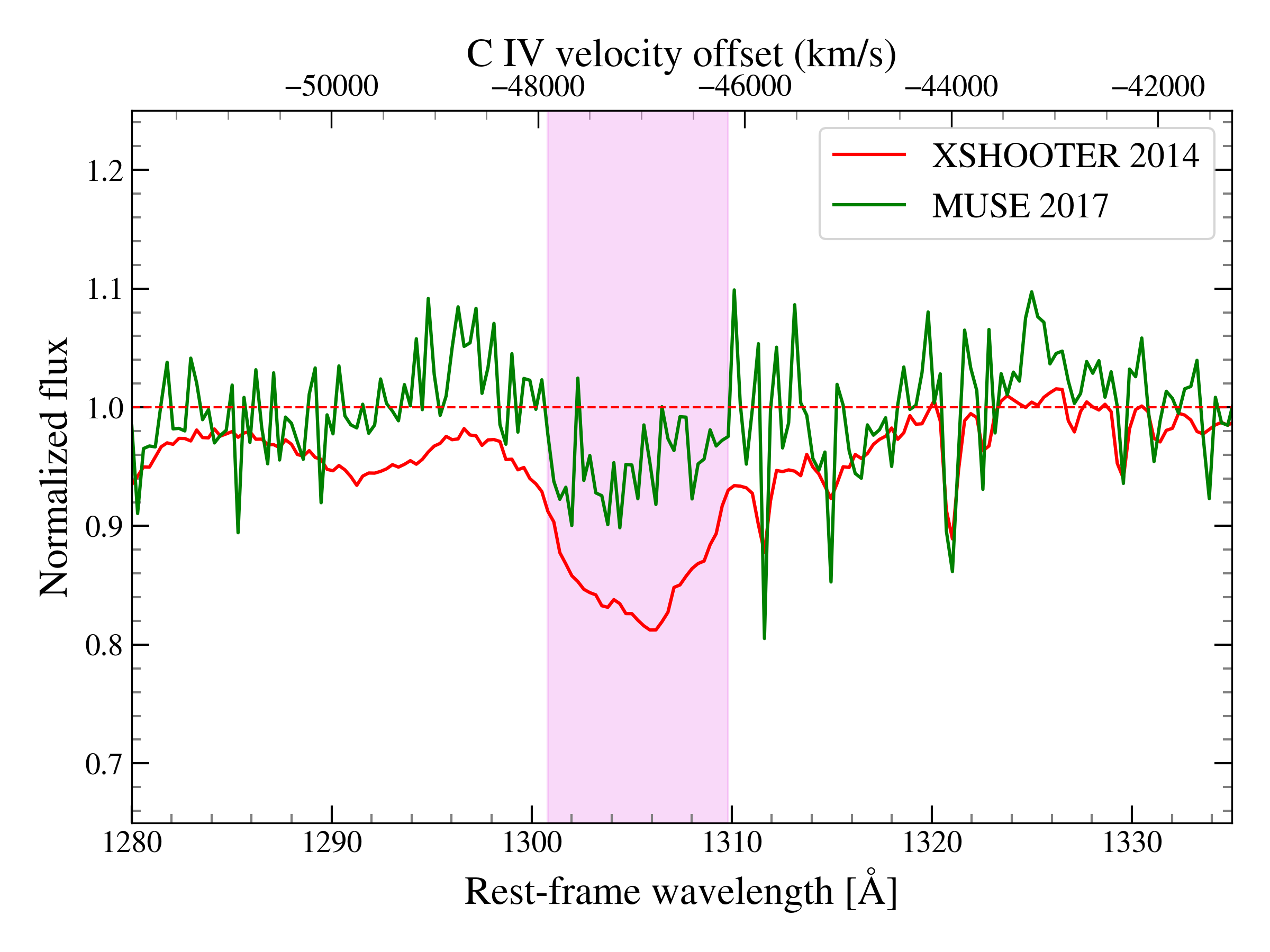}
    \includegraphics[width=0.33\linewidth]{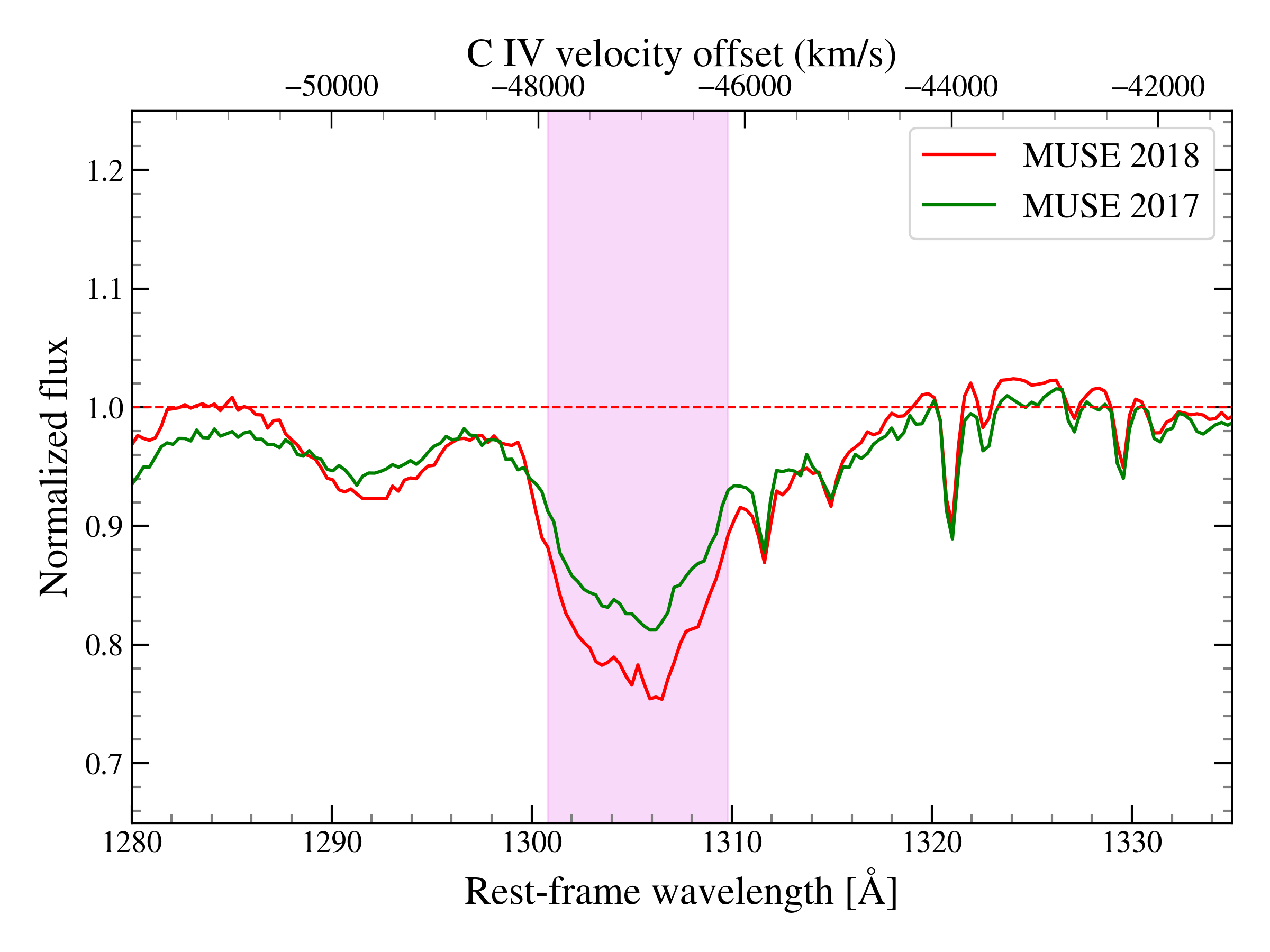}
    \includegraphics[width=0.33\linewidth]{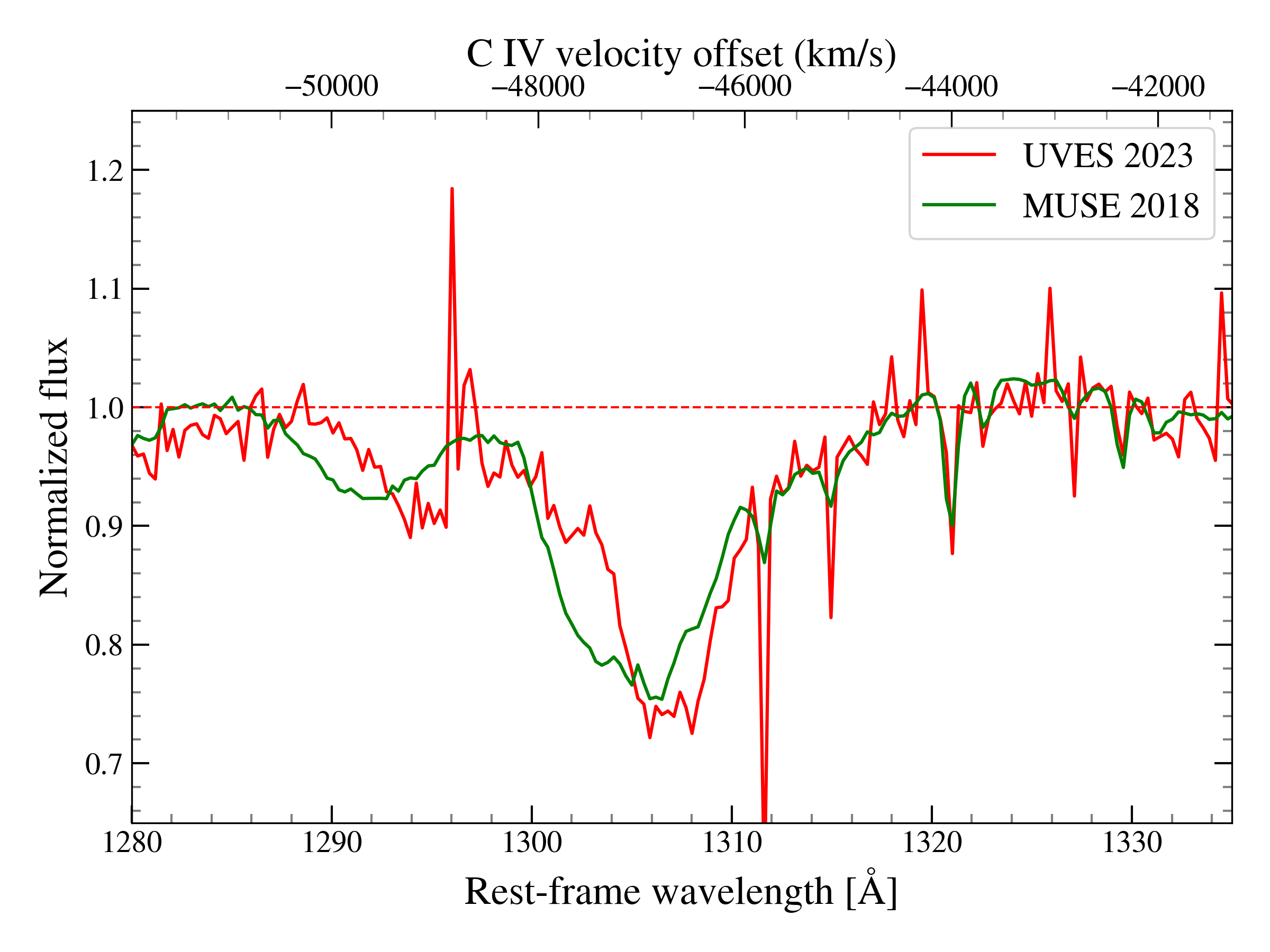}
    \caption{Comparison between the spectra of two consecutive epochs over the \civ\ velocity interval of the trough B of WISSH53, whose variability is thought to be driven by gas transverse motion. The magenta-shaded region marks where significant spectral changes occur and is used to estimate the absorption strength variation $\Delta A_{\rm S}$ in Sect. \ref{sec:radius_gas_motion}. In the case of UVES 2023 and MUSE 2018 comparison (right plot), no variations broader than 1200 km s$^{-1}$ were detected. The dashed line represents the continuum level.}
    \label{fig:J1249_delta_as}
\end{figure}

\clearpage
\section{WISSH56}

\begin{figure*}[ht!]
    \centering
\includegraphics[width=\linewidth]{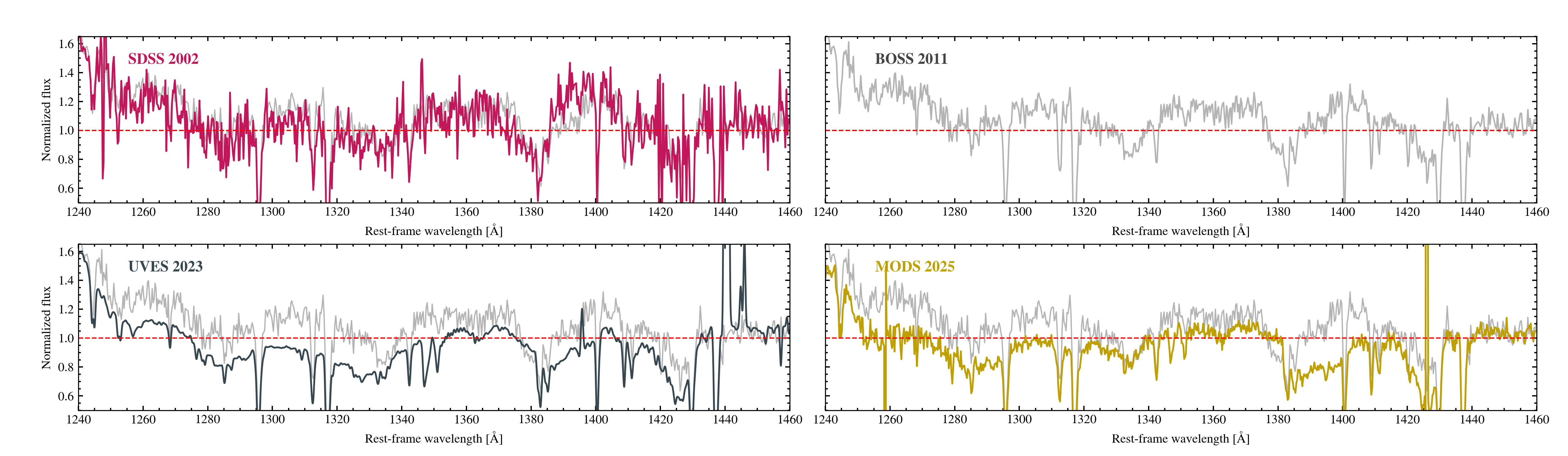}
    \caption{Same as Fig. \ref{fig:J1249_profiles_troughs}, but for WISSH56.}
    \label{fig:J1305_profile_troughs}
\end{figure*}

\begin{figure*}[ht!]
    \includegraphics[width=0.33\linewidth]{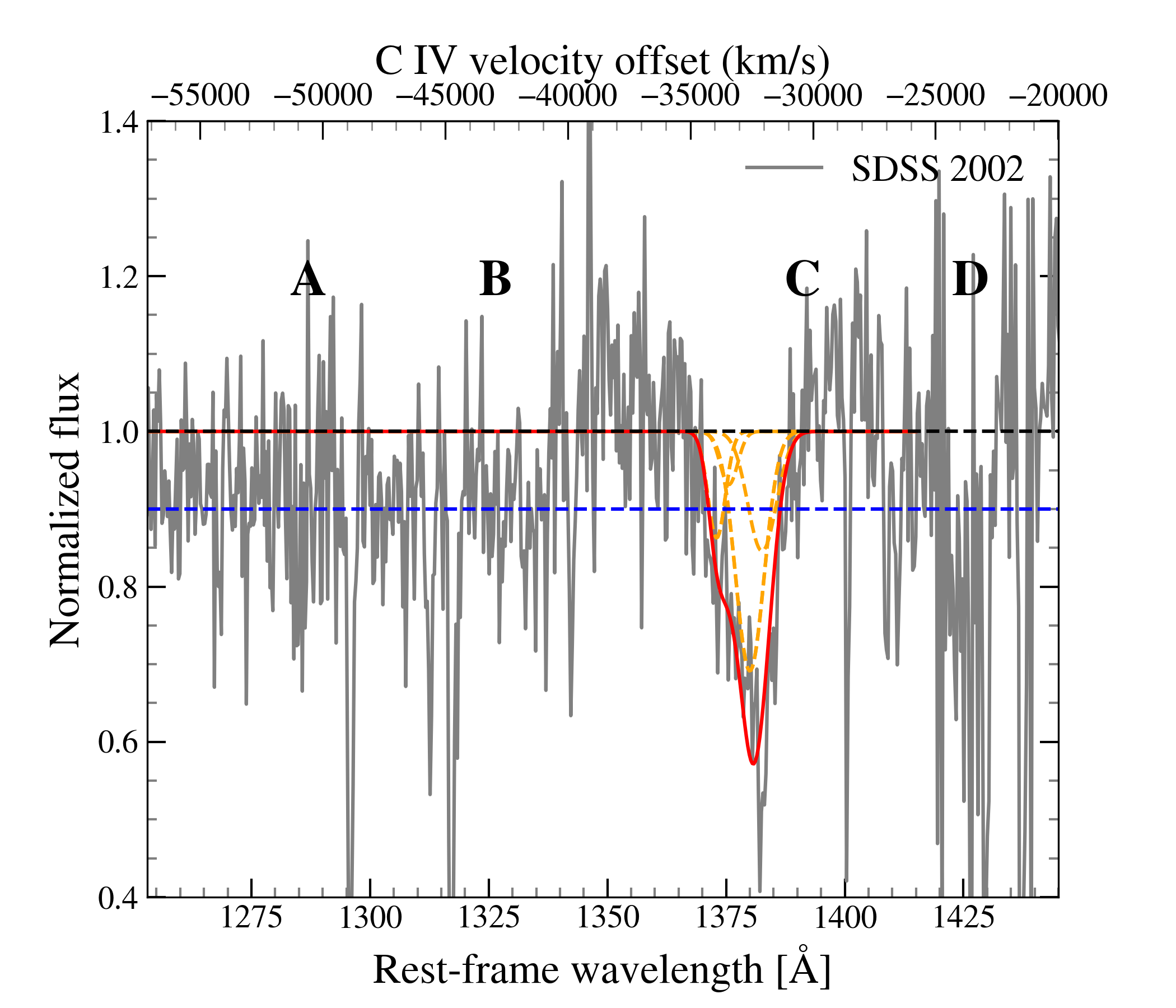}
    \includegraphics[width=0.33\linewidth]{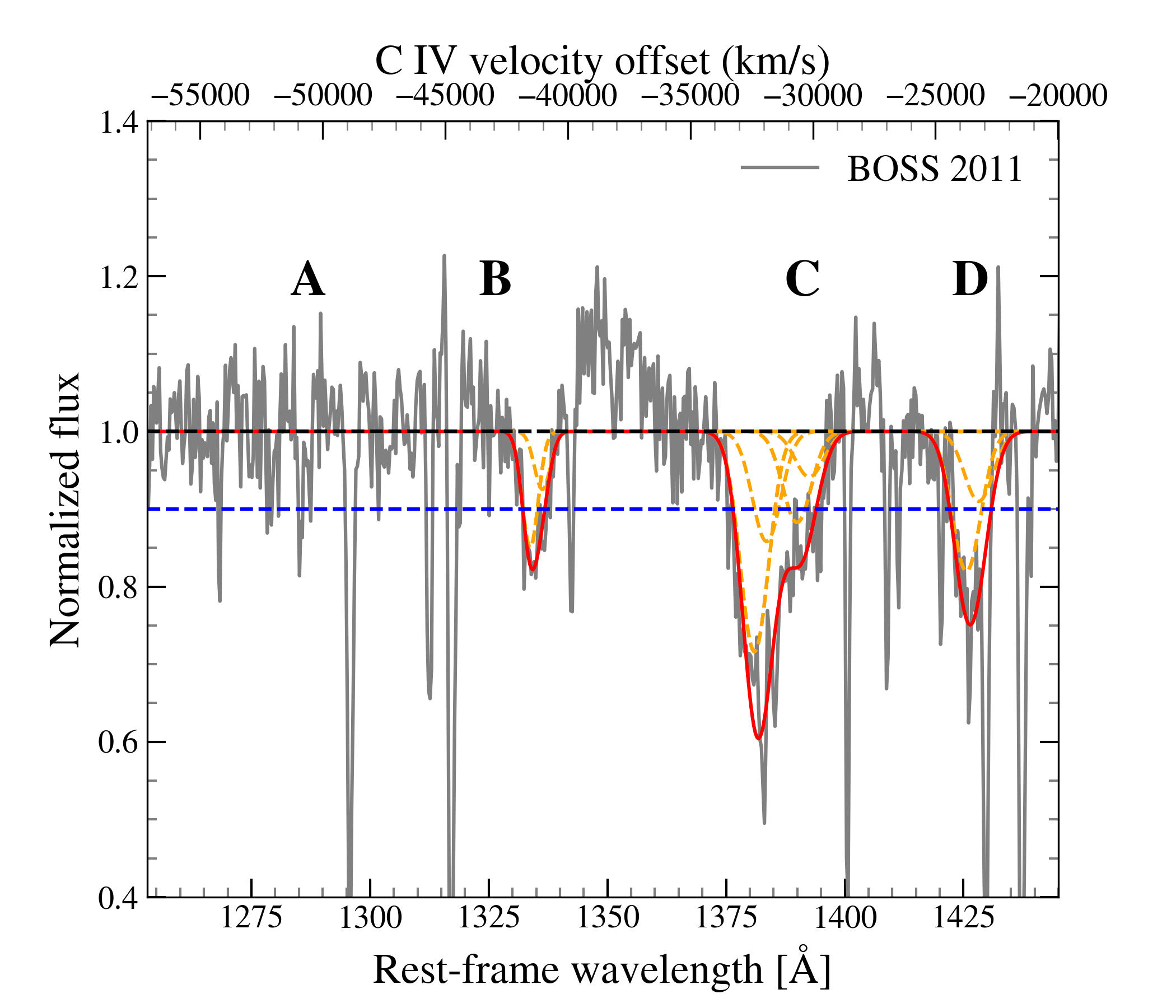}
    \includegraphics[width=0.33\linewidth]{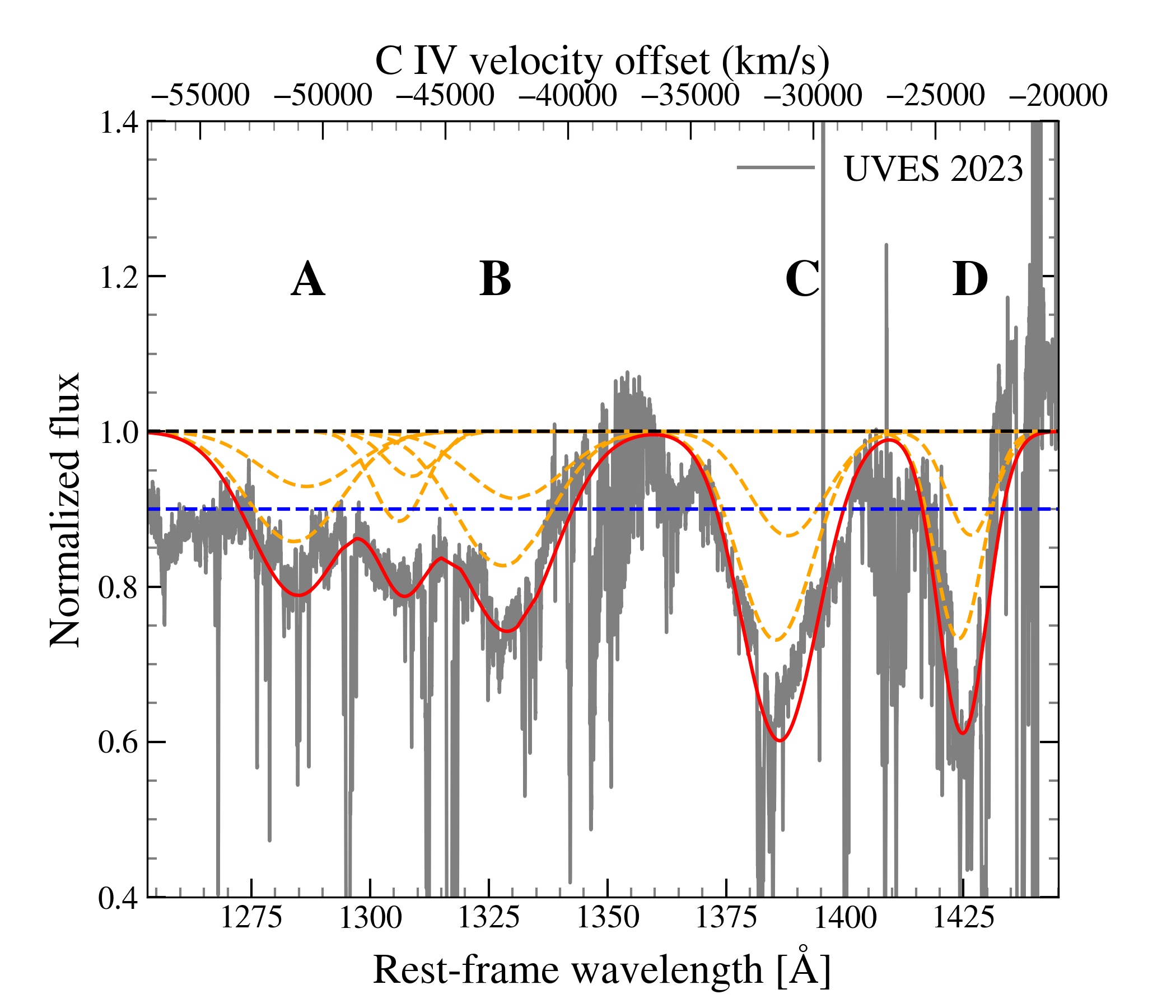}
    \includegraphics[width=0.33\linewidth]{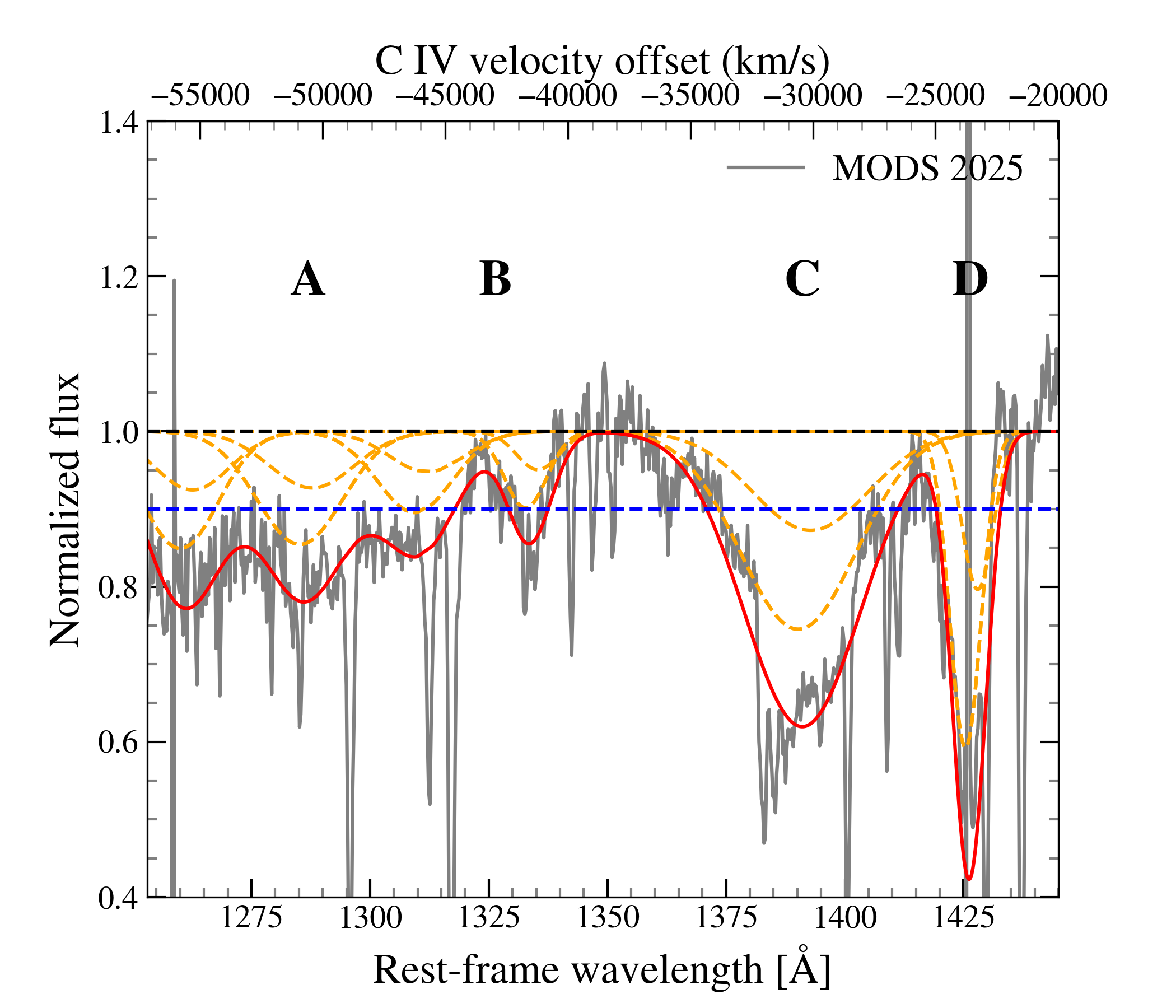}
    \caption{Same as Fig. \ref{fig:WISSH53_absorptions}, but for WISSH56.}
    \label{fig:WISSH56_absorptions}
\end{figure*}

\clearpage
\section{WISSH71}

\begin{figure*}[h!]
    \centering
    \includegraphics[width=\linewidth]{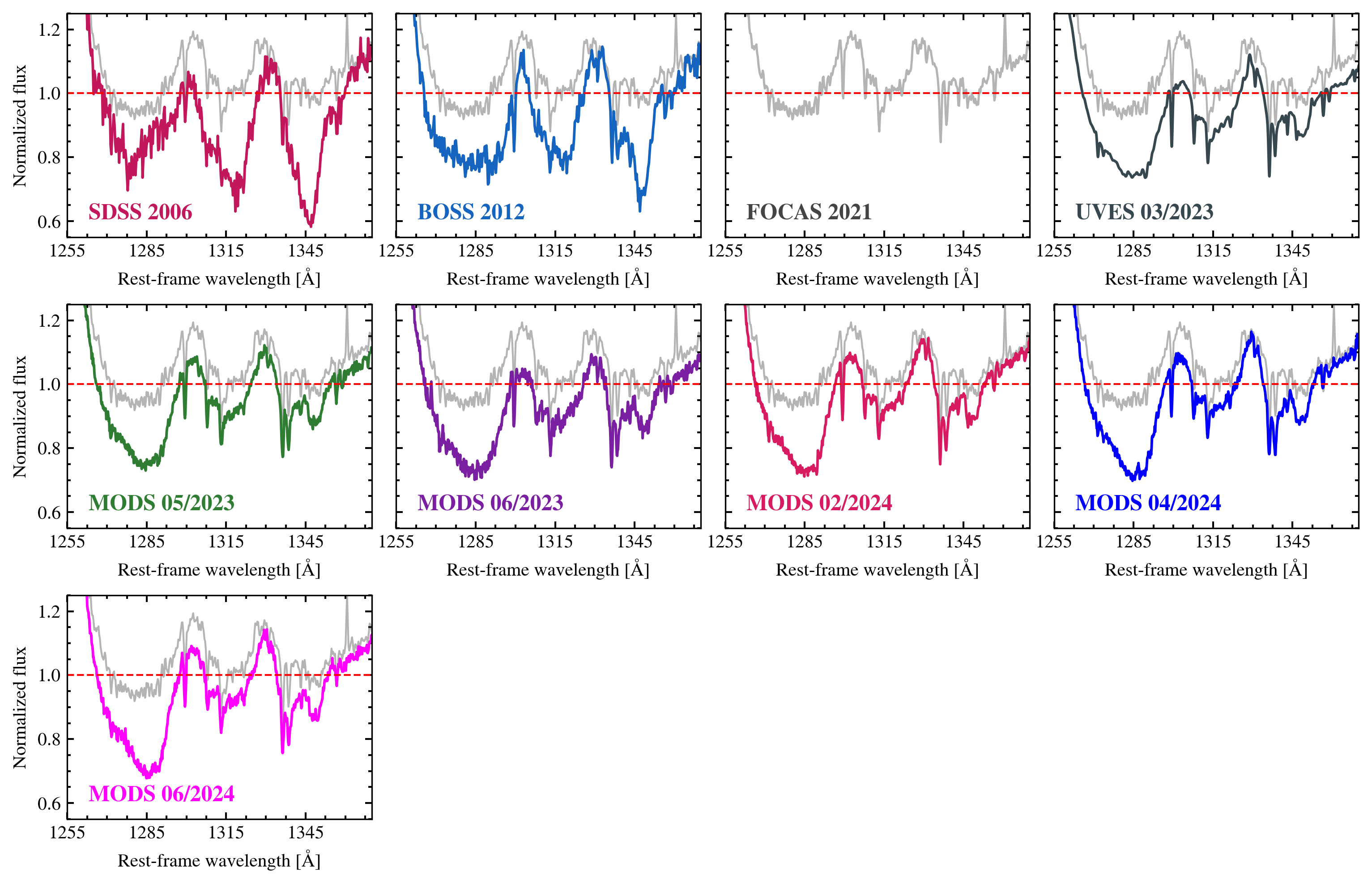}
    \caption{Same as Fig. \ref{fig:J1249_profiles_troughs}, but for WISSH71.}
    \label{fig:J1538_profiles_troughs}
\end{figure*}

\begin{figure}[h!]
    \includegraphics[width=0.33\linewidth]{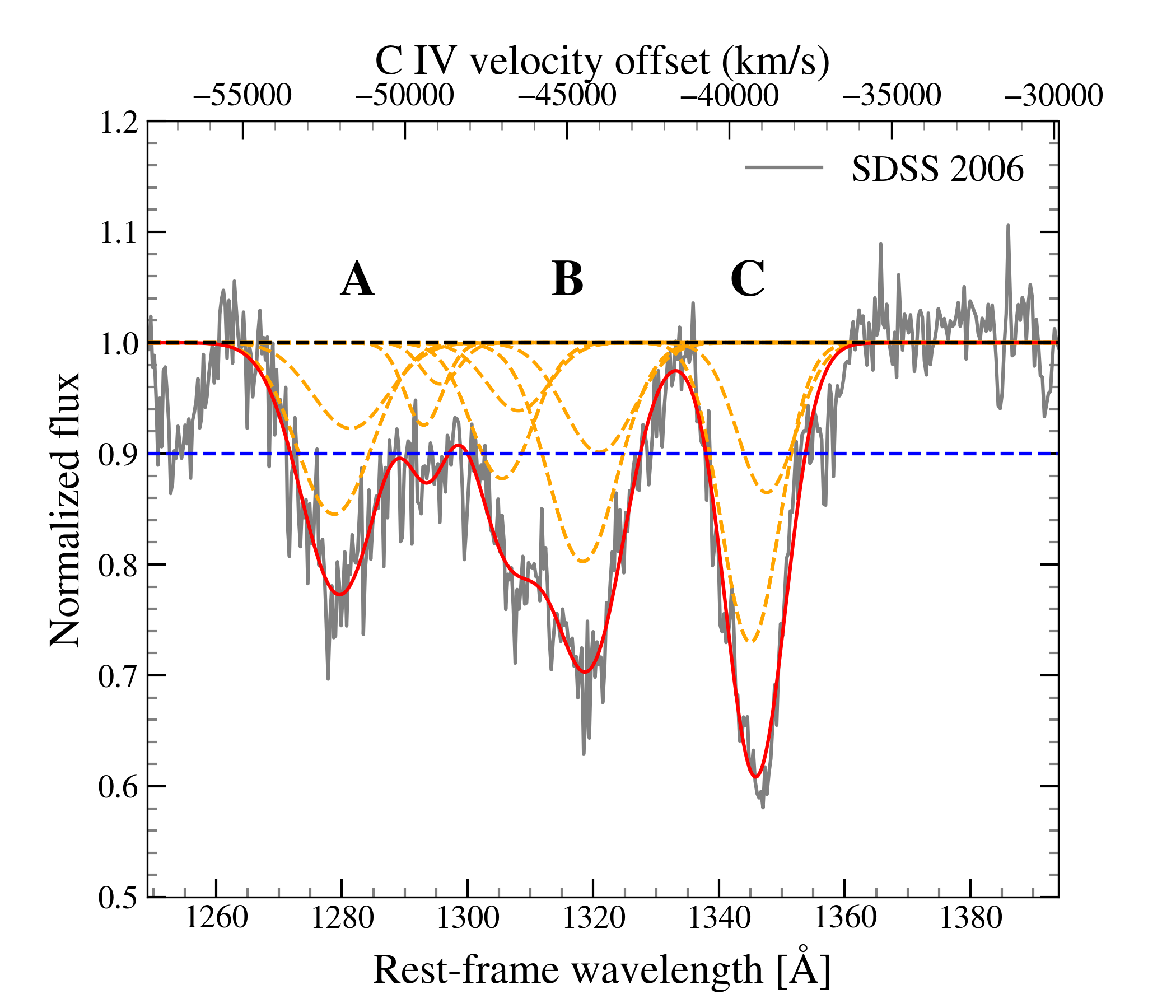}
    \includegraphics[width=0.33\linewidth]{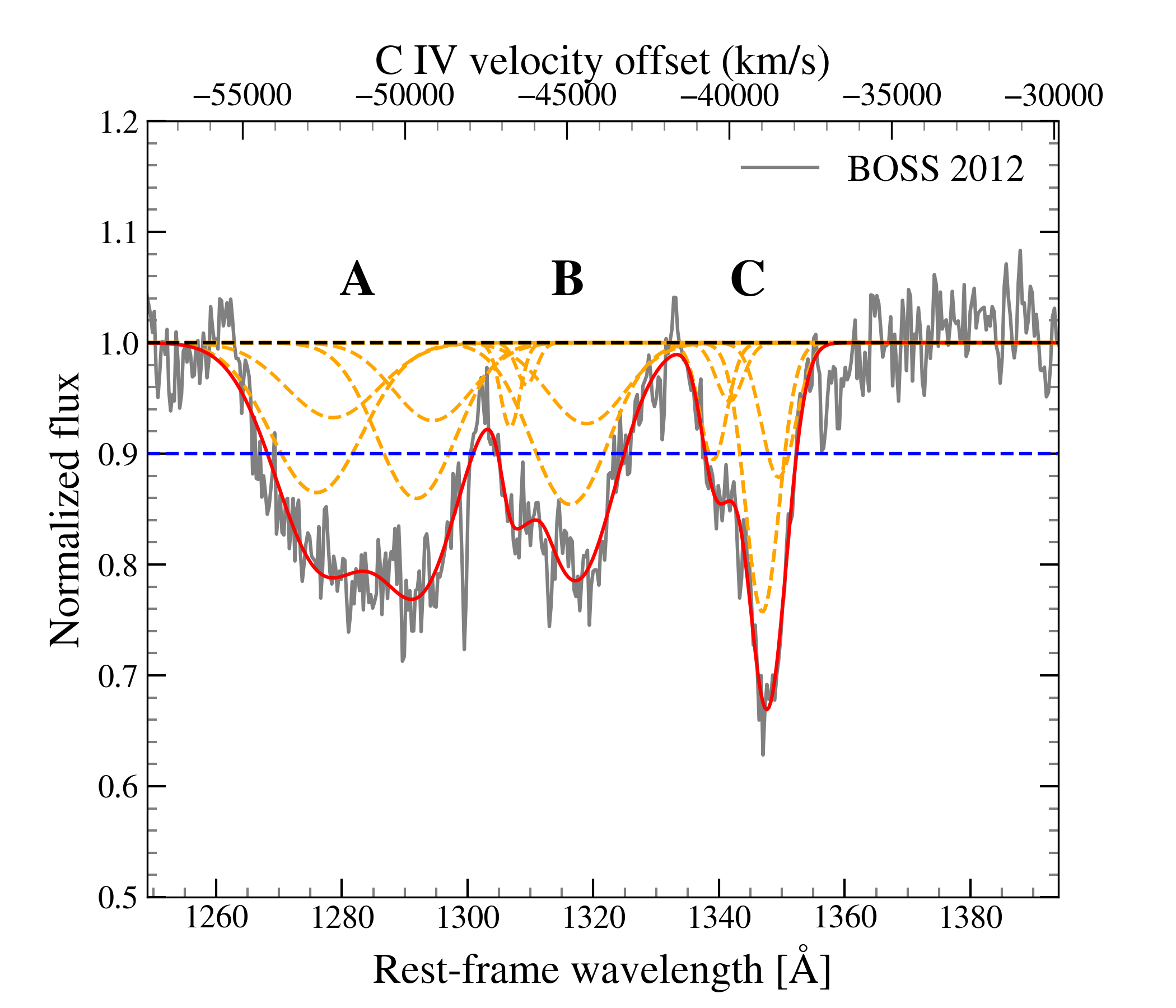}
    \includegraphics[width=0.33\linewidth]{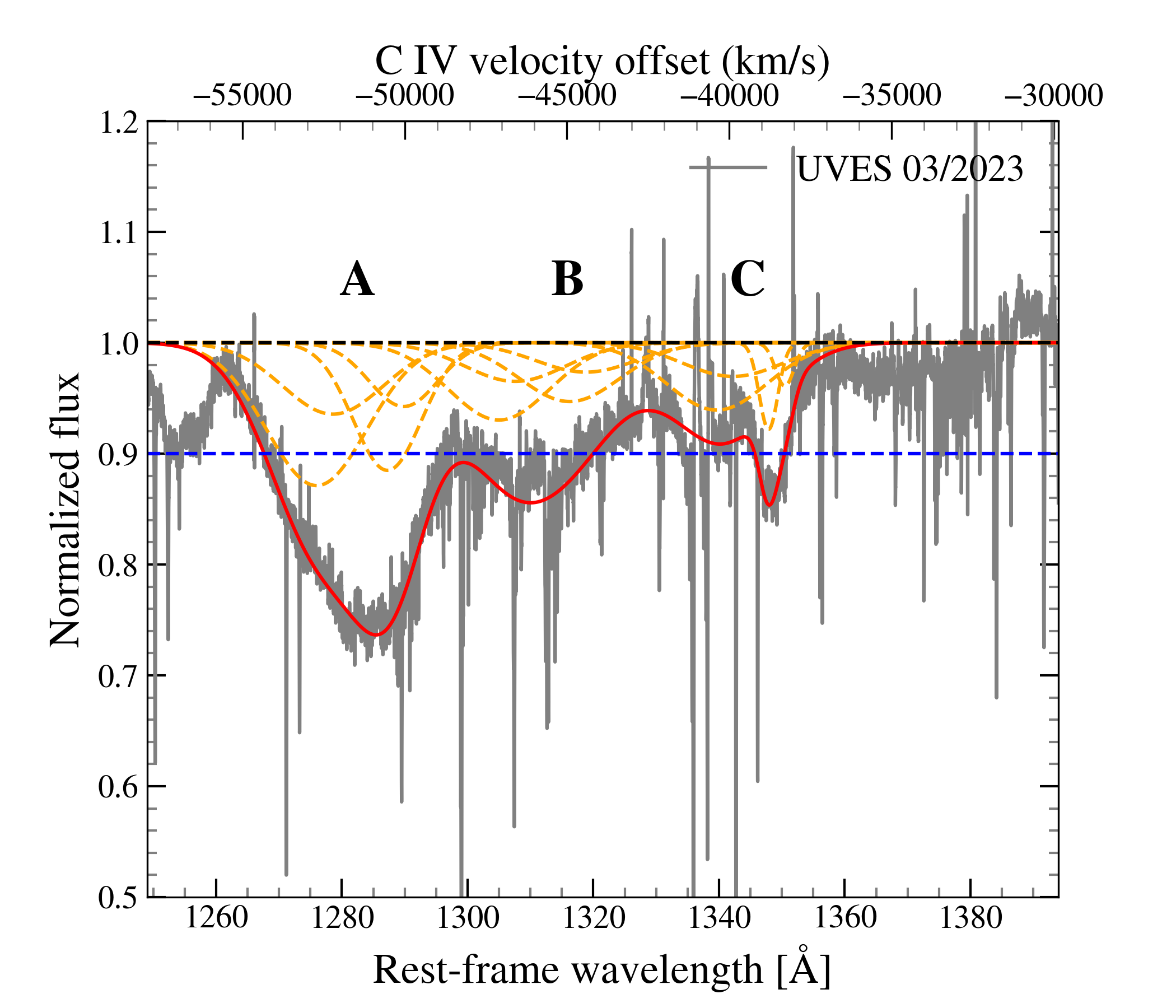}
    \includegraphics[width=0.33\linewidth]{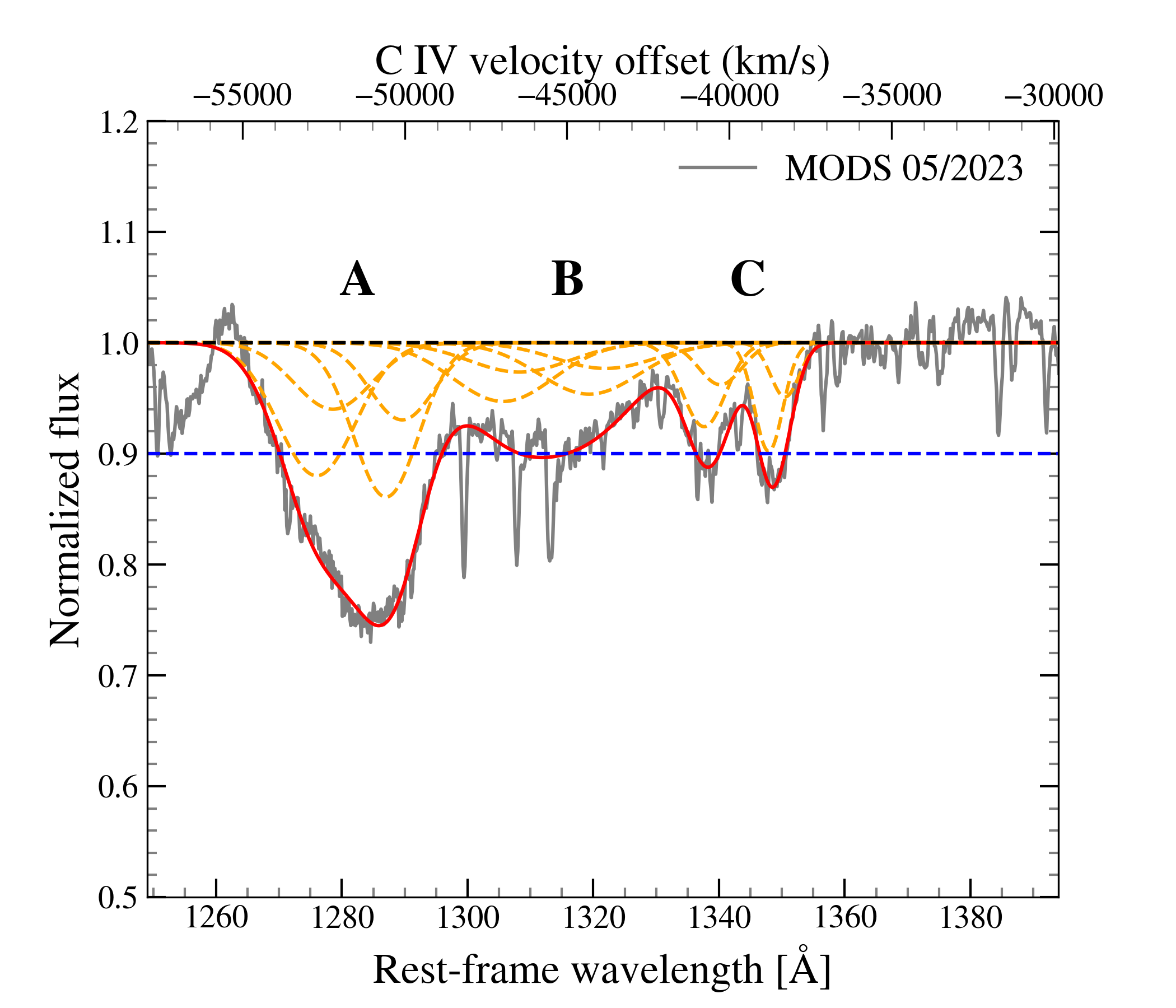}
    \includegraphics[width=0.33\linewidth]{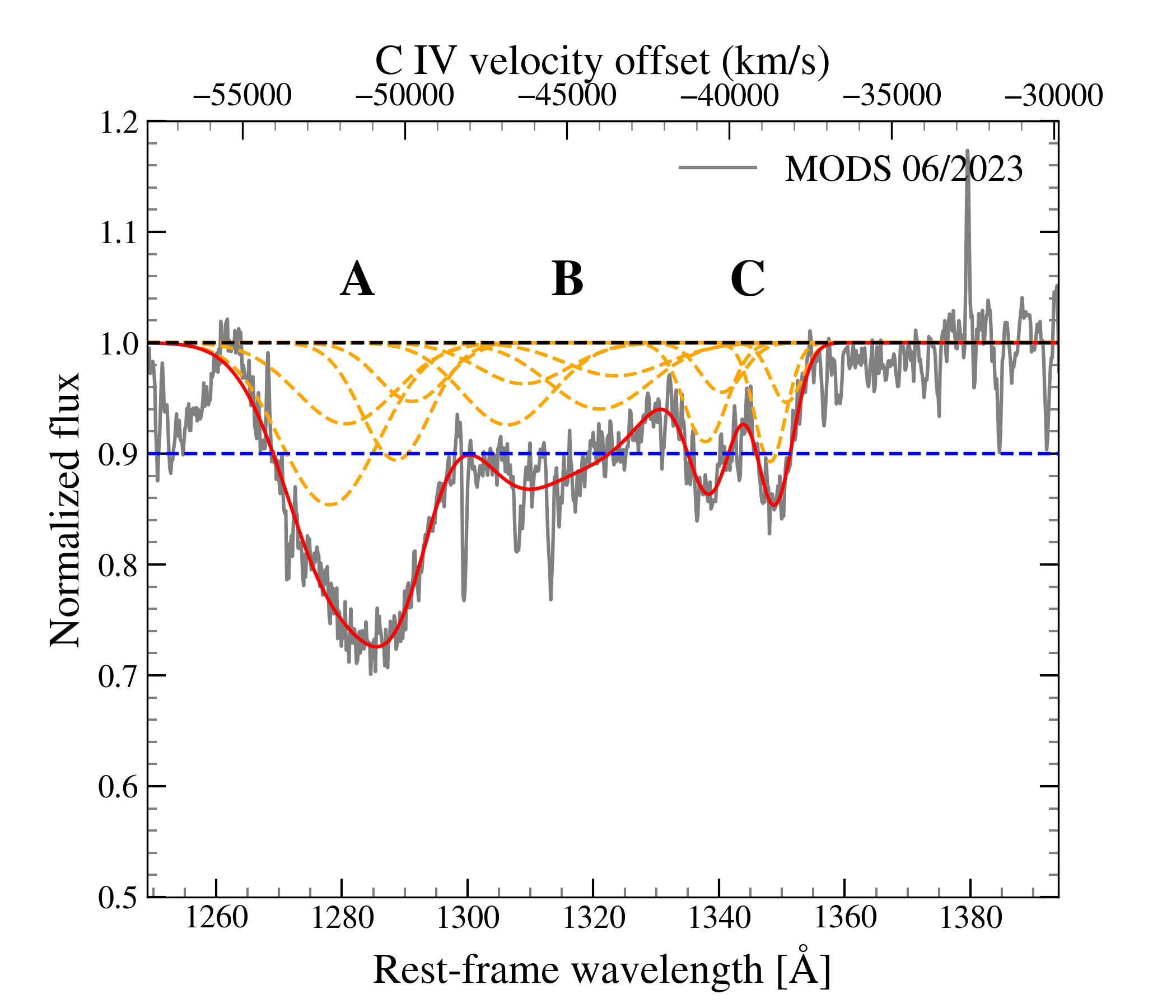}
\includegraphics[width=0.33\linewidth]{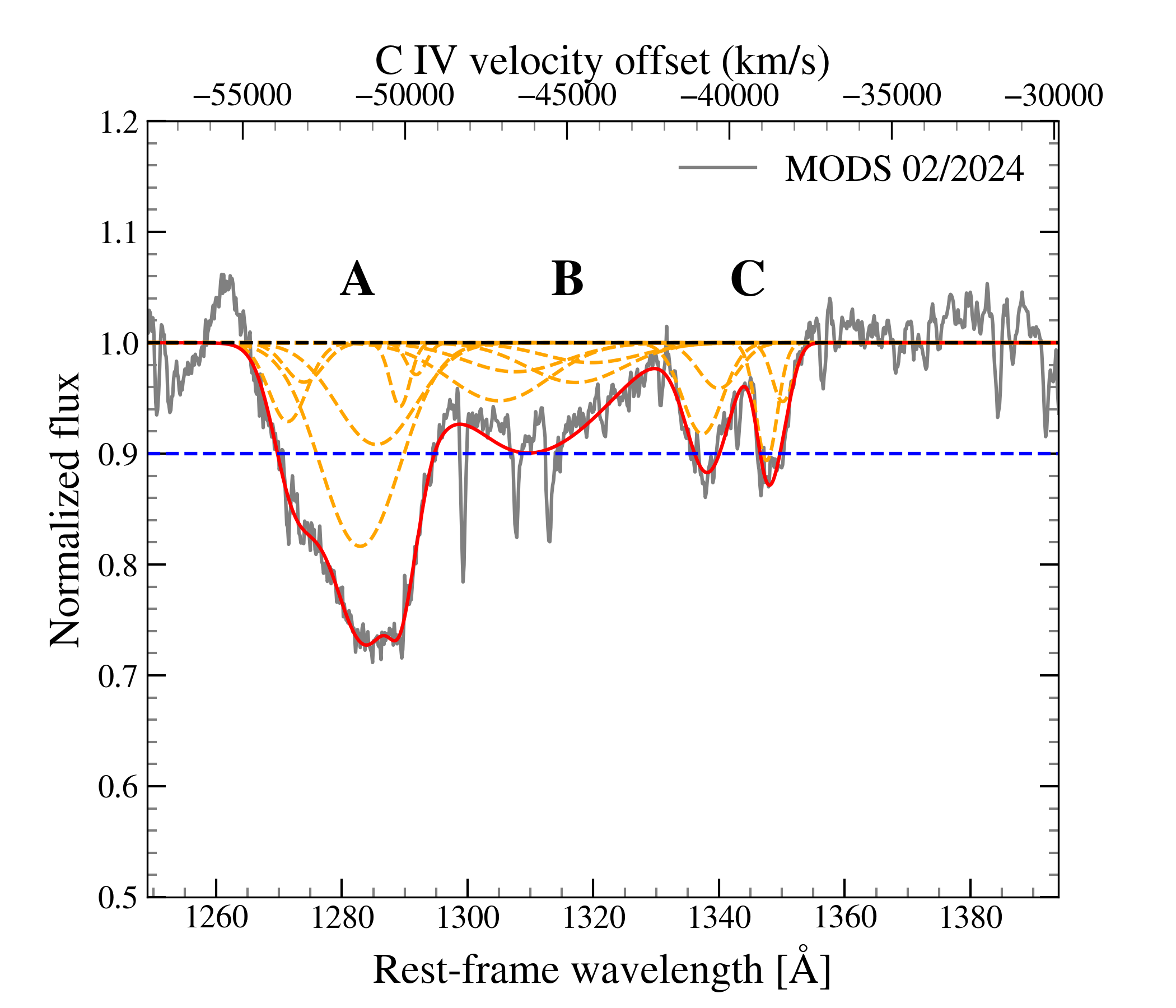}
\includegraphics[width=0.33\linewidth]{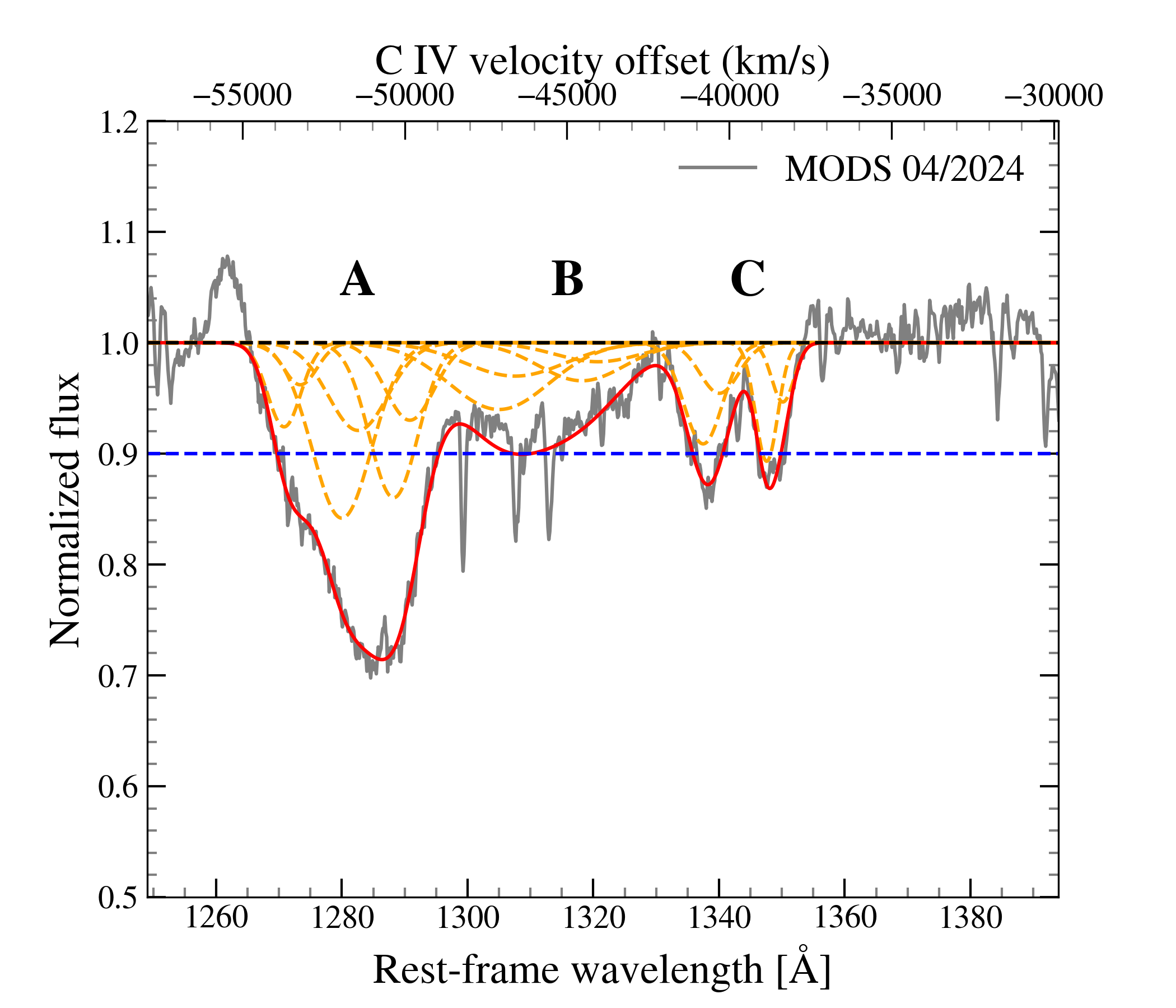}
\includegraphics[width=0.33\linewidth]{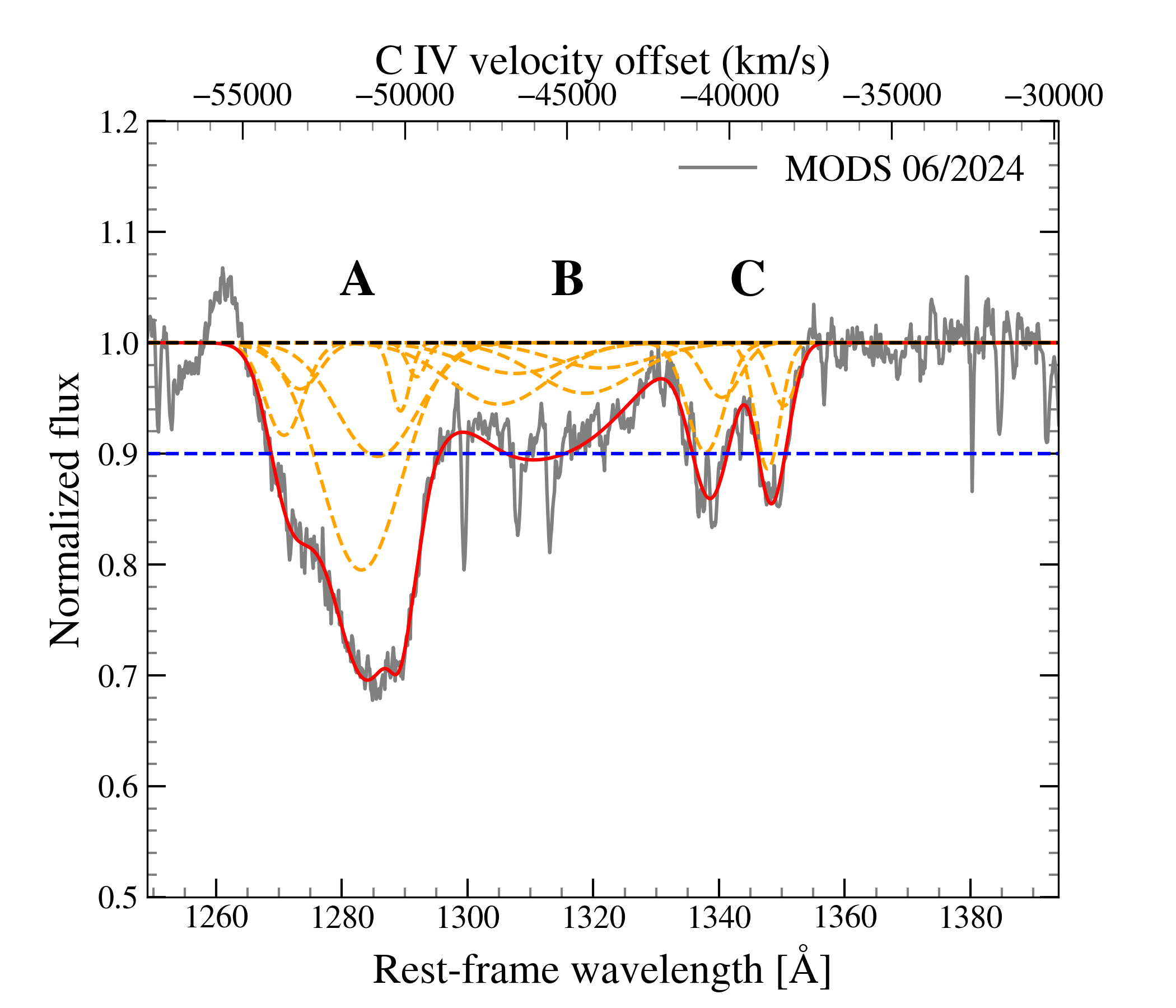}
    
    \caption{Same as Fig. \ref{fig:WISSH53_absorptions}, but for WISSH71.}
    \label{fig:WISSH71_absorptions}
\end{figure}

\begin{figure}[th!]
    \includegraphics[width=0.33\linewidth]{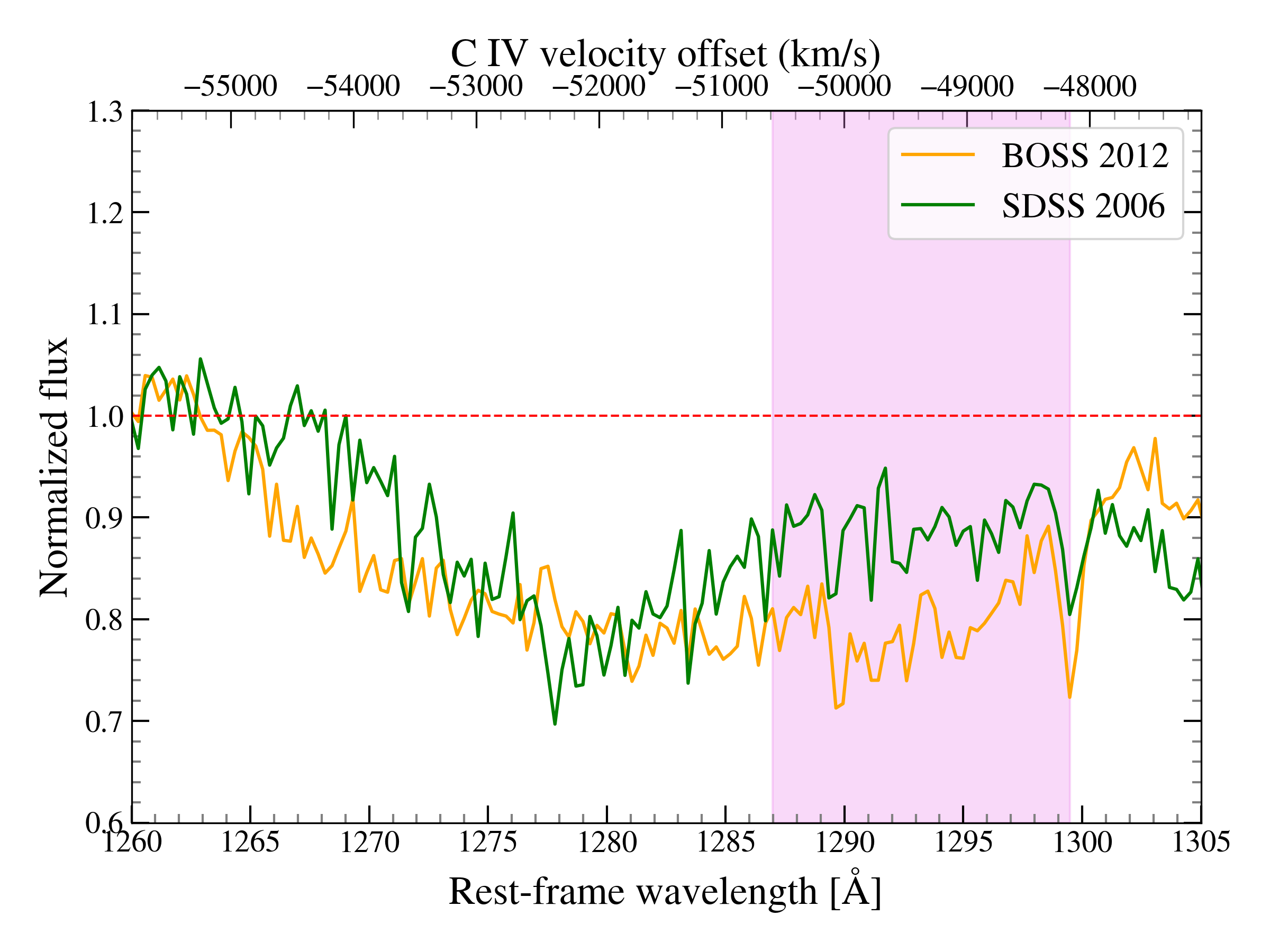}
    \includegraphics[width=0.33\linewidth]{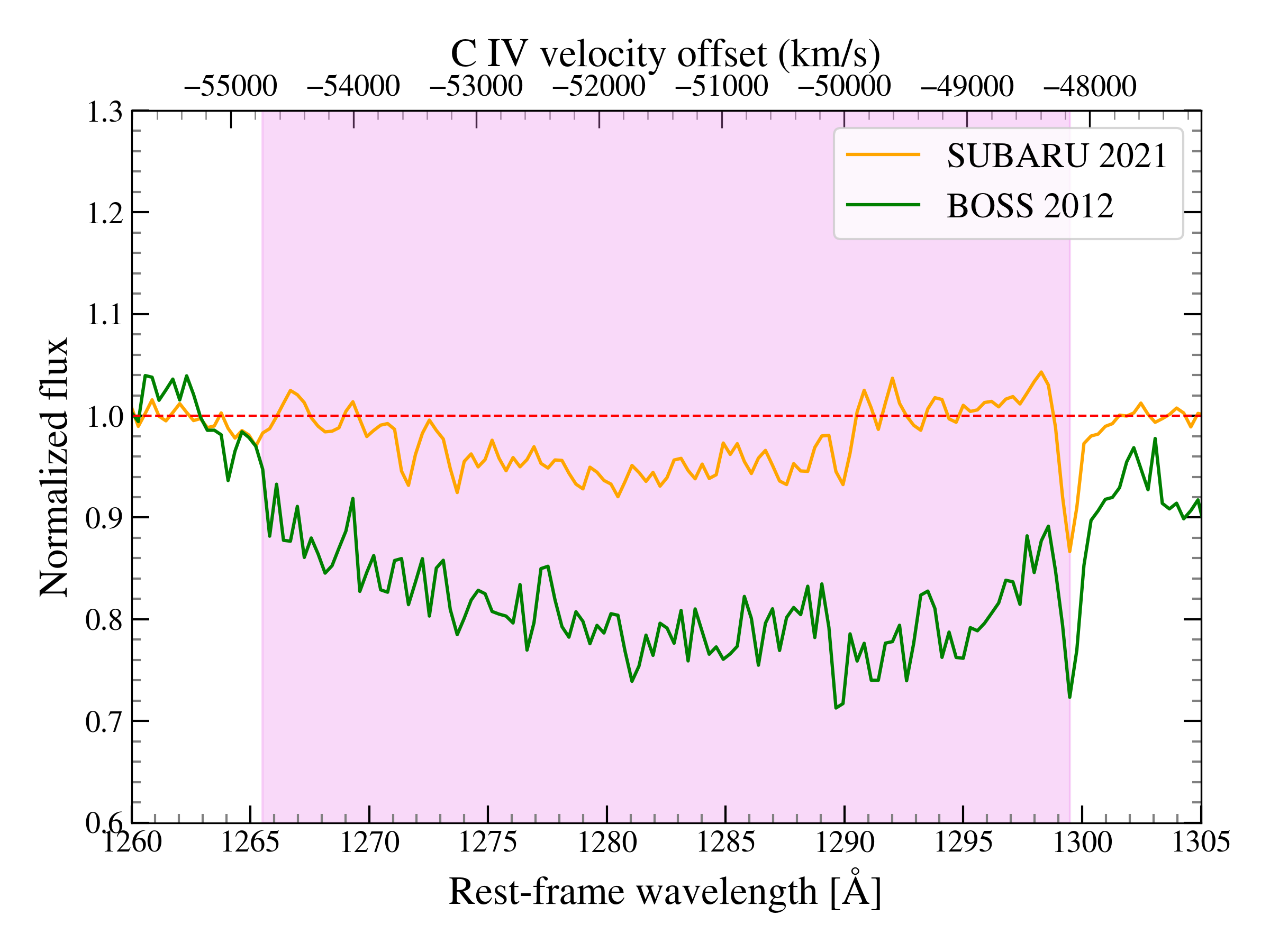}
    \includegraphics[width=0.33\linewidth]{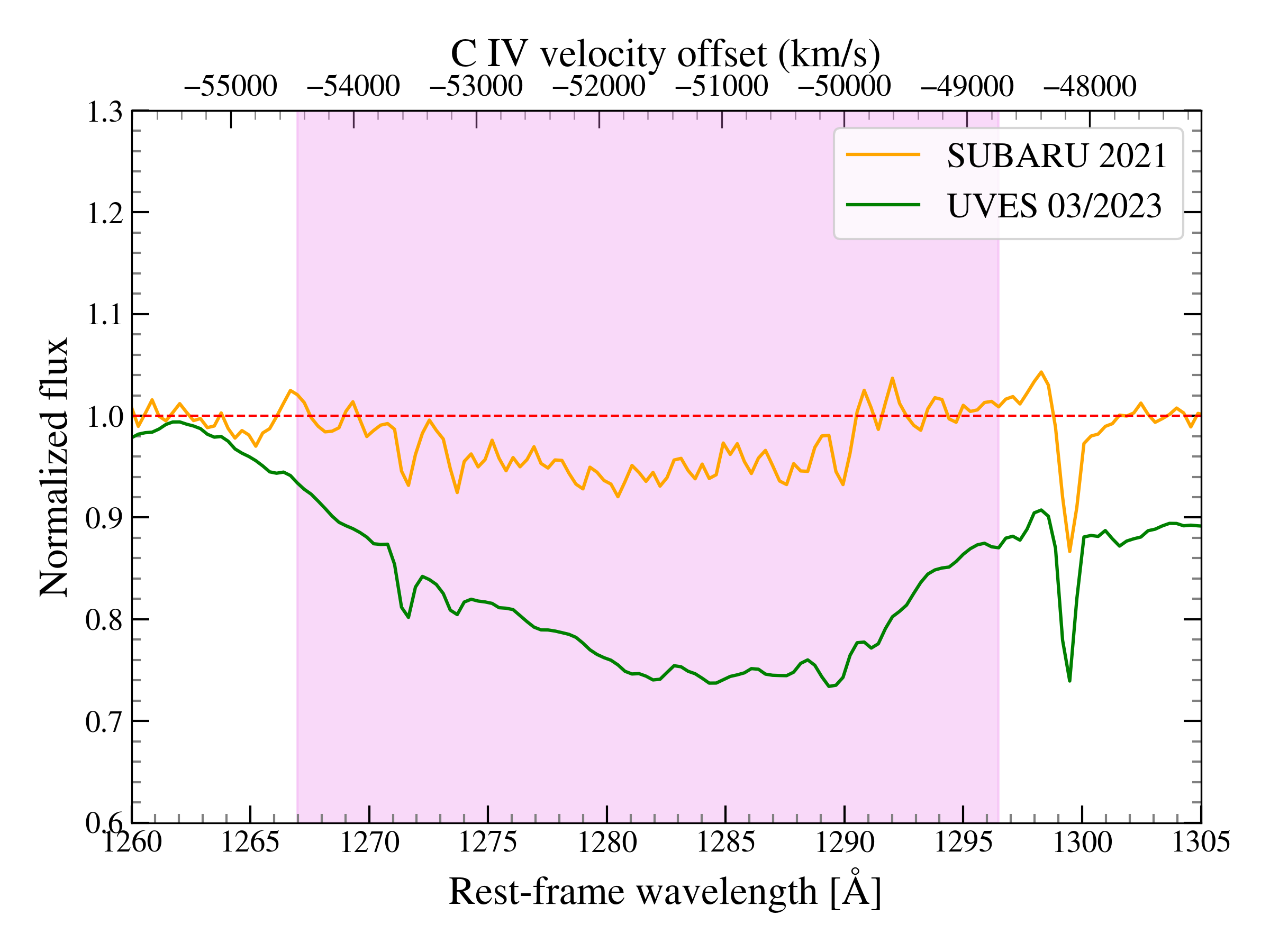}
    \includegraphics[width=0.33\linewidth]{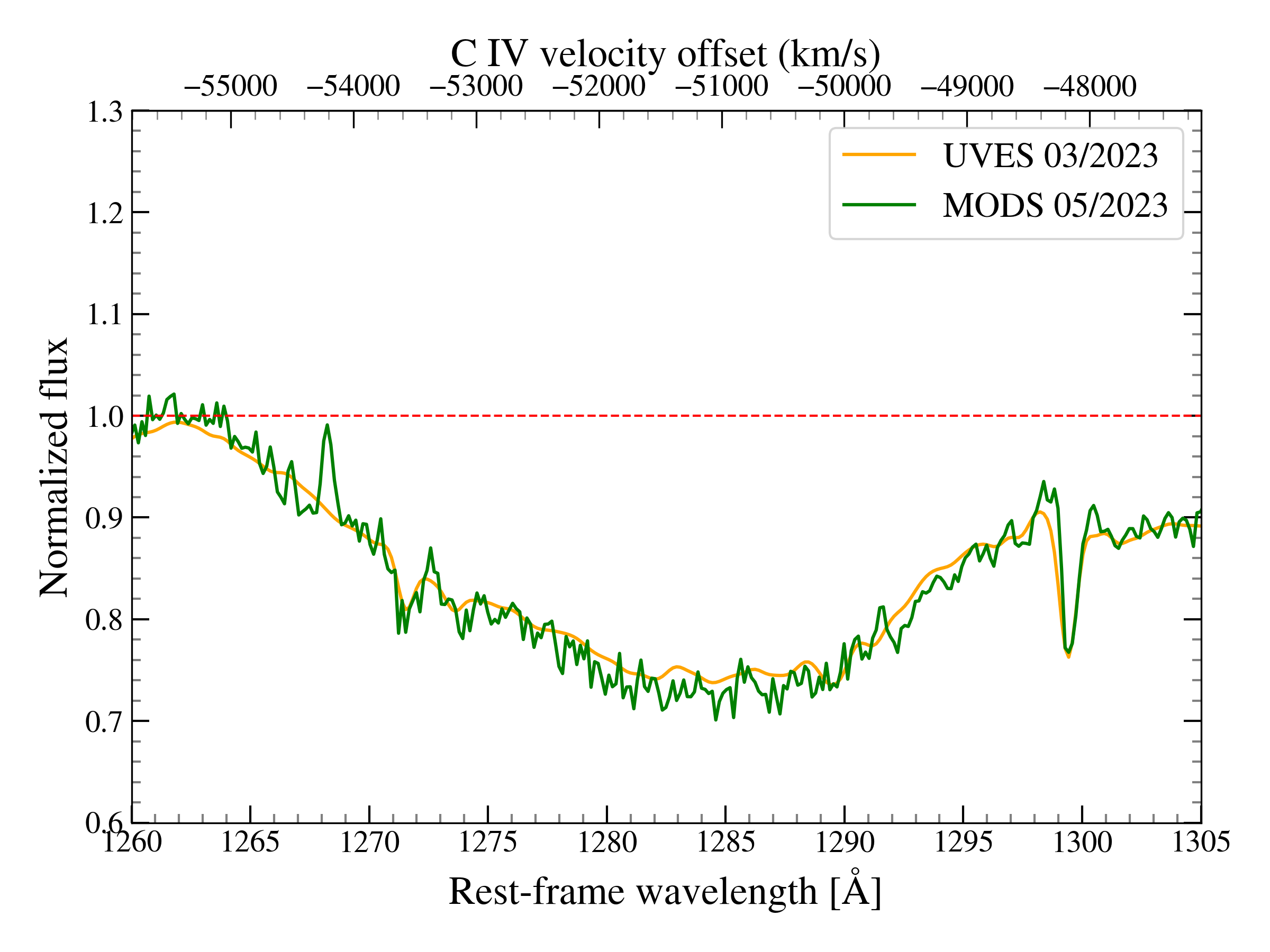}
    \includegraphics[width=0.33\linewidth]{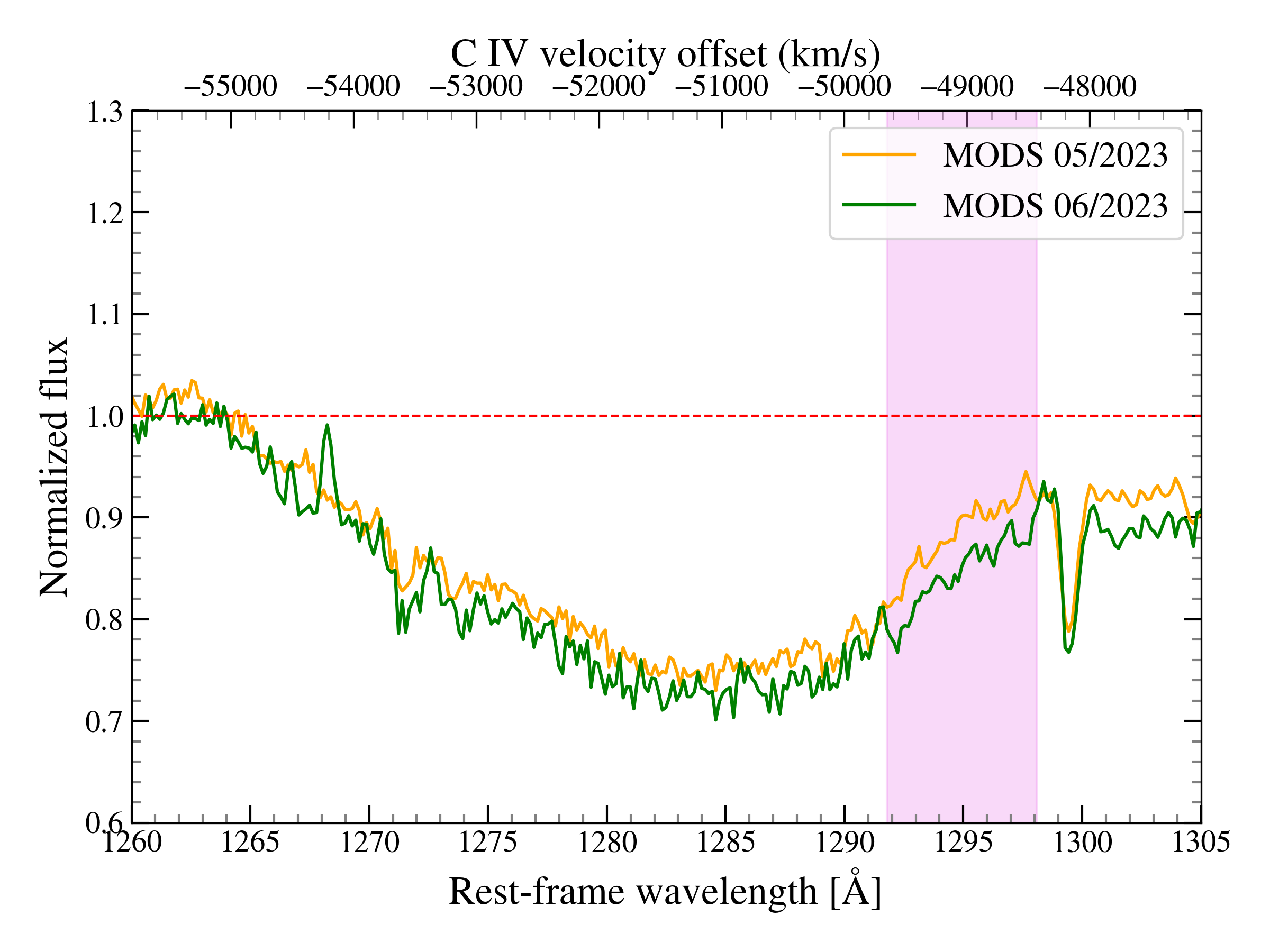}
    \includegraphics[width=0.33\linewidth]{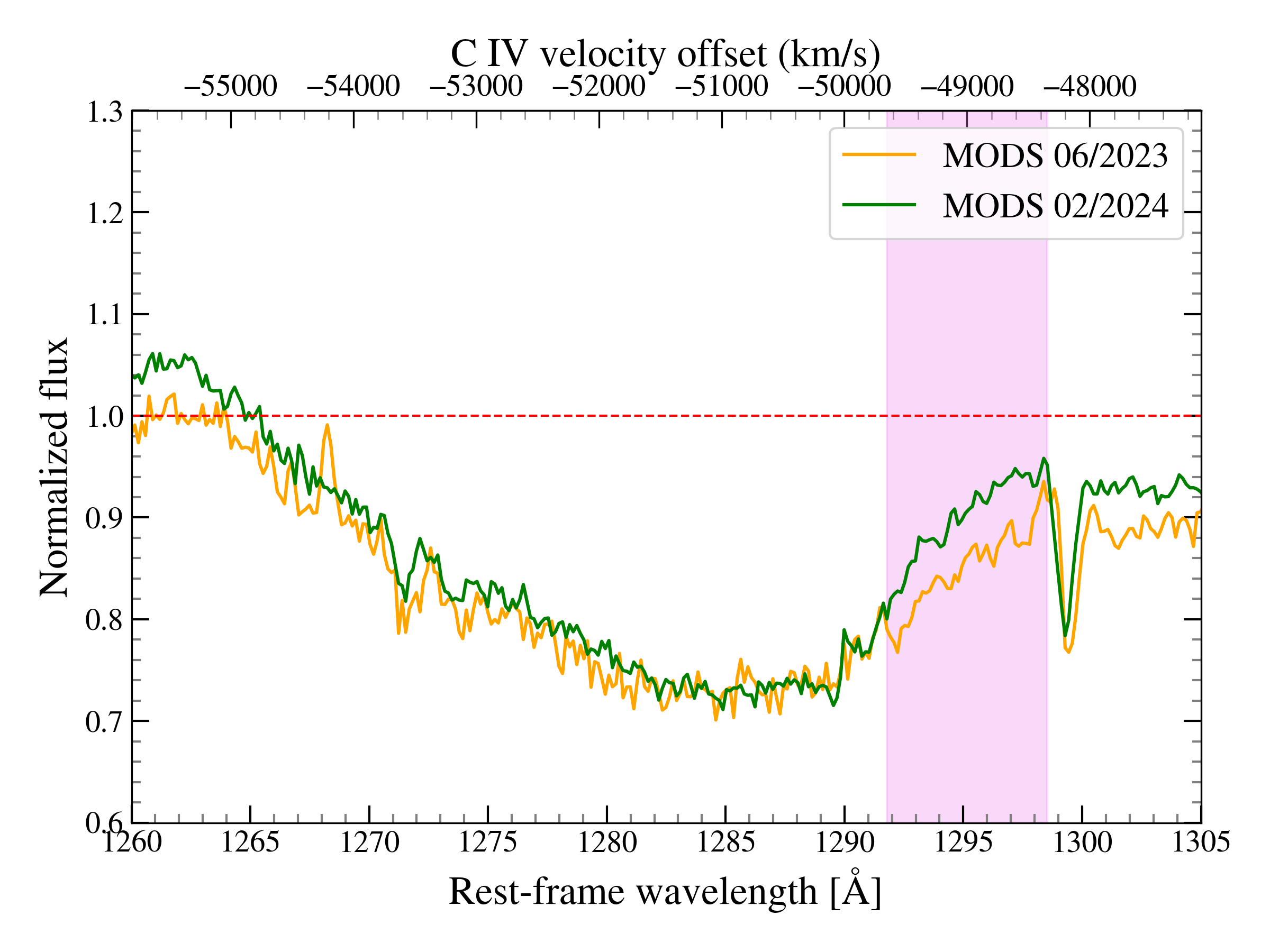}
    \includegraphics[width=0.33\linewidth]{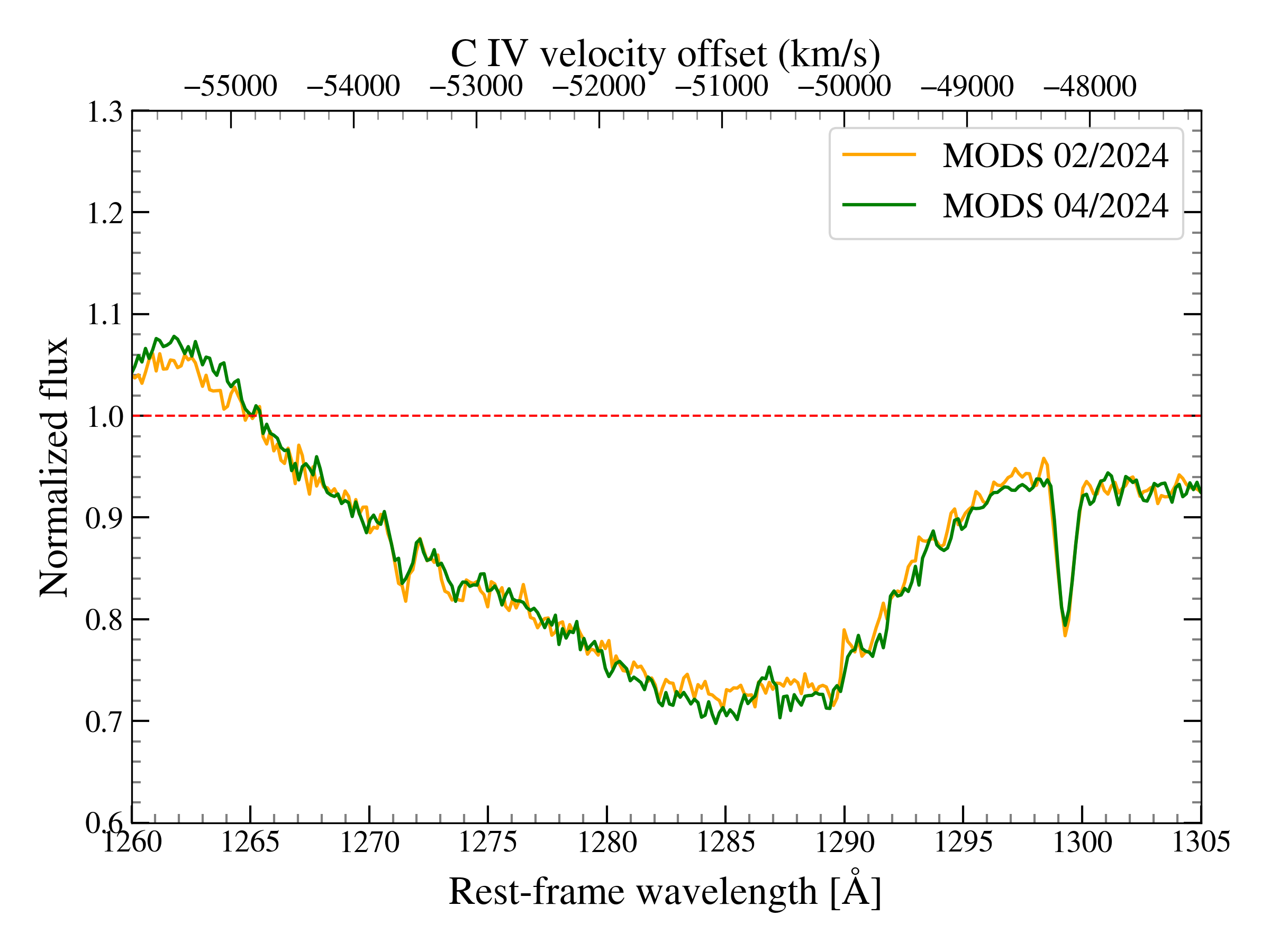}
    \includegraphics[width=0.33\linewidth]{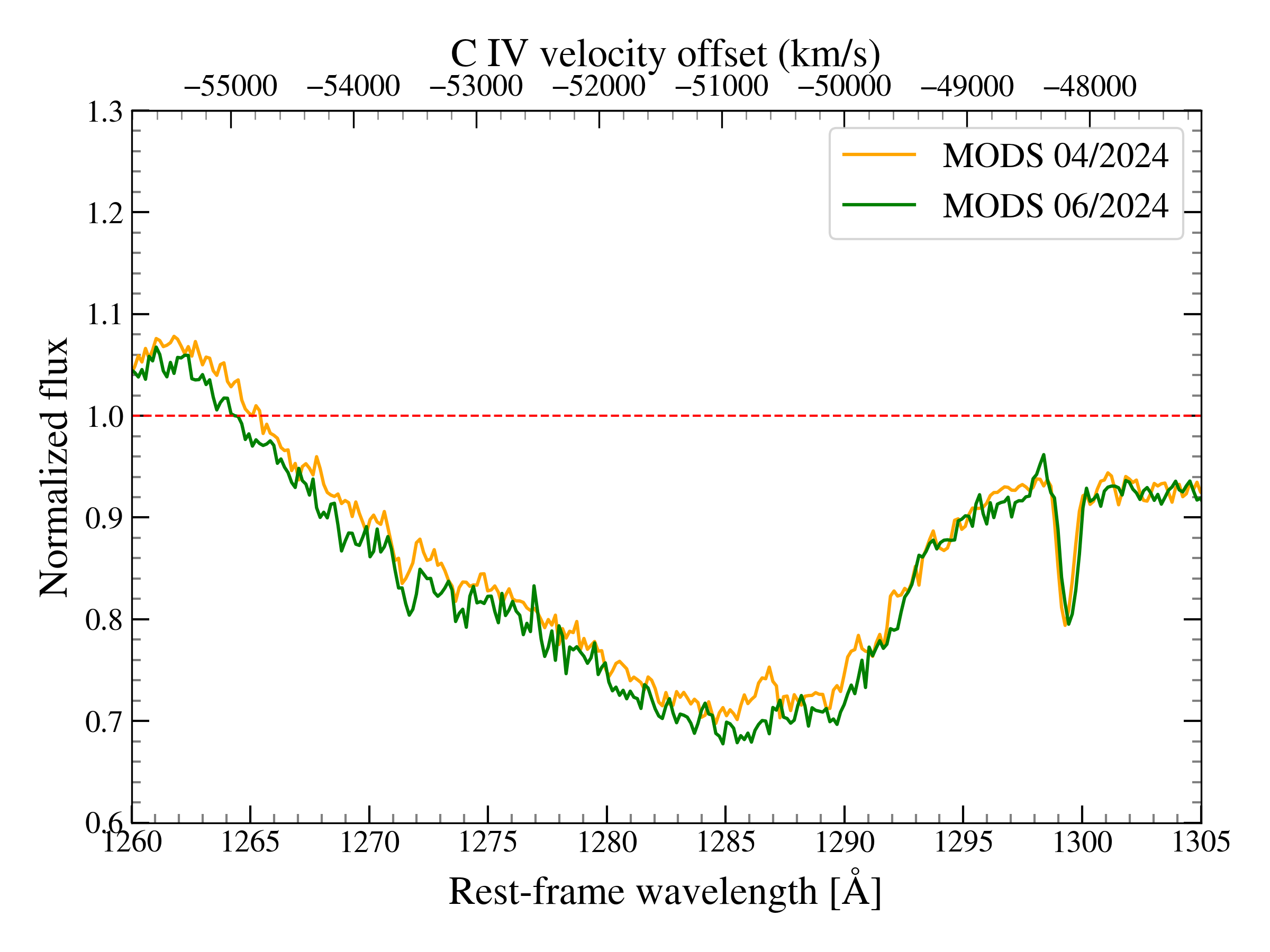}
    \caption{Same as Fig. \ref{fig:J1249_delta_as}, but for WISSH71.}
    \label{fig:J1538_delta_as}
\end{figure}

\end{appendix}

\end{document}